\documentclass{pas}

\usepackage{multirow}
\usepackage{aas-macros}
\usepackage{xcolor}
\usepackage{hyperref}
\usepackage{rotating}

\RequirePackage{etex}

\newcommand{\NII}{\mbox{[N\,\textsc{ii}]}}

\usepackage{multirow}
\usepackage{graphicx}
\usepackage[stable]{footmisc}

\defcitealias{babyk18}{B18a}
\defcitealias{lakhchaura18}{L18}
\defcitealias{donahue26}{D26}

\begin{document}

    \lefttitle{Publications of the Astronomical Society of Australia}
    \righttitle{Tamhane et al.}

    \jnlPage{1}{4}
    \jnlDoiYr{2026}
    \doival{10.1017/pasa.xxxx.xx}

    \articletitt{Research Paper}

    \title{NGC~4936: abundant multi-phase medium in the central weak X-ray cool core and optically dark H\,\textsc{i} clouds in the group}

    \author{\gn{Prathamesh} \sn{Tamhane}$^{1}$,
				 \gn{Ming} \sn{Sun}$^{1}$,
                 \gn{Hao}
                 \sn{Chen}$^{2}$,
                 \gn{Thomas}
                 \sn{Connor}$^{3}$,
                 \gn{Megan}
                 \sn{Donahue}$^{4}$,
                 \gn{Mark}
                 \sn{Voit}$^{4}$,
                 \gn{Mpati}
                 \sn{Ramatsoku}$^{5,6}$,
                 \gn{Kokoro} \sn{Hosogi}$^{1}$,
                 \gn{Tim} \sn{Edge}$^{1}$
                 }

    \affil{$^{1}$Department of Physics and Astronomy, University of Alabama in Huntsville, 301 Sparkman Drive,             Huntsville, AL 35899, USA\\
    $^{2}$Research Center for Astronomical Computing, Zhejiang Laboratory, Hangzhou 311100, China\\
    $^{3}$Center for Astrophysics $\vert$ Harvard \& Smithsonian, 60 Garden Street, Cambridge, MA 02138, USA\\
    $^{4}$Michigan State University, Physics and Astronomy Department, East Lansing, MI 48824-2320, USA\\
    $^{5}$Department of Physics and Electronics, Rhodes University, PO Box 94, Makhanda, 6140, South Africa \\
    $^{6}$INAF -- Osservatorio Astronomico di Cagliari, Via della Scienza 5, I-09047 Selargius (CA), Italy \\
        }

\corresp{P. Tamhane, Email: \href{mailto:pdt0003@uah.edu}{pdt0003@uah.edu}; M. Sun, Email: \href{mailto:ms0071@uah.edu}{ms0071@uah.edu}}


\history{(Received xx xx xxxx; revised xx xx xxxx; accepted xx xx xxxx)}

\begin{abstract}
    We present a multiwavelength analysis of NGC 4936, the central galaxy of a weak cool-core galaxy group hosting an unusually large cold/warm gas reservoir. The galaxy hosts $\sim2\times10^8$ M$_\odot$ of molecular gas in a rotating disk of radius $\sim1$ kpc. MeerKAT observations do not detect H\,\textsc{i} in NGC 4936, placing a $3\sigma$ upper limit of $1.13\times10^7$ M$_\odot$ on the atomic gas mass within a radius of $\sim 2.5$ kpc. This H$_{2}$/H\,\textsc{i} mass lower limit of $\sim$ 17, the highest known lower limit in X-ray cool cores, is consistent with predominantly molecular medium in high pressure environments of central galaxies.
    SOAR narrow-band imaging data and MUSE integral-field spectroscopy reveal ionised gas extending to at least $\sim 15$ kpc with a complex filamentary morphology that broadly follows molecular gas kinematics while extending well beyond the CO disc. Optical imaging reveals dust filaments extending to $\sim 14$ kpc including structures that appear to wrap around a companion galaxy located only $\sim 2.5$ kpc away in projection. Stellar population modelling indicates a stellar mass ratio of approximately 20:1 between NGC 4936 and the companion, while the star formation rate in NGC 4936 is $0.29$ M$_\odot$ yr$^{-1}$. Additionally, the nuclear X-ray and millimetre continuum measurements provide evidence of variability on timescales of days.
    We favour atmospheric condensation as the primary origin of the molecular gas, although externally acquired gas may also contribute.
    Our thermodynamic profiles, recalculated from previous published works using a consistent methodology reach $t_{\rm cool}/t_{\rm ff}\sim25$ in the innermost region, placing NGC~4936 at the edge of the regime commonly associated with multiphase gas condensation. The present interaction may perturb the atmosphere and promote condensation while influencing the morphology and kinematics of the multiphase medium, rather than directly supplying most of the central cold gas. NGC~4936 therefore provides an example of cold-gas accumulation in a group-central galaxy where atmospheric cooling and environmental perturbations may act together. Additionally, we detect 20 H\,\textsc{i} structures across the surrounding group environment with masses of $\sim10^7$--$10^8$ M$_\odot$; four coincide with compact blue optical sources, while ten have no obvious optical counterparts.

\end{abstract}


\maketitle

\section{Introduction}

    Galaxy groups represent a critical environment for understanding the interplay between cooling, feedback, and galaxy evolution. As the most common dark matter halos in the Universe, they contain a large fraction of baryons in hot X-ray emitting atmospheres \citep[e.g.,][]{mulchaey00,giodini09,sun12}, while also hosting a significant amount of multiphase gas and AGN activity in the low-redshift Universe \citep[e.g.,][]{osullivan18,olivares22}. The cold gas reservoirs are believed to be formed by the condensation of hot gas in the intragroup medium. However, the processes governing the condensation of hot gas into the cold phase remain poorly constrained in group-scale halos compared to massive clusters \citep{eckert21}.
    
    In the standard cooling feedback framework, the onset of multiphase gas condensation is expected to occur when radiative cooling becomes sufficiently efficient compared to the dynamical timescale, often parametrised by the cooling time ($t_{\rm cool}$) to free-fall time ($t_{\rm ff}$) ratio $t_{\rm cool}/t_{\rm ff} \lesssim 10$--20 \citep{sharma12,gaspari12,voit15a,gaspari17}. Systems with higher values are generally expected to remain stable against widespread condensation of cold gas. While the $t_{\rm cool}/t_{\rm ff}$ ratio provides a useful empirical diagnostic, local condensation may also be promoted by turbulence and active galactic nucleus (AGN) driven uplift, which can create favourable conditions for cooling \citep[e.g.,][]{gaspari17,mcnamara16}. Indeed, most well-studied cool-core clusters with extended molecular gas reservoirs lie near or below this threshold \citep{babyk18,lakhchaura18,pulido18,olivares19}. In this work, we study NGC 4936, a system with $t_{\rm cool}/t_{\rm ff} \gtrsim 25$, yet hosting abundant multiphase gas.
    
    Previous studies of NGC 4936 (R.A.: 13:04:17.126, decl.: $-$30:31:34.62) have largely focused on the hot intragroup medium and the large-scale environment of the group \citep[e.g.,][]{lovisari15,babyk18,lakhchaura18,donahue26}. These studies have established NGC 4936 as a weak X-ray group cooling core with a short gas cooling time, but having $t_{\rm cool}/t_{\rm ff} \gtrsim 25$. While in the optical, it has been described as a shell galaxy or a galaxy with a disturbed morphology \citep{malin83,wilkinson87,reid94}. However, none of the studies have explored the multiphase gas reservoir in this system. In this work, we present a detailed multiwavelength study of NGC 4936 that provides an interesting test of the standard precipitation framework.
    Despite exhibiting elevated $t_{\rm cool}/t_{\rm ff} \gtrsim 25$, the system hosts one of the most massive cold molecular gas reservoirs ($\sim 2 \times 10^8$ M$_\odot$) detected in a galaxy group environment to date. This includes a centrally concentrated molecular gas component, extended H$\alpha$ emission to at least $\sim$ 15 kpc, and large-scale dust structures that trace interactions on similar scales.

    \begin{figure*}
    \centering
        \includegraphics[width=0.9\textwidth]{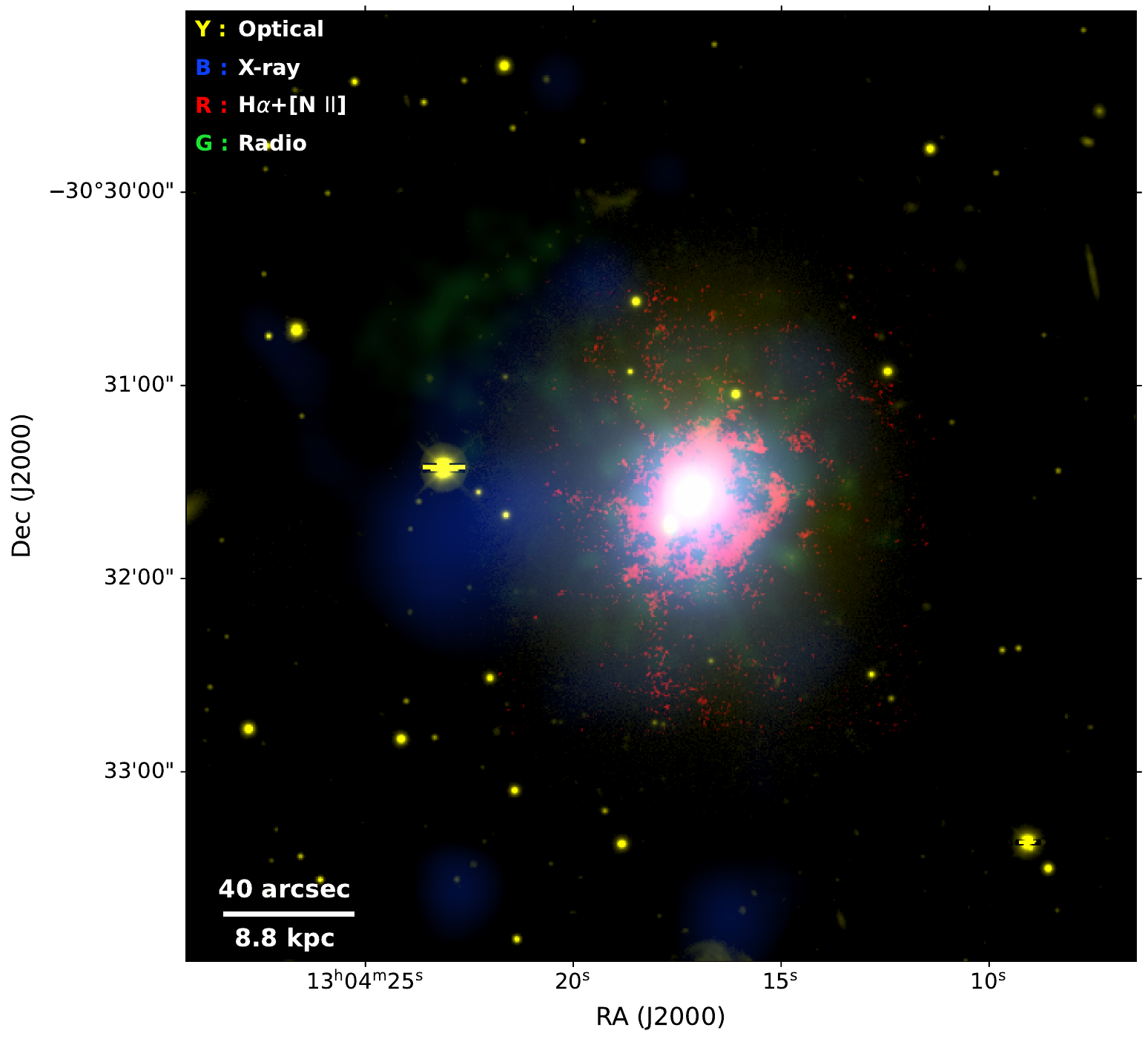}
        \caption{Colour composite image of NGC 4936 showing a multiwavelength view of the galaxy and its surroundings. The Dark Energy Camera Legacy Survey (DECaLS) $r$-band image, Chandra X-ray emission, SOAR H$\alpha$+[N~\textsc{ii}] emission, and MeerKAT radio continuum emission are shown in yellow, blue, red, and green, respectively.
        }
        \label{fig:composite}
    \end{figure*}
    
    The presence of such a substantial cold gas reservoir in a thermodynamically stable hot atmosphere raises important questions about the origin and survival of cold gas in group environments. Possible explanations include externally driven gas accretion, past or ongoing galaxy interactions, or local departures from hydrostatic equilibrium that are not captured by global thermodynamic diagnostics.
    
    In this paper, we combined Atacama Large Millimeter/submillimeter Array (ALMA) CO observations, Multi-Unit Spectroscopic Explorer (MUSE) optical spectroscopy, MeerKAT radio data and $Chandra$ data together with optical imaging from Hubble Space Telescope (HST), Southern Astrophysical Research Telescope (SOAR) and Dark Energy Camera (DEC) to investigate the morphology, kinematics, and multi-phase structure of the gas in NGC 4936. We aim to (i) characterise the distribution and dynamics of molecular gas, (ii) assess the connection between cold gas, ionised gas, and dust on galaxy-group scales, and (iii) explore whether the system represents in-situ cooling or externally acquired cold material. We also report the detection of a large number of H\,\textsc{i} clouds beyond NGC 4936, including 10 clouds that are apparently optically dark.
    
    We adopt flat $\Lambda$CDM cosmology with $H_0 = 70$ km s$^{-1}$ Mpc$^{-1}$ and $\Omega_M = 0.3$. The redshift of NGC 4936 determined from the MUSE spectrum as discussed in Section~\ref{sec:muse} is $z = 0.010742 \pm 0.000016$, corresponding to $cz = 3220 \pm 5$ km s$^{-1}$ and a luminosity distance of 46.39 Mpc. At this distance, $1'' = 220$ pc.

\section{Data Reduction and Analysis}

    \subsection{Chandra}

    NGC~4936 was observed with the \emph{Chandra} Advanced CCD Imaging Spectrometer (ACIS) on 9 and 15 February 2004 (ObsIDs 04997 and 04998). These observations were previously analysed by \citet{babyk18} (hereafter \citetalias{babyk18}), \citet{lakhchaura18} (hereafter \citetalias{lakhchaura18}), and \citet{donahue26} (hereafter \citetalias{donahue26}). We use their published deprojected temperature and density profiles to recalculate the thermodynamic properties of the hot atmosphere under a common set of assumptions, adopting the \citetalias{babyk18} profiles as our reference dataset. The resulting profiles are presented in Section~\ref{sec:xray}, while differences between the published analyses and details of our calculations are discussed in Appendix~\ref{sec:discrepancy}.
    
    We also reprocessed the archival data using Chandra Interactive Analysis of Observations (CIAO) version 4.18.0 \citep{fruscione06} and Calibration DataBase (CalDB) version 4.12.3 following standard procedures. Periods affected by background flares were identified and excluded using \textsc{deflare}, leading to a total cleaned exposure time of 26.88 ks. Exposure-corrected and background-subtracted images were generated in the 0.5--2.0 keV band for comparison with the optical and molecular gas distributions.

    \subsection{ALMA}
    
    NGC~4936 was observed with ALMA in Band~3 (Cycle~11; project 2024.1.01206.S; PI: Tamhane) in two execution blocks with the 12~m array in November 2024 and five execution blocks with the 7~m array between August and September 2025, targeting the redshifted CO(1--0) line at 114.122~GHz. The total on-source integration time was 64.92 minutes with the 12 m array and 223.8 minutes with 7 m array. The observations used a single pointing centred on the BCG nucleus with a primary beam of 65 arcsec for 12 m observations and 54.2 arcsec at 7 m observations. The 12 m observations used 41--43 antennas with baselines ranging from 15 -- 313.7 m, and the 7 m observations used 8--9 antennas with baselines from 8.9 -- 48.95 m. The frequency division correlator mode was used for the CO spectral line observations, providing a 1.875 GHz bandwidth with 976.562 kHz frequency resolution. This corresponds to a velocity resolution of 2.56 km s$^{-1}$, although the data were smoothed to coarser velocity channels for subsequent analysis. An additional three basebands with the time division correlator mode, each with a 2 GHz bandwidth and frequency resolution of 15.625 MHz, were employed for each spectral line in order to measure the sub-mm continuum.
    
    The observations were calibrated in Common Astronomy Software Applications (CASA) version 6.6.1.17 using the pipeline reduction scripts \citep{casa22}. Continuum-subtracted data cubes were created using UVCONTSUB and tCLEAN. Additional phase self-calibration did not provide a meaningful increase in the signal-to-noise of the continuum at CO(1-0). Images of the line emission were reconstructed using Briggs weighting with a robust parameter of 0.5, chosen to balance angular resolution and sensitivity. The maximum recoverable scale was 21.5$''$ and 36.5$''$ for 12 m and 7 m observations, respectively. The synthesised beams of the final CO(1-0) data cubes were $3.02 \times 2.20$ arcsec (P.A. $-82.7\deg$) and $12.43 \times 8.44$ arcsec (P.A. $-84\deg$), respectively, for 12 m and 7 m observations. The data cubes were binned to velocity resolutions of 20 km s$^{-1}$ and the RMS noise in line-free channels were 1.53 and 8.9 mJy beam$^{-1}$, for 12 m and 7 m observations.
    
    We also combined the calibrated and continuum subtracted 12 m and 7 m data visibilities and imaged it using tCLEAN with Briggs weighting with a robust parameter of 0.5. The resulting data has a synthesised beam size of $3.07 \times 2.23$ arcsec (P.A. $-82.7$ deg) and RMS noise of 1.2 mJy per channel in 20 km s$^{-1}$ channels. 
    
    \subsection{MUSE}
    \label{sec:muse}
    
    The galaxy was observed with MUSE on the very large telescope (VLT) in Wide Field Mode on 09 January 2015 in two observing blocks, with a total on-source exposure time of 2460 s (Program ID 094.A-0859, PI: Hamer). The observations cover the wavelength range $\sim$4750–9350~\AA\ at a spectral resolution of $R \sim 3000$. The data were reduced using the MUSE pipeline (version 2.8.9; \citealt{weilbacher20}).
    
    The reduced datacube was analysed using the TARDIS data analysis pipeline\footnote{\url{https://gitlab.com/francbelf/ifu-pipeline}}, which is based on the PHANGS-MUSE data analysis pipeline \citep{emsellem22} and employs the penalised pixel-fitting (pPXF) method \citep{cappellari23}. Stellar kinematics were derived after applying Voronoi binning \citep{cappellari03} to achieve a target signal-to-noise ratio of 40 per bin. The fits were performed using the MILES stellar library \citep{miles}. The pipeline also produced spatially resolved maps of stellar kinematics and ionised gas emission-line properties across the field of view.
    
    We additionally extracted a nuclear spectrum from the original calibrated datacube within a circular aperture of radius 1.5$''$ and fitted it with pPXF to determine the systemic velocity. We measure a systemic velocity of $3220 \pm 5$ km s$^{-1}$, corresponding to a redshift of $z = 0.010742 \pm 0.000016$.

    \subsection{SOAR}
    
     NGC~4936 was observed with the 4.1 m SOAR Telescope on Cerro Pachon on June 13, 2015. The observations were made with the SOAR Optical Imager (SOI). The night was photometric and seeing was $\sim 1.1''$. We took three 600 sec exposures with the CTIO 6649/76 filter (central wavelength of 6650 \AA{} and FWHM of 77 \AA{}) to cover the redshifted H$\alpha$ and \NII{} lines, and three 600 sec exposures with the CTIO 6563/75 filter (central wavelength of 6563 \AA{} and FWHM of 75 \AA{}) for the continuum. Both narrowband filters are 2 inch filters so the non-vignetted field is only $\sim 3.5'$ across. Each image was reduced using the standard procedures with the IRAF MSCRED package. The pixels were binned 2 $\times$ 2, for a scale of 0.154$''$ per pixel. Dome flats were used. The spectrophotometric standards were LTT~7379.

    \subsection{HST}
    \label{sec:hst_analysis}

    We used HST WFPC2/PC F555W image of NGC~4936 from the Hubble Legacy Archive\footnote{\url{https://hla.stsci.edu/}}. The data were obtained as part of the programme 6587 (PI: Richstone).

    To reveal the dusty features in the central regions, we used {\sc photutils} \citep{photutils} to iteratively fit and subtract elliptical isophotes of the galaxy stellar continuum following the instructions in their sample notebooks\footnote{\url{https://github.com/astropy/photutils-datasets/tree/main/notebooks/isophote}}. Two iterations were performed, with the dust features identified in the first iteration masked during the second iteration to reveal fainter structures. We also masked the companion galaxy in both iterations. The resulting residual maps are shown in Figure~\ref{fig:hst}.

    \begin{figure*}
        \centering
        \includegraphics[width=\textwidth]{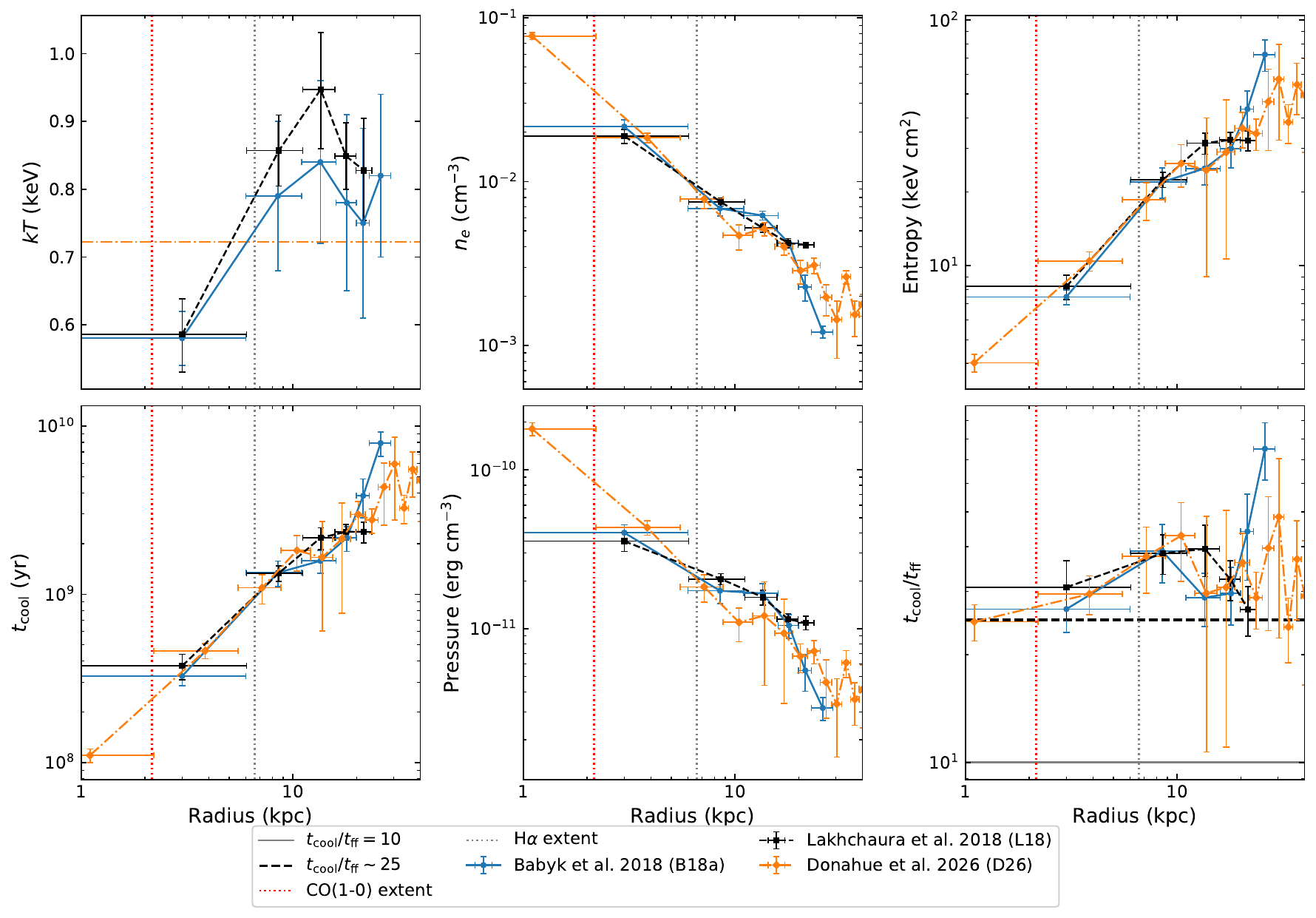}
        \vspace{-0.8cm}
        \caption{Thermodynamic radial profiles for NGC 4936 showing (from top left to bottom right) temperature ($kT$), density, entropy, cooling time ($t_{\rm cool}$), pressure and cooling time to freefall time ratio ($t_{\rm cool}/t_{\rm ff}$). The temperature and $n_e$ values were taken from \cite{babyk18} (\citetalias{babyk18}), \citet{lakhchaura18} (\citetalias{lakhchaura18}) and \citet{donahue26} (\citetalias{donahue26}) and we estimated pressure, entropy, cooling time and $t_{\rm cool}/t_{\rm ff}$ ratio as described in Section~\ref{sec:xray}. The solid grey line in the bottom right panel shows the $t_{\rm cool}/t_{\rm ff} = 10$ threshold typically adopted \citep[e.g.,][]{voit15a,sharma12} to predict onset of multiphase gas. The ashed black line shows the min($t_{\rm cool}/t_{\rm ff} \sim 25$) reported by \cite{pulido18} below which cold molecular gas is observed. The maximum extent of H$\alpha$ and CO(1-0) emission from the nucleus is shown by grey and red dotted lines in each panel, respectively.
        }
        \label{fig:profiles}
    \end{figure*}
    
    \subsection{MeerKAT}
    \label{sec:meerkat}
    
    NGC 4936 was observed with the MeerKAT array in L band for 6 hours in total on 14 January and 08 April, 2023, in the 32K mode using 60-61 antennas (PI: Sun, Project: SCI-20220822-MS-02). The MeerKAT L-band receivers cover a frequency range of 856-1712 MHz. In the 32K mode, the data are divided into 32768 channels with a channel width of 26.1 kHz which equates to roughly ${\rm 5.5\ km\ s^{-1}}$.
    
    The continuum imaging was performed with Obit using multi-term and multi-frequency synthesis using CLEAN deconvolution in 10 sub-bands spanning 0.89--1.71 GHz, centred at 1.284 GHz within the 856 MHz L-band correlator mode. The resulting image has a synthesised beam of $6.9'' \times6.9''$ and an RMS noise of 6.5 $\mu$Jy.

    To search for H\,\textsc{i} 21 cm emission/absorption associated with NGC 4936, the calibrated visibilities from the MeerKAT archive were reduced using the CARACal pipeline \citep{jozsa20a,jozsa20b} on the ilifu computer cluster hosted by the Inter-University Institute for Data Intensive Astronomy (IDIA). We did continuum subtraction, and imaged into a spectral-line datacube (with robust 0 and taper 15$''$) spanning the redshifted H\,\textsc{i} line at a velocity resolution of 5.57 km s$^{-1}$ per channel. The resulting cube has an RMS noise of 0.152 mJy per channel per beam (with the beam size of $25.1'' \times20.3''$, P.A. = $-46.9^\circ$). We searched the data cube for line emission using the Source Finding Application \citep[\textsc{SoFiA};][]{serra15,westmeier21}. Sources were identified using the Smooth+Clip algorithm, in which the cube is smoothed over a range of spatial and spectral scales and spaxels exceeding the local signal-to-noise threshold of 5 are retained. Detected spaxels were subsequently linked into three-dimensional source candidates and filtered using the \textsc{SoFiA} reliability criteria. Detected sources were visually verified and spurious sources were rejected. 3D masks were created from valid sources which were then used to construct the H\,\textsc{i} moment maps. No significant H\,\textsc{i} emission or absorption was detected in the centre of NGC 4936. However, we detected spatially extended H\,\textsc{i} emission on the outer edge of NGC 4936 comprised of multiple clouds spanning a total length of $\sim 200$ kpc. The properties of the extragalactic clouds and H\,\textsc{i} detections associated with other galaxies are discussed in Appendix~\ref{sec:blue_sources}. The results of the central region of NGC 4936 are described in Section~\ref{sec:HI}.

\section{RESULTS}

    \subsection{X-ray Properties of the Group}
    \label{sec:xray}

    Archival XMM-Newton observations analysed by \citet{lovisari15} reveal extended X-ray emission from the galaxy group, centred on NGC~4936 and extending to the edge of the XMM-Newton field of view. From these observations, they derived a system temperature of $0.85\pm0.03$ keV, a total luminosity of $(4.2\pm0.4)\times10^{42}$ erg s$^{-1}$ in the 0.1--2.4 keV band, and $M_{500}$\footnote{\label{fn:m500} $M_{500}$ is the mass enclosed within $R_{500}$, the radius within which the mean density is 500 times the critical density of the Universe at the group's redshift.} of $\sim2.1\times10^{13}$ M$_{\odot}$ (note that they used $z$=0.012).
    The corresponding $R_{500}$\footref{fn:m500} is $\sim$ 376 kpc (or 27.7$'$, adjusted for our redshift and cosmology), which is much larger than the XMM field of view and the radial range with spectral analysis. \cite{lovisari15} also noted a very flat X-ray surface brightness within the XMM field, $\beta = 0.32\pm$0.01.
    \citetalias{babyk18} derived a total mass of $M_{\rm tot}=(2.3 \pm 0.4 \pm 0.6)\times10^{12}$ M$_\odot$ and an X-ray luminosity of $L_X=(1.2\pm0.5)\times10^{41}$ erg s$^{-1}$ within a radius of 69 kpc, corresponding to five effective radii ($5r_e$) of NGC 4936. They also derived a core radius of 1.1 kpc from a $\beta$-model fit to the X-ray surface brightness profile.

    The thermodynamic properties of the hot atmosphere have previously been studied by \citetalias{babyk18}, \citetalias{lakhchaura18}, and \citetalias{donahue26}. Their temperature profiles broadly agree, and their density profiles are consistent after accounting for the possible difference in density convention discussed in Appendix~\ref{sec:discrepancy}. However, the published cooling time profiles differ appreciably. Differences in the adopted metallicities, solar abundance tables, and calculation methods complicate a direct comparison. To obtain an independent estimate of the central gas metallicity, we therefore extracted and jointly fitted spectra from both \emph{Chandra} ObsIDs within a projected radius of 10 kpc.

    We extracted spectra in the 0.5--6 keV energy range range from both \emph{Chandra} ObsIDs within a circular aperture of radius 10 kpc and fitted them jointly in \textsc{xspec} version 12.15.1, using AtomDB version 3.1.3 and data calibrated with CALDB version 4.12.3. We removed point sources using {\tt wavdetect} before spectral extraction. Following \citetalias{babyk18}, we adopted the model \texttt{phabs(apec+powerlaw+mekal+powerlaw)}, in which the \texttt{apec} component represents the hot gas and the additional components account for unresolved stellar X-ray emission. The parameters of the stellar components were fixed to the values given in \citetalias{babyk18} except for normalisations, which were kept free. The Galactic hydrogen column density ($N_{\rm H}$) was fixed to $5.91\times10^{20}$ cm$^{-2}$, as adopted by \citetalias{babyk18} and close to the value of $5.97\times10^{20}$ cm$^{-2}$ used by \citetalias{donahue26}. We could not find the adopted $N_{\rm H}$ value in \citetalias{lakhchaura18}. Adopting the solar abundance table of \citet{anders89}, we obtained a temperature of $0.92^{+0.15}_{-0.11}$ keV and a metallicity of $0.20^{+0.37}_{-0.12}$ Z$_\odot$ consistent with the value of $0.22\pm0.09$ derived by \citetalias{babyk18} in a larger aperture. The unabsorbed rest frame 0.01--100 keV luminosity of the hot gas component is $L_{\rm X}=6.42^{+0.40}_{-0.39}\times10^{40}$ erg s$^{-1}$.

    We also estimated the gas mass from the best-fitting \texttt{apec} normalisation, assuming that the emitting gas uniformly fills a sphere of radius 10 kpc. Using $n_{\rm H}=0.83n_e$, the normalisation gives the electron density through
    \begin{equation}
    {\rm norm} =
    \frac{10^{-14}}{4\pi[D_A(1+z)]^2}\,
    0.83n_e^2V,
    \end{equation}
    where $D_A$ is the angular diameter distance and $V$ is the assumed emitting volume. The gas mass is then $M_{\rm gas}=\mu_e m_p n_eV$, where $\mu_e = 1.16$ is the mean mass per electron. We obtained $M_{\rm gas}=7.70^{+0.04}_{-0.03}\times10^8$ M$_\odot$ and an aperture-averaged cooling time of $t_{\rm c,10kpc}=1.68^{+0.08}_{-0.09}\times10^9$ yr. These imply a nominal cooling rate of $\dot{M}_{\rm Xray}=M_{\rm gas}/t_{\rm c,10kpc}=0.46^{+0.05}_{-0.04}$ M$_\odot$ yr$^{-1}$.
    
    We then recalculated the pressure, entropy, cooling time, and $t_{\rm cool}/t_{\rm ff}$ profiles from the published deprojected temperature and electron density profiles, using common definitions and density conversions. For the cooling time calculation, we adopted a spatially constant metallicity of $0.20$ Z$_\odot$, as measured from our aperture spectrum, and evaluated the bolometric cooling function using the same abundance table and atomic database as in our spectral analysis which was also used by \citetalias{babyk18} and \citetalias{donahue26}. \citetalias{lakhchaura18}, however, used \cite{grevesse98} abundance table. This places the derived profiles on a common basis, although the published densities retain their dependence on the assumptions adopted in the original spectral fits. The calculation methods and comparison with the published profiles are presented in Appendix~\ref{sec:discrepancy}. Figure~\ref{fig:profiles} shows the consistently recalculated profiles.
    
    We estimated the free-fall time assuming an isothermal gravitational potential, for which $t_{\rm ff}=r/\sigma_\star$. From the stellar velocity-dispersion map in Figure~\ref{fig:stellar}, we measured $\sigma_\star=241\pm17$ km s$^{-1}$, excluding the companion and the surrounding region of enhanced dispersion. The recalculated profiles reveal a cool central atmosphere, with temperatures of $\sim0.7$ keV in the inner 10 kpc, a minimum cooling time of $\sim10^8$ yr, and a minimum entropy of $\sim10$ keV cm$^2$. The central $t_{\rm cool}/t_{\rm ff}$ ratios reach $\sim25$. For comparison with previous studies, we also estimated unabsorbed rest-frame 0.2-5 keV luminosity from a spectrum extracted from a radius of 20.5 kpc corresponding to deprojected cooling time of 3 Gyr and fitted in the same way as described above. We derive a luminosity of $L_{\rm Xray} = (9.5 \pm 0.3) \times 10^{40}$ erg s$^{-1}$, $\sim10$ times lower than the luminosity of NGC 5044 in a slightly larger aperture in the same energy band \citep{babyk18} and also lower than the brightest groups in the \cite{sun09} sample.
    
    We also produced a background-subtracted 0.5--2~keV \emph{Chandra} image of the group, shown in the left panel of Figure~\ref{fig:soar}. The image reveals a bright X-ray core within the central $\sim 1$ kpc surrounded by more extended diffuse emission. A faint filamentary feature extends to the east and may trace either a cooling filament or a disturbance associated with a past merger event. However, the relatively shallow \emph{Chandra} data do not allow a definitive interpretation. No clear X-ray cavities, shocks, or sloshing structures are detected in the current observations.

    \begin{figure*}
        \centering
        \includegraphics[width=\textwidth]{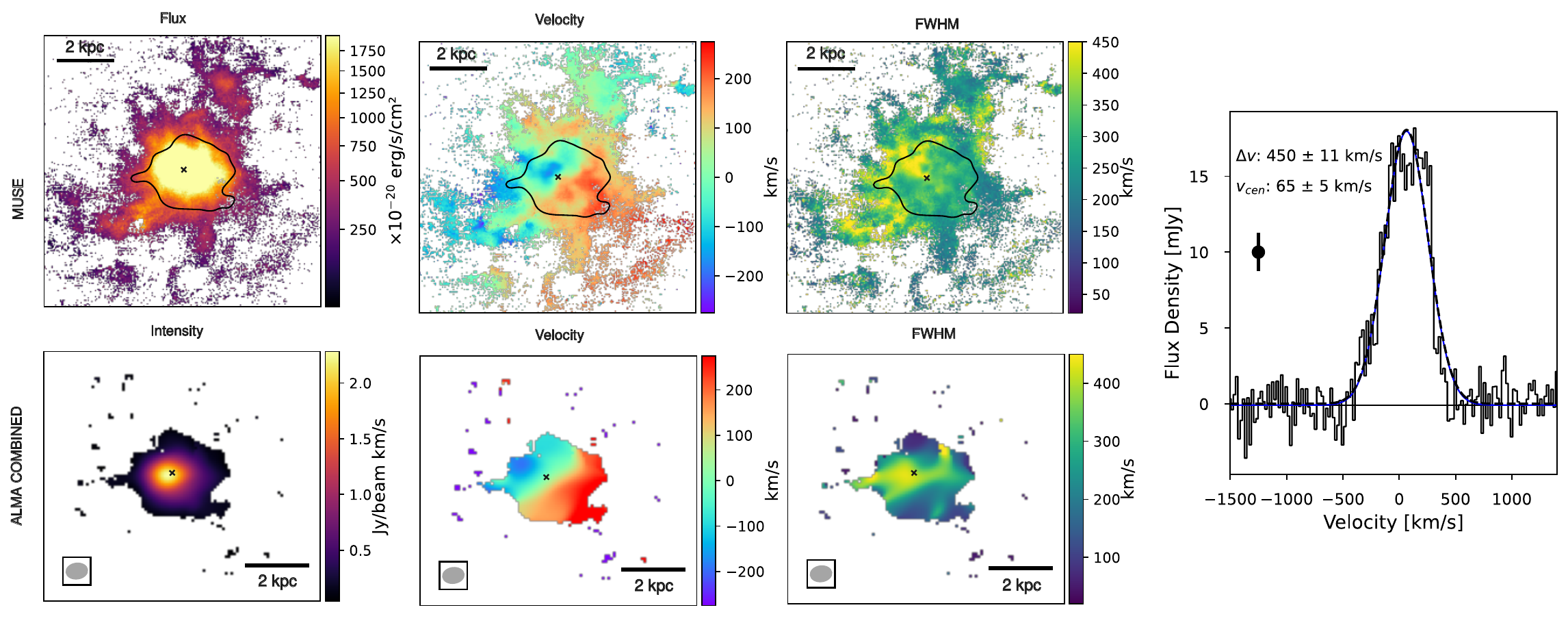}
        \caption{Moment maps and integrated spectra of NGC 4936. The rows show the MUSE H$\alpha$ (top) and combined ALMA 12 m$+$7 m (bottom)  observations. The columns show (from left to right) the integrated intensity (or H$\alpha$ line flux), velocity field and FWHM. The far right panel shows the CO(1-0) integrated spectrum, with the black point with errorbar representing  average error. The spectrum is fitted with Gaussian profile, and the best-fitting central velocity ($v_{\rm cen}$) and FWHM ($\Delta v$) are indicated. MUSE H$\alpha$ flux, velocity and FWHM maps show spaxels with $>5 \sigma$ detection. The black contours in MUSE maps show the outline of the molecular gas emission traced by CO(1-0) in the combined image. The synthesised beam is shown in the lower-left corner of each ALMA map. The `X' marks the location of the central continuum source. North is up and east is to the left.
        }
        \label{fig:alma}
    \end{figure*}

    \subsection{Molecular Gas Distribution and Kinematics}
    \label{sec:co_kin}

    All of the molecular gas detected in NGC 4936 is concentrated within the central $\sim2$ kpc surrounding the nucleus and is distributed asymmetrically. Figure~\ref{fig:alma} shows the molecular gas distribution and kinematics derived from the 12 m, and 12 m + 7 m combined data cubes. The CO(1--0) emission peaks at the position of the central continuum source and extends $\sim2$ kpc toward the southwest and $\sim0.7$--0.9 kpc toward the west-northwest. The molecular gas is spatially coincident with the dust features observed in the HST images shown in Figure~\ref{fig:hst}.
    
    The velocity field is broadly consistent with rotational motion about the nucleus, with line-of-sight velocities ranging from $-200$ to $200$ km s$^{-1}$. However, an extended redshifted component is present west of the nucleus, reaching velocities of up to $\sim300$ km s$^{-1}$. The molecular gas line widths vary across the system, with FWHM values of $\sim80$ km s$^{-1}$ north of the nucleus and $\sim110$ km s$^{-1}$ in the western regions. The nuclear gas exhibits a substantially broader line width, with a FWHM of 434 km s$^{-1}$. A single gaussian fit to the total spectrum gives a FWHM of $450 \pm 11$ km s$^{-1}$ for the combined datacube.
    
    \subsubsection{Rotating Disk Modelling}
    \label{sec:rotation}

    \begin{figure*}
        \centering
        \includegraphics[width=0.8\textwidth]{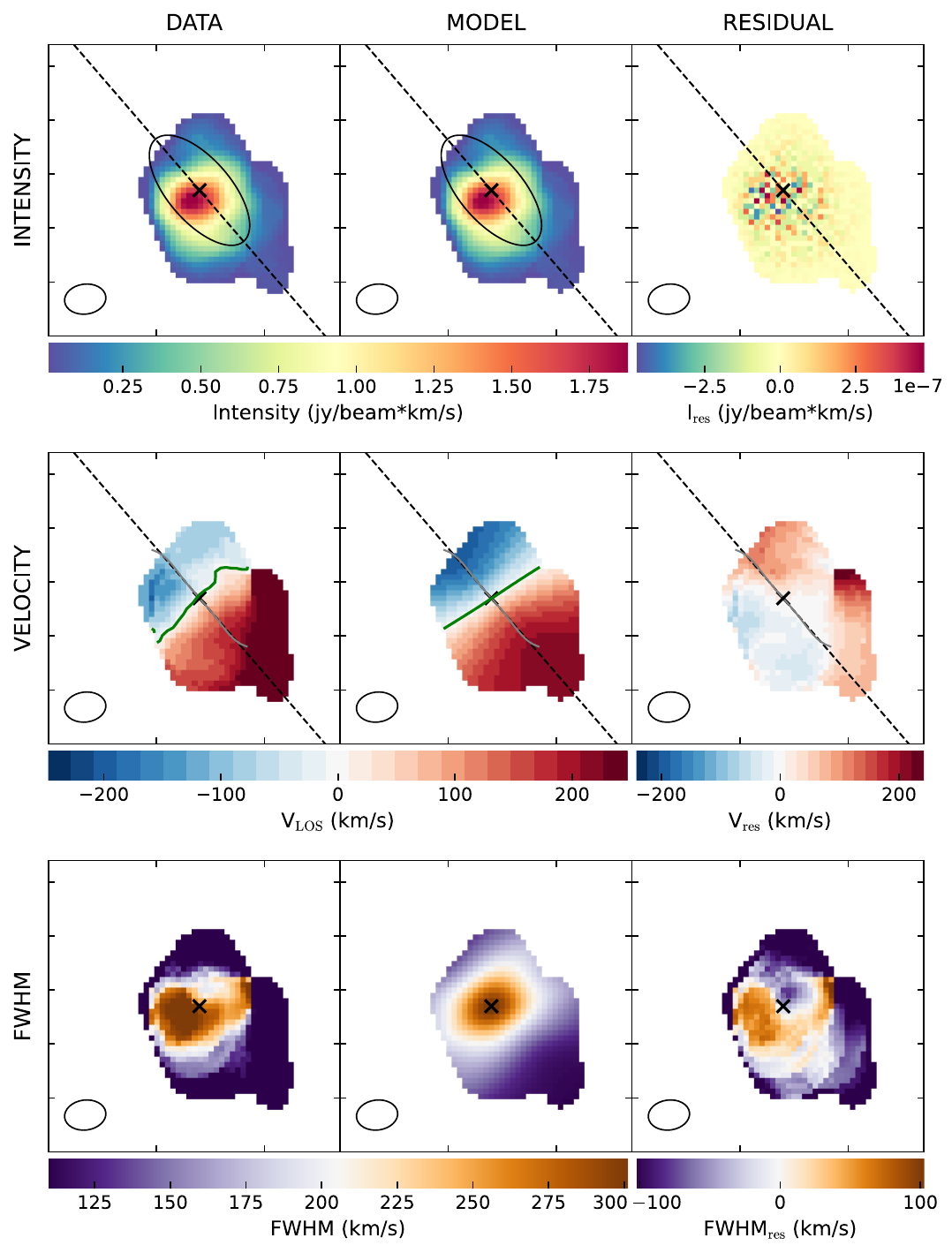}
        \caption{Moment maps, disk model, and residuals generated by the {\sc 3D-Barolo} fit to the combined data cube. The left, middle, and right columns show the data, model, and residuals, respectively. The black dashed line indicates the fitted kinematic major axis, the black ellipse marks the outermost fitted ring, and the green line traces the systemic-velocity contour. The `X' denotes the position of the sub-mm continuum peak.}
        \label{fig:Bbarolo}
    \end{figure*}
    
    We modelled the molecular gas kinematics using {\sc 3D-Barolo}, which fits three-dimensional tilted-ring models directly to emission-line data cubes \citep{di15}. The combined data cube was modelled using 13 rings with a radial spacing of 0.4$''$. The best-fitting model and residual velocity field are shown in Figure~\ref{fig:Bbarolo}. The best-fitting model yields an inclination of $61.8\pm1.6^\circ$ and a position angle of $221.1\pm3.4^\circ$. The molecular disk extends to a radius of $\sim1.1$ kpc and has a rotational velocity of $\sim200$ km s$^{-1}$. The velocity dispersion decreases from $\sim90$ km s$^{-1}$ in the central regions to $\sim43$ km s$^{-1}$ at larger radii. The FWHM residuals are strongly asymmetric, with local deviations of up to $\sim 100$ km s$^{-1}$ from the model. This indicates that the line width distribution is not fully described by the axisymmetric tilted-ring model and likely reflects localised non-circular motions and/or unresolved kinematic components.
    
    The residual velocity map reveals a coherent north-south filamentary structure west of the nucleus that spatially coincides with a linear dust feature seen in the HST residual image (see Figure~\ref{fig:hst}). The molecular and ionised gas show a common rotational pattern, suggesting that the different phases are dynamically coupled. The asymmetric swirl-like dust morphology, with more extinction south of the nucleus, together with the orientation of the molecular disk, may indicate that the southern side of the disk is closer to line-of-sight. Under this geometry, the observed residual velocities are consistent with radial motions directed toward the nucleus, suggesting that the filament may represent inflowing material. However, the available data do not uniquely distinguish inflow from other non-circular motions, such as streaming motions or tidal perturbations.

    \subsection{Molecular Gas Mass}
    \label{sec:mass_co}

    We estimated the total molecular gas mass from the CO(1--0) line. The integrated CO(1--0) fluxes measured from the 12 m array, 7 m array, and combined data cube are $8.04 \pm 0.16$, $8.08 \pm 0.60$, and $8.65 \pm 0.21$ Jy km s$^{-1}$, respectively. Although the fluxes measured from the 12 m and 7 m arrays are comparable, combining the two data sets recovers $\sim8\%$ more flux. Given the near-solar metallicity of the hot gas and the lack of evidence for starburst-like conditions, we adopted a Galactic CO-to-H$_2$ conversion factor of $X_{\rm CO,Gal}=2\times10^{20}$ cm$^{-2}$ (K km s$^{-1}$)$^{-1}$, which is commonly assumed for molecular gas in nearby massive galaxies and cool-core systems (e.g., \citealt{edge01}; \citealt{russell19}). We then converted the integrated CO(1--0) flux into molecular gas mass using the relation

    \begin{equation} \begin{split} M_{\rm mol} = & 1.05 \times 10^{4} \Bigg(\frac{X_{\rm CO}}{X_{\rm CO,Gal}}\Bigg) \bigg(\frac{1}{1+z}\bigg)\\ & \bigg(\frac{S_{\rm CO} \Delta v}{\rm Jy\, km\, s^{-1}}\bigg) \bigg(\frac{D_{\rm L}}{\rm Mpc}\bigg)^{2}\, {\rm M_{\odot}}, \end{split} \end{equation}
    
    where $S_{\rm CO}\Delta v$ is the integrated CO(1--0) flux, $D_{\rm L}$ is the luminosity distance, and $z$ is the galaxy redshift \citep{solomon05,bolatto13}.
    
    The resulting molecular gas masses are $(1.80\pm0.04)\times10^8 \, {\rm M_\odot}$, $(1.81\pm0.13)\times10^8 \, {\rm M_\odot}$, and $(1.93\pm0.05)\times10^8 \, {\rm M_\odot}$ for the 12 m, 7 m, and combined data cubes, respectively. This corresponds to an average molecular hydrogen column density of $N_{\rm H_2} \approx 1.33 \times 10^{21}$ cm$^{-2}$ for an emission region of 133.7$''^2$ and $\mu_{\rm H_2} = 2.8$ \citep{kauffman08}.

    \subsection{Ionised Gas Distribution and Kinematics}
    \label{sec:ha_kin}

    \begin{figure*}
        \centering
        \includegraphics[width=\textwidth]{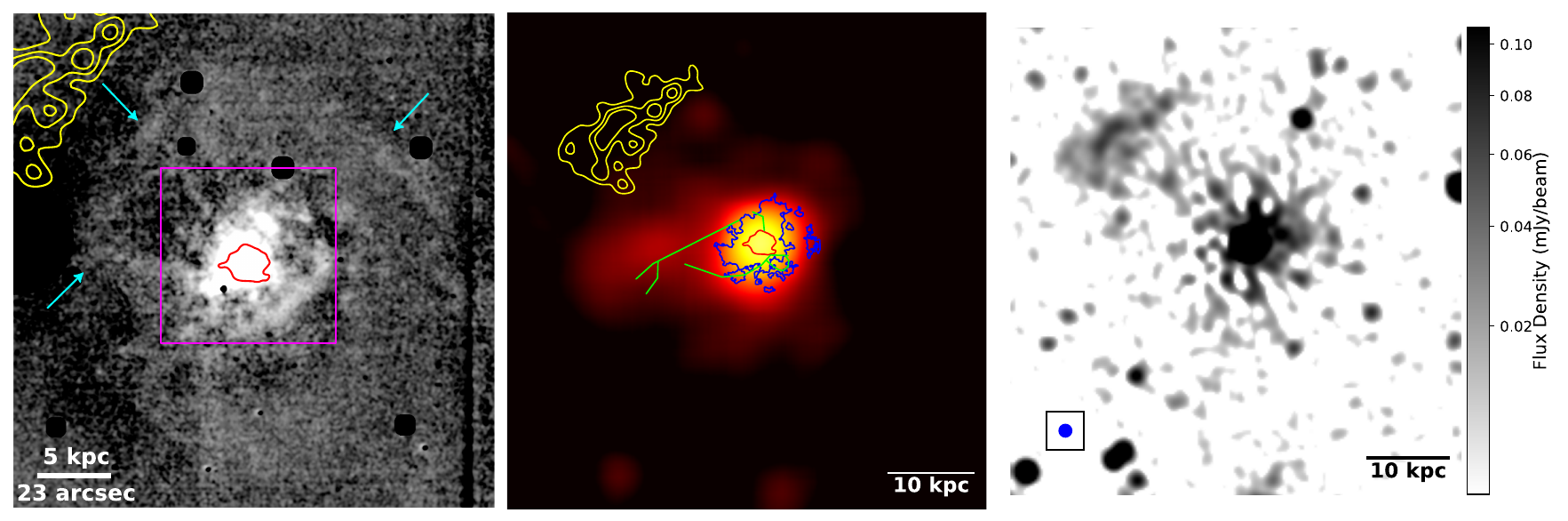}
        \caption{$Left~panel$: the net H$\alpha +$[N~\textsc{ii}] emission map of NGC 4936 derived from the SOAR narrow-band images. The map was produced by subtracting the scaled continuum image from the narrow-band image. Bright stars are masked. Cyan arrows indicate filamentary structures extending to projected distances of $\sim 15$ kpc, beyond the MUSE coverage, shown by the magenta square. Red contour indicates the extent of molecular gas traced by CO from combined ALMA datacube. The yellow contours show the extended ratio emission to the north east. Point sources are masked for better visibility. $Middle~panel$: Chandra 0.5--2 keV image of the NGC 4936 group smoothed by a gaussian kernel with a $\sigma=10''$. The blue contours show extent of the H$\alpha$ emission at 1.5$\times 10^{-18}$ erg s$^{-1}$ cm$^{-2}$ level and red region shows the CO(1-0) emission as shown in Figure~\ref{fig:alma}. The yellow contours show radio halo-like emission detected in the MeerKAT image with contours at 15, 27.96 and 40.93 $\mu$Jy beam$^{-1}$. The green lines show the rough outline and the extent of dusty filaments. $Right~panel$: MeerKAT 1.28 GHz image of the NGC 4936 group, in the same region as the central panel. The central galaxy has a nuclear continuum source but shows an extended radio halo-like emission to the NE. The beam size is shown by the blue ellipse. North is up and east is to the left in all panels.
        }
        \label{fig:soar}
    \end{figure*}

    \begin{figure*}
        \centering
        \includegraphics[width=\textwidth]{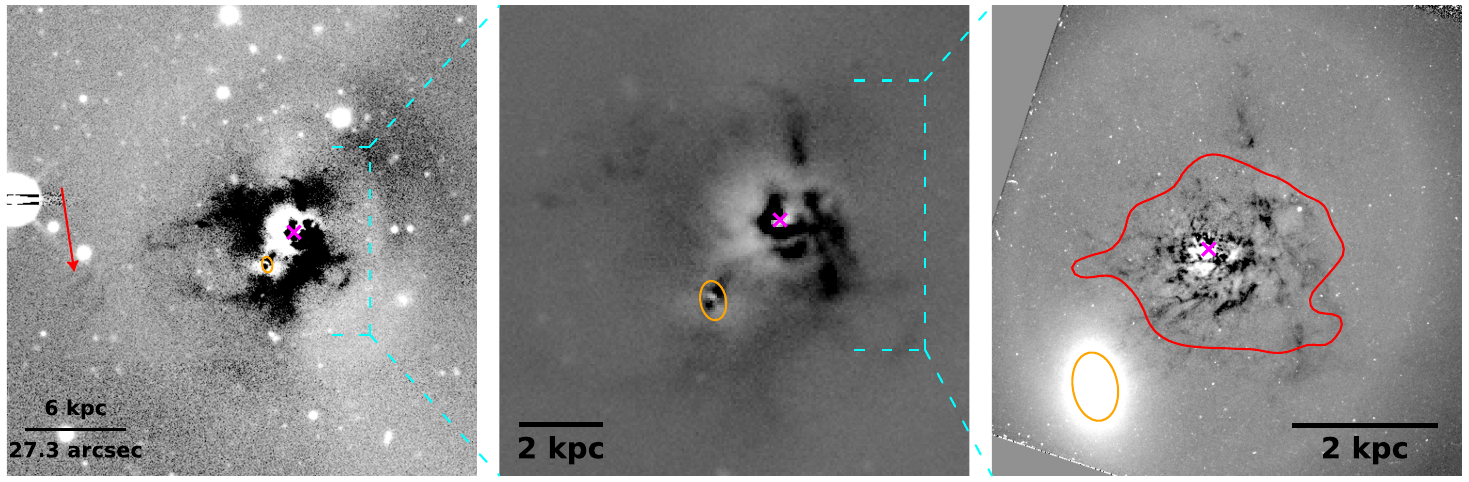}
        \vspace{-0.7cm}
        \caption{$Left~panel$: Stellar continuum subtracted DECaLS $g$-band image displayed with a stretch chosen to emphasize faint, filamentary dust structures extending to projected radii of $\sim14$ kpc from the nucleus, indicated by the red arrow. Darker regions correspond to negative residuals relative to the smooth stellar model and trace the locations of prominent dust absorption features. The cyan box outlines the field of view shown in the middle panel. $Middle~panel$: Zoomed view of the central region shown using a different image stretch to highlight the inner dust morphology. The cyan box indicates the field of view of the right panel. $Right~panel$: Stellar continuum subtracted HST F555W image showing dust seen in absorption against the stellar continuum. The red contour outlines the molecular gas emission traced by CO(1--0) from the combined image shown in Figure~\ref{fig:alma}.  In all images, `X' denotes the nucleus of NGC 4936 and the orange ellipse denotes the position of the companion galaxy. North is up and east is to the left.
        }
        \label{fig:hst}
    \end{figure*}
    
    Figure~\ref{fig:soar} shows the morphology of H$\alpha$+[N~\textsc{ii}] gas, with the narrow-band imaging data from SOAR. 
    The ionised gas is mostly concentrated in the inner $\sim 6$ kpc from the centre. However, there are some faint filaments extending $\sim 15$ kpc east, north and possibly to the south. The ionised gas extends $\sim15$ kpc from the nucleus in the SOAR map and is significantly more extended than the molecular gas. The right column of Figure~\ref{fig:alma} shows the H$\alpha$ flux, velocity and FWHM maps of NGC 4936 derived from the MUSE datacube. The total H$\alpha$ flux estimated from the MUSE map is $(1.16 \pm 0.02) \times 10^{-13}$ erg s$^{-1}$ cm$^{-2}$ corresponding to an H$\alpha$ luminosity of $(3.00 \pm 0.06) \times 10^{40}$ erg s$^{-1}$, uncorrected for intrinsic extinction. We estimate the contribution from H$\alpha$ emission beyond the MUSE field of view by comparing the SOAR H$\alpha$+[N\,\textsc{ii}] flux within the MUSE footprint with that measured over the full SOAR field, after masking bright stars and negative pixels. Assuming that the [N\,\textsc{ii}]/H$\alpha$ ratio is similar inside and outside the MUSE field, this scaling suggests that the total H$\alpha$ flux could be larger by 26\% than that measured within the MUSE field.
    More than $\sim80$ per cent of this flux originates within the central 0.6 kpc radius region around the nucleus. The surface brightness of the extended filamentary emission within the MUSE FOV is 10--30 times lower than in the central region and for the filaments outside the FOV of MUSE, it is up to 100 times lower than in the central $\sim1$ kpc region.

    The velocity structure in the inner region is similar to that of the molecular gas, with rotational velocities ranging from $-200$ to $200$ km s$^{-1}$. The eastern filamentary extension shows blueshifted velocities ranging from $-200$ to $-100$ km s$^{-1}$, while the western filamentary emission is redshifted with velocities between 200 and 300 km s$^{-1}$ (see the middle-right panel of Figure~\ref{fig:alma}). The northern filament has velocities closer to the systemic value, between 0 and 50 km s$^{-1}$. The FWHM of the eastern filament is high, at 400--500 km s$^{-1}$, while the western and northern filaments have FWHM of $\sim250$ km s$^{-1}$. The central region has FWHM in the range 250--400 km s$^{-1}$ (see the bottom-right panel of Figure~\ref{fig:alma}). The FWHM are generally elevated indicating multiple velocity components, unresolved velocity structure or shocks. Given the similar large-scale morphology and kinematics of the ionised and molecular gas in the central region, they likely trace a common origin and reservoir of cold gas.

    The BPT diagnostic diagrams presented in Figure~\ref{fig:bpt} show that most of the ionised gas has line ratios consistent with LINER-like excitation, with a smaller fraction of spaxels showing AGN-like ionisation. The LINER-like emission is distributed across both the central region and the extended filaments. The presence of LINER-like emission in the extended structures suggests that the ionised gas is likely associated with low-ionisation processes in the diffuse medium rather than purely nuclear activity or shocks.

    \subsubsection{Gas phase metallicity}
    We extracted the spectrum from a 5$''$ radius aperture around the nucleus and fitted it with pPXF to subtract the stellar continuum and obtain emission line fluxes. Since the region enclosed has predominantly LINER-like ionisation (see Figure~\ref{fig:bpt}), we used the calibration from \citet{kumari19} for LINER like systems to estimate the Oxygen abundance using
    \begin{equation}
        12+{\rm log(O/H)} = 7.673+0.22 \sqrt{25.25 - 9.073\, {\rm O3N2 + 0.127 \, O3}},
    \end{equation}
    where, O3N2 = log([O\,\textsc{iii}]$\lambda 5007/$H$\beta)\,-\,$log([N\,\textsc{ii}]$\lambda$6583/H$\alpha$) and O3 = log([O\,\textsc{iii}]$\lambda 5007/$H$\beta$). Adopting the Solar oxygen abundance $12+\log({\rm O/H})_\odot=8.69$ from \citet{asplund09}, this corresponds to (O/H)/(O/H)$_\odot$=1.15.

    \subsection{Dust}
    \label{sec:dust}

    To reveal the dust structures in NGC 4936, we produced a stellar continuum subtracted HST F555W image as described in Section~\ref{sec:hst_analysis}. We also obtained a Dark Energy Camera Legacy Survey (DECaLS) $g$-band image and subtracted the stellar continuum using \textsc{galfit} by simultaneously modelling NGC 4936 and its companion galaxy. The resulting images are presented in Figure~\ref{fig:hst}.

    The left panel of Figure~\ref{fig:hst} shows the stellar continuum subtracted DECaLS $g$-band image. It reveals filamentary dust structures that extend to projected radii of $\sim14$ kpc eastward from NGC 4936 and wrap around the northern and southern sides of the companion galaxy. This large-scale morphology suggests a possible physical connection between the two galaxies, potentially indicating that the dust lanes are associated with a past or ongoing interaction. Such structures could arise from cooling and condensation of gas in tidally or shock-compressed regions, or from the stripping of dust-rich material from the companion galaxy. The eastern dust filament is also roughly co-spatial with the extended X-ray emission shown in Figure~\ref{fig:soar}, suggesting a possible connection between the cold and hot gas phases. The middle panel shows the dust morphology within the central 5--6 kpc as seen in the DECaLS image, while the right panel shows the higher-resolution HST view. The broad dust lanes apparent in the DECaLS image are resolved by HST into a complex, clumpy network exhibiting a swirl-like morphology around the galaxy centre that is roughly co-spatial with the molecular gas distribution traced by CO. In addition, a north--south dust filament appears to connect to the large-scale filamentary structures visible in the left panel.
    
\begin{figure}
        \centering
        \includegraphics[width=0.48\textwidth]{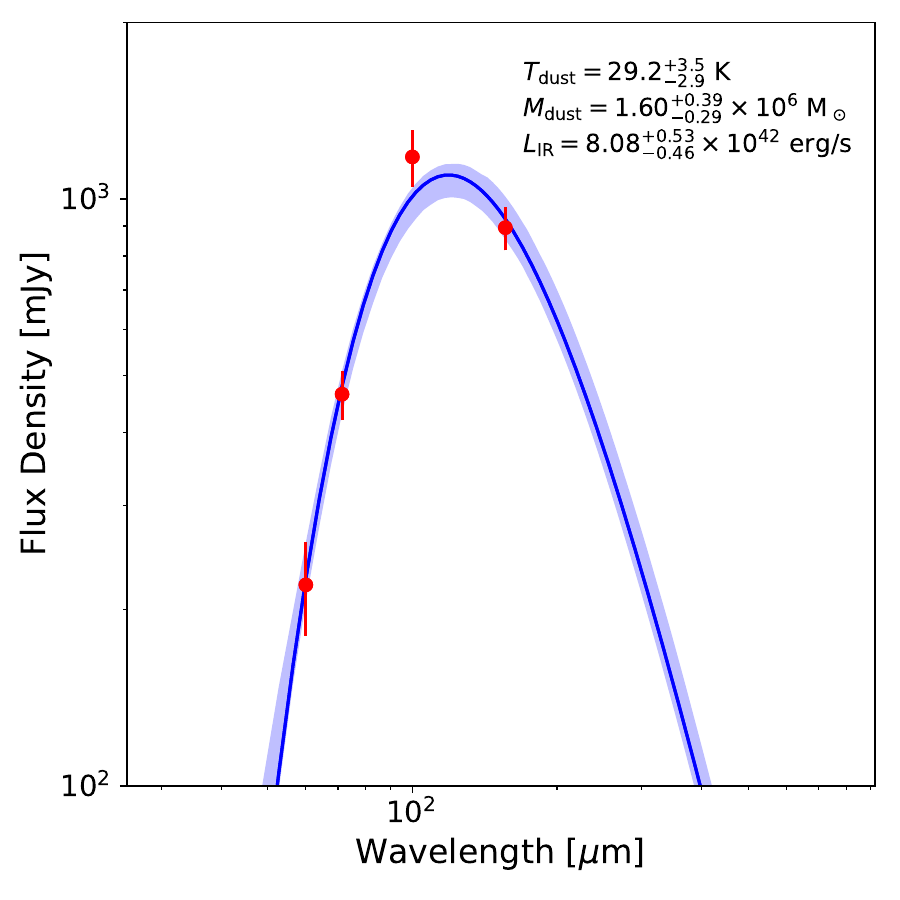}
        \caption{The $Spitzer$ IR SED of NGC 4936 fitted with a dust continuum model (modified blackbody) with $\beta = 2$. The shaded blue region shows the 1$\sigma$ uncertainty.}
        \label{fig:ir_sed}
    \end{figure}

    We obtained $Spitzer$ infrared fluxes for NGC 4936 from \citet{temi09} and fitted the spectral energy distribution with a modified blackbody model using {\sc mbb\_emcee} \citep{conley16}, assuming dust temperatures in the range $T_{\rm dust}=10$--40 K and a fixed emissivity index $\beta=2$. The IR SED and best-fit model are shown in Figure~\ref{fig:ir_sed}. We derive a dust temperature of $T_{\rm dust} = 29.2^{+3.5}_{-2.9}$ K, a dust mass of $M_{\rm dust} = 1.60^{+0.39}_{-0.29} \times 10^6\,{\rm M_\odot}$, and a total infrared luminosity of $L_{\rm TIR} = 8.08^{+0.53}_{-0.46} \times 10^{42}\,{\rm erg\,s^{-1}}$.

    Using the molecular gas mass derived from CO observations (Section~\ref{sec:mass_co}), we obtain a dust-to-gas ratio of 0.008, which is closer to the value for star forming spiral galaxies ($\sim 0.01)$. For comparison, in another group central galaxy with large amount of cold gas, NGC 5044, the dust-to-gas ratio is $\sim 0.003$ \citep{tamhane26}. The ratio is typically $\sim 0.001$ in central galaxies in groups and clusters \citep[e.g.,][]{edge01}.

    We also estimated the extinction in dusty clumps and filaments seen in the residual HST image as $A_{555} = 2.5 \, {\rm Log ((F_{555\, mod} - abs (F_{555\, res}))/F_{555 \, mod})}$, where F$_{\rm 555 \, mod}$ and F$_{\rm 555 \, res}$ are fluxes extracted from the model image of the galaxy's continuum and residual image, respectively, within dust absorption regions. The dust absorbed regions were identified by constructing contours around negative residuals and selecting a contour level that encompassed the majority of visually identified dust clumps. The visual extinction $A_V$ was then estimated using the \citet{cardelli89} extinction law and $R_V = 3.1$. We found an average visual extinction $A_V = 0.10  \pm 0.04$.

    \subsection{Star Formation}
    \label{sec:sf}
    
    NGC 4936 has far- and near-ultraviolet (FUV and NUV) magnitudes of 19.28 and 16.92 AB mag, respectively, from Galaxy Evolution Explorer (GALEX), corrected for foreground Galactic extinction using \citet{schlegel98} dust maps \citep{vaddi16}, indicating the presence of star formation. Since there is a significant amount of dust in the centre as discussed Section~\ref{sec:dust}, we combined FUV and IR luminosities to estimate the star formation rate ($SFR$) using the relation from \cite{hao11} as follows:
    \begin{equation}
        SFR = 10^{-43.17} \,  (L_{\rm FUV} + 0.48 \, L_{\rm IR}),
    \end{equation}
    where $L_{\rm FUV} = 3.56 \times 10^{41}$ erg s$^{-1}$ and L$_{IR} = 8.08 \times 10^{42}$ erg s$^{-1}$. This relation assumes Salpeter initial mass function with continuous star formation rate over 100 Myr. This gives a star formation rate of 0.29 M$_\odot$ yr$^{-1}$. This corresponds to molecular gas depletion time of $6.74 \times 10^8$ yr assuming that the current star formation rate remains unchanged during this period.

    We also compared the star formation rate with the X-ray derived gas cooling rate of $0.46$ M$_\odot$ yr$^{-1}$ within 10 kpc region (Section~\ref{sec:xray}). The corresponding ratio between the star formation rate and the cooling rate is $SFR/\dot{M}_{\rm Xray}\approx0.63$.

    \subsection{Central AGN}

\begin{figure}
        \centering
        \includegraphics[width=\linewidth]{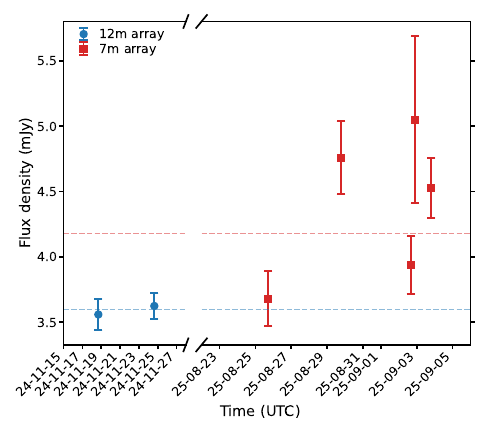}
        \caption{
        ALMA Band 3 continuum flux density at 107.3 GHz as a function of observing time for NGC 4936, measured from the 12 m array (blue circles) and the 7 m array (red squares). Error bars show the $1\sigma$ uncertainty. The horizontal dashed lines mark the error-weighted mean flux for each array configuration (3.60 mJy for the 12 m data and 4.18 mJy for the 7 m data). The time axis is broken between the two observing campaigns. The 7 m data show flux variations of up to $\sim$30\% between epochs separated by only a few days.
        }
        \label{fig:variability}
    \end{figure}

    We constructed ALMA continuum images from line-free spectral channels across all basebands, detecting a source coincident with the BCG nucleus (R.A.: 13:04:17.126, decl.: $-$30:31:34.62). Measurements from individual observing epochs at 107.3 GHz are listed in Table~\ref{tab:submm_cont} and shown in Fig.~\ref{fig:variability}. The two 12 m observations in November 2024 have consistent flux densities, with an error-weighted mean of 3.60 mJy. The five 7 m observations in August--September 2025 have flux densities from $3.68 \pm 0.21$ to $5.05 \pm 0.64$ mJy, with an error-weighted mean of 4.18 mJy. In particular, the flux density increased by $\sim$30\% between August 25 and 29. This difference corresponds to $3.1\sigma$ using the quoted statistical uncertainties without including flux calibration uncertainties for each epoch. These measurements show evidence of nuclear sub-mm continuum variability. Similar variability is also observed in the nucleus of NGC 5044, another central galaxy of a nearby group which also hosts low-luminosity X-ray AGN \citep{schellenberger24}. No continuum emission is detected from the nearby companion galaxy. Spectral imaging at a velocity resolution of 2.56 km s$^{-1}$ reveals no detectable line absorption against the nuclear continuum source.

    \begin{table}
    \centering
    \caption{
    ALMA Band 3 continuum flux density measurements of NGC~4936, extracted from the 12 m and 7 m array visibilities in linefree channels using the standard ALMA calibration pipeline.
    }
    \label{tab:submm_cont}
    \begin{tabular}{lccc}
        \hline
        \hline
        Array & Date \& time & Frequency & Flux density \\
         & (UTC) & (GHz) & (mJy) \\
        \hline
        12 m & 2024 Nov 18, 16:29:00 & 107.2700 & $3.56 \pm 0.12$ \\
        12 m & 2024 Nov 24, 15:21:53 & 107.2700 & $3.63 \pm 0.10$ \\
        \hline
        7 m  & 2025 Aug 25, 16:58:25 & 107.3014 & $3.68 \pm 0.21$ \\
        7 m  & 2025 Aug 29, 18:25:42 & 107.3014 & $4.76 \pm 0.28$ \\
        7 m  & 2025 Sep 02, 16:06:25 & 107.3014 & $3.94 \pm 0.22$ \\
        7 m  & 2025 Sep 02, 21:18:19 & 107.3014 & $5.05 \pm 0.64$ \\
        7 m  & 2025 Sep 03, 19:30:03 & 107.3014 & $4.53 \pm 0.23$ \\
        \hline
    \end{tabular}
    \end{table}

    The MeerKAT image shows a nuclear continuum source with a flux density of $39.76 \pm 0.01$ mJy (see Section~\ref{sec:radio_cont}). \cite{wilkinson87b} report flux densities of $41.0 \pm 0.24$ mJy and $33.3 \pm 0.14$ mJy at 1.5 and 5 GHz (beam sizes $\sim 21''$ and $7''$), respectively, using VLA data in C configuration. In contrast, \cite{grossova22} report a significantly lower 1.5 GHz flux density of $3.33 \pm 0.14$ mJy from VLA A-configuration data (beam size $\sim 3''$), implying that $\sim$92\% of the flux is resolved out at high angular resolution. \cite{disney77} report higher flux densities of $80 \pm 3$ mJy at 5 GHz and $97 \pm 10$ mJy at 2.7 GHz from Parkes observations. However, their $\sim4'$ beam encompasses a much larger area containing multiple radio sources, and these measurements are therefore likely contaminated.

    The radio spectral index of the continuum source is $\alpha \approx 0.17$ \citep[$S_\nu \propto \nu^{-\alpha}$;][]{wilkinson87b}, consistent with a flat-spectrum core. Combining radio and millimetre measurements (excluding the value of the VLA configuration A), we derive a spectral index of $\alpha = 0.57 \pm 0.09$, consistent with the optically thin synchrotron emission from a low-luminosity AGN. The systematic increase in recovered flux with larger beam sizes, observed in both ALMA and VLA data, indicates that the continuum emission consists of both a compact core and an extended component, likely associated with diffuse synchrotron emission either from AGN-related structures such as jets or star formation in the centre.

    \subsubsection{X-ray and optical counterparts}
    
    We also identify a possible hard X-ray source in the 2--7 keV \textit{Chandra} image spatially coincident with the radio continuum peak in ObsID 04998. We corrected the astrometry of both observations using CIAO's \texttt{fine\_astro} tool. ObsID 04997 required a comparatively large correction of more than four pixels. Using a $1.5''$ radius source aperture and a local background annulus with radii $1.5''$--$3''$, we measure 11.7$\pm$3.7 background subtracted net counts associated with the nuclear source in ObsID 04998, corresponding to a count rate of $(8.0\pm2.5) \times 10^{-4}$ cts s$^{-1}$. Whereas in ObsID 04997, obtained six days earlier, only $1.7 \pm 1.5$ counts were detected in 2--7 keV range, suggesting possible short-timescale variability, consistent with variability seen in ALMA 7 m observations. 
    We converted the background-subtracted 2--7 keV count rate to an intrinsic rest-frame 2--10 keV flux using \textit{Chandra} RMF and ARF extracted at the source position in \textsc{xspec}, assuming a power-law spectrum with $\Gamma=1.9$ and Galactic absorption of $N_{\rm H}=8.84\times10^{20}$ cm$^{-2}$. The flux ($F$) was converted to luminosity ($L$) using $L = F \times 4 \pi D_L^2$, where $D_L$ is the luminosity distance. For no intrinsic absorption, this corresponds to a 2--10 keV AGN luminosity of $L_{\rm 2-10 \, keV} = (5.9\pm 1.9) \times 10^{39}$ erg s$^{-1}$. We measured a $N_{\rm H_2} = 3.55 \times 10^{21}$ cm$^{-2}$ using the combined ALMA CO data within a beam sized region around the central radio continuum peak. Converting it to $N_{\rm H} = 2N_{\rm H_2} = 7.1 \times 10^{21}$ cm$^{-2}$ gives an intrinsic luminosity of $L_{\rm 2-10 \, keV} = (6.4\pm2) \times 10^{39}$ erg s$^{-1}$.

    The HST optical image does not reveal a clear point-like nucleus at the position of the hard X-ray source. Instead, the nuclear region is obscured by dust, which may suppress optical emission from the central source. The spatial coincidence between the hard X-ray excess, radio continuum emission, and dust-obscured optical nucleus supports the presence of a weak nuclear AGN.

    \subsection{Radio Continuum}
    \label{sec:radio_cont}
    
    The MeerKAT 0.9 -- 1.7 GHz continuum map is shown in the right panel of Figure~\ref{fig:soar}. The radio emission is centred on the nucleus and appears slightly extended relative to the synthesised beam.
    The total radio flux density in the central region is $39.76 \pm 0.01$ mJy, which corresponds to a spectral power of
    1.02 
    $\times 10^{22}$ W Hz$^{-1}$ at 1.28 GHz.
    Thus, the radio AGN in NGC~4936 is weak \citep[e.g.,][]{sun12}.
    
    In addition to the compact nuclear radio source, we identify diffuse radio emission located approximately 22 kpc northeast of the nucleus. The feature forms a detached arc-shaped structure that extends over $\sim 13$ kpc and has an integrated flux density of $1.05 \pm 0.04$ mJy. To our knowledge, this diffuse radio feature has not been reported previously. No corresponding emission is detected in the optical, X-ray, H$\alpha$ or CO images (see Figure~\ref{fig:composite}), but its inner edge just encloses the outer edge of the H$\alpha$ emission detected.

    Its detached morphology and lack of a multiwavelength counterpart instead point toward other possible origins such as a relic synchrotron plasma from an earlier episode of nuclear activity, magnetised plasma displaced by group-scale dynamical processes, potentially related to the interaction with the companion galaxy discussed in Section~\ref{sec:merger} or an unrelated background radio source lacking an obvious optical counterpart. The present data do not allow us to distinguish between these possibilities. Deeper multi-frequency radio observations, together with more sensitive X-ray imaging, will be required to determine the nature of this feature.

    \subsection{H\,\textsc{i}}
    \label{sec:HI}
    \subsubsection{H\,\textsc{i} in NGC 4936}
    
    The MeerKAT observations do not reveal any detection of H\,\textsc{i} in either absorption or emission. We derived an upper limit on H\,\textsc{i} gas mass from the MeerKAT data using \citep{meyer17}

    \begin{equation}
        \frac{M_{\rm H I}}{h^{-2} \, M_\odot} \simeq \frac{2.35 \times 10^5}{(1 + z )^2} \Bigg(\frac{D_L}{h^{-1} \, {\rm Mpc}}\Bigg)^2 \Bigg( \frac{S^{V_{Obs}}}{{\rm Jy \, km \, s^{-1}}} \Bigg),
        \label{eqn:hi_mass}
    \end{equation}

    where $D_L$ is the luminosity distance and $S^{V_{\rm Obs}}$ is the velocity-integrated flux density in the observed frame. We estimated the uncertainty on $S^{V_{\rm Obs}}$ from the per-channel rms noise of the H\,\textsc{i} data cube, assuming a velocity width of $\Delta v = 450$ km s$^{-1}$ corresponding to the width of the CO(1--0) emission. The integrated flux uncertainty is given by

    \begin{equation}
        S^{V_{\rm Obs}} = \sigma_{\rm RMS}\sqrt{\Delta v\,\delta v},
    \end{equation}
    
    where $\delta v$ is the channel width. For $\delta v = 5.57$ km s$^{-1}$ and a per-channel rms noise of $\sigma_{\rm RMS}=0.152$ mJy, we obtain a $1\sigma$ uncertainty of $S^{V_{\rm Obs}} = 7.6\times10^{-3}$ Jy km s$^{-1}$. This corresponds to a $3\sigma$ upper limit on the atomic gas mass of $M_{\rm H\,I} < 1.13 \times 10^7$ M$_\odot$. We can use these quantities to also derive the H\,\textsc{i} column density upper limit. The upper limit on the per channel optical depth is $\sigma_\tau = \sigma_{\rm RMS}/S_{\rm cont}$. Therefore, a $3\sigma$ upper limit on the integrated optical depth of $\int\tau(v)\,dv < 3\sigma_\tau \sqrt{\Delta v\,\delta v} = 0.27$ km s$^{-1}$ assuming absorption linewidth $\Delta v = $ 100 km s$^{-1}$. The H\,\textsc{i} absorption column density is then estimated as,

    \begin{equation}
    \frac{N_{\rm H\,I,\, abs}}{\rm cm^{-2}} = 1.82\times10^{18}\left(\frac{T_s}{100~{\rm K}}\right)\left(\frac{f_c}{1}\right)^{-1} \int\tau(v)\,dv,
    \end{equation}

    \noindent where $T_s$ is the spin temperature and $f_c$ is the covering factor. This gives $N_{\rm H\,I,\, abs} < 4.9 \times 10^{17}$ cm$^{-2}$ for $T_s = 100$ K and $f_c = 1$.

    Previous studies have reported the presence of a substantially larger H\,\textsc{i} reservoir. \citet{reid94} quote an unpublished measurement corresponding to an H\,\textsc{i} mass of $1.1 \times 10^9$ M$_\odot$. \citet{oosterloo07} reported the same H\,\textsc{i} mass and associated it with an H\,\textsc{i} cloud offset from the galaxy by $\sim80$ kpc based on a private communication. Inspection of the H\,\textsc{i} Parkes All Sky Survey (HIPASS) data reveals H\,\textsc{i} emission in a pointing centred at R.A.: $13^{\rm h}04^{\rm m}08.25^{\rm s}$ and decl.: $-30^\circ35'16.41''$, corresponding to a projected distance of $\sim54$ kpc southwest of NGC 4936. However, the large HIPASS beam is large and cannot resolve individual sources.

    Nevertheless, the low H\,\textsc{i} optical depth and H\,\textsc{i} mass upper limit which is an order of magnitude lower than the molecular gas mass derived from the CO(1--0) observations, imply that the cold gas reservoir in NGC 4936 is strongly dominated by molecular gas. The resulting lower limit on the molecular to atomic gas mass ratio is $M_{\rm H_2}/M_{\rm H~I}>17$. 
    Such large molecular fractions are often associated with high-pressure environments in BCGs, where increasing ambient pressure favours the conversion of H\,\textsc{i} to H$_2$ and the survival of molecular gas \citep[e.g.,][]{babyk23,tamhane26}.
    For comparison, NGC 5044, another nearby cool-core group with mostly centrally concentrated molecular reservoir, has a comparable total thermal pressure inside $\sim 10$ kpc to NGC 4936 of $P \gtrsim 2\times10^{-11}$ erg cm$^{-3}$. This corresponds to $P/k_{\rm B}\gtrsim1.5\times10^{5}$ K cm$^{-3}$, substantially higher than the characteristic mid-plane pressures of $P/k_{\rm B}\sim(2$--$5)\times10^{4}$ K cm$^{-3}$ at which the ISM in nearby disk galaxies transitions from H\,\textsc{i} to H$_2$ dominated \citep{blitz06,leroy08}. Although the thermal pressure of the hot atmosphere is not directly equivalent to the hydrostatic pressure used in these relations, the high confining pressure may nevertheless favour the formation and survival of molecular gas and contribute to the large molecular to atomic gas mass ratio in NGC~4936 and in the population of central galaxies in groups and clusters in general.

    \begin{figure*}
        \centering
        \includegraphics[width=\textwidth]{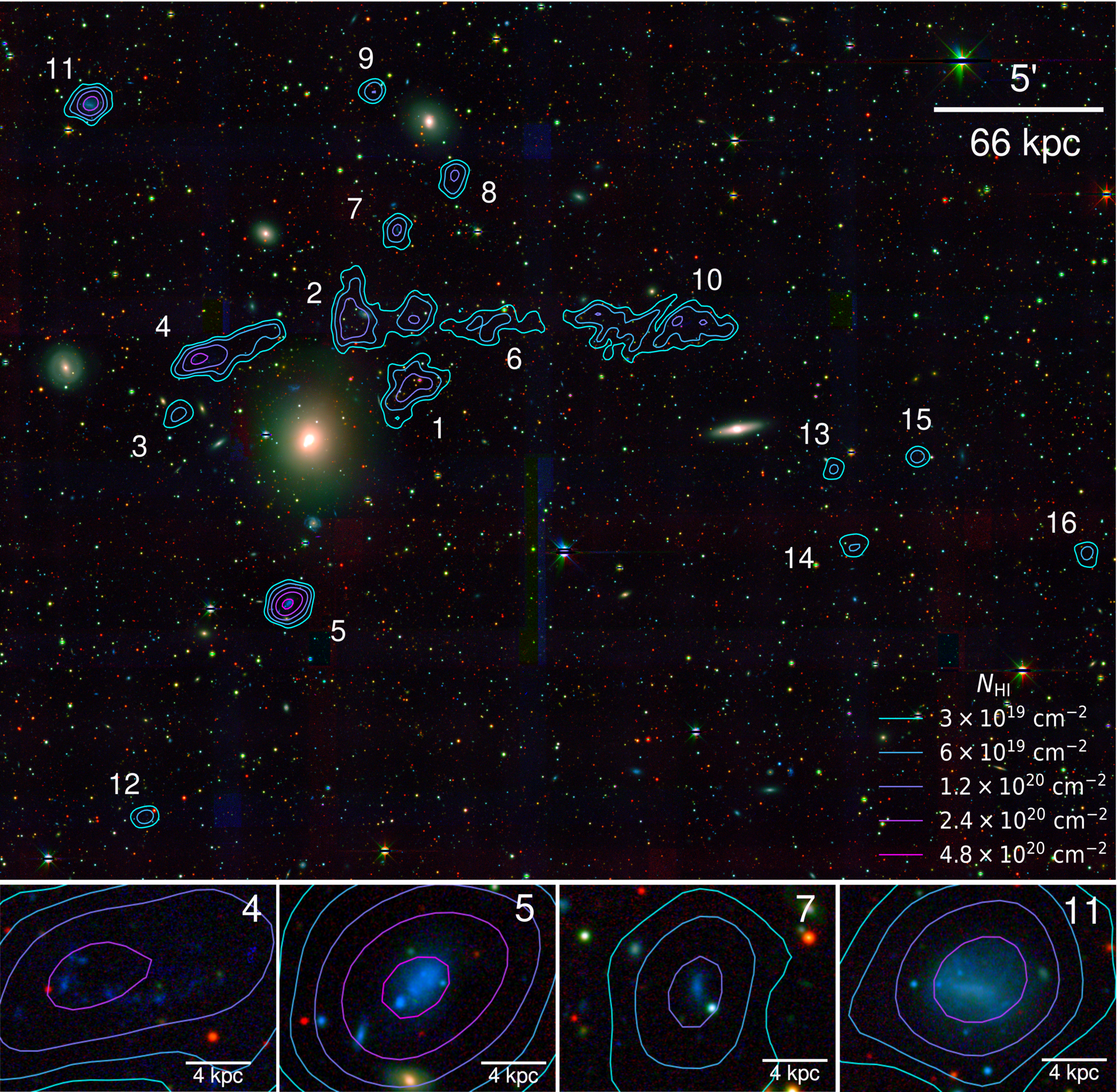}
        \caption{RGB colour composite of NGC 4936 constructed from DECaLS $g$ (blue), $r$ (green), and $z$ (red) images. The contours show H\,\textsc{i} column density $N_{\rm HI}$ from $(0.3 - 4.8) \times 10^{20}$ cm$^{-2}$, increasing by a factor of 2 as shown in the legend. IDs of H\,\textsc{i} clouds as reported in Table~\ref{tab:hi_sources} are also shown. The lower panels show zoomed-in cutouts of H\,\textsc{i} clouds that have associated optical blue sources. The H\,\textsc{i} datacube beam size is $25.1'' \times20.3''$ with P.A. = $-46.9^\circ$. North is up and east is to the left.
        }
        \label{fig:HI_sources}
    \end{figure*}

    \subsubsection{H\,\textsc{i} in the broader group environment}
    As shown in Figure~\ref{fig:HI_sources}, a blind search of the MeerKAT cube found 20 H\,\textsc{i} structures across the wider environment of NGC~4936, at projected distances of 43 to $\sim500$ kpc. The detections span a range of morphologies, including irregular, elongated, and tail-like structures, as well as more compact clouds. All but one lie within $\pm300$ km s$^{-1}$ of the systemic velocity of NGC~4936. Their H\,\textsc{i} masses range from $\sim4\times10^7$ to $6\times10^8$ M$_\odot$, with column densities ranging from $\sim3\times10^{19}$ cm$^{-2}$ in the more diffuse structures to $\sim10^{21}$ cm$^{-2}$ in spiral galaxies. Several of the narrower clouds have Gaussian-equivalent velocity dispersions of only $\sim6$--$15$ km s$^{-1}$, whereas the more extended and morphologically complex structures have velocity dispersions of $\sim20$--$65$ km s$^{-1}$. The properties of these sources are described in Table~\ref{tab:hi_sources}.

    Nine H\,\textsc{i} detections coincide with catalogued galaxies or compact blue optical sources, while others have no obvious counterpart in the DECaLS imaging. Three blue counterparts are also detected in GALEX ultraviolet imaging, consistent with recent or ongoing star formation. The extended, asymmetric, and partially optically dark H\,\textsc{i} distribution suggests that at least some of the atomic gas has been redistributed through tidal interactions and/or stripping within the wider group environment. A detailed source-by-source discussion is presented in Appendix~\ref{sec:blue_sources}. Despite these detections in the wider group environment, no H\,\textsc{i} is detected in individual group member galaxies within a projected distance of $\sim300$ kpc from NGC~4936.

    The main properties of NGC~4936 derived throughout the preceding sections are summarised in Table~\ref{tab:properties}.

    \begin{table*}
    \vspace{0.2cm}
    \centering
    \caption{Properties of NGC 4936 and its multiphase core.}
    \label{tab:properties}
    \begin{tabular}{lll}
    \hline
    Property & Value & Notes \\
    \hline
    \multicolumn{3}{c}{\textbf{Hot Gas}} \\
    \hline
    Cooling time & $1.68^{+0.08}_{-0.09}$ Gyr & at 10 kpc \\
    $t_{\rm cool}/t_{\rm ff}$ & 41 (25) & at 10 kpc (minimum at $R\sim1$ kpc from \cite{donahue26}) \\
    Cooling rate & $0.46^{+0.05}_{-0.04}$ M$_\odot$ yr$^{-1}$ & within $R = 10$ kpc \\
    Hot gas mass $M_{\rm Xray}$ & $7.70^{+0.04}_{-0.03} \times 10^9$ M$_\odot$ & within $R = 10$ kpc, corresponding to $t_{\rm cool}=1.68$ Gyr \\
    Bolometric X-ray luminosity & $6.42^{+0.40}_{-0.39}$ erg s$^{-1}$ & unabsorbed within $R = 10$ kpc \\
    \hline
    \multicolumn{3}{c}{\textbf{Molecular Gas}} \\
    \hline
    CO(1--0) flux & $8.65 \pm 0.21$ Jy km s$^{-1}$ & Combined (12 m $+$ 7 m) cube within $6''$ aperture\\
    Molecular gas mass, $M_{\rm mol}$ & $(1.93 \pm 0.05)\times10^{8}$ M$_\odot$ &  \\
    \hline
    \multicolumn{3}{c}{\textbf{Atomic Gas}} \\
    \hline
    H\,\textsc{i} mass & $<1.13\times10^{7}$ M$_\odot$ & 3$\sigma$ upper limit from MeerKAT (beam $\sim 25''  \times 20''$) \\
    \hline
    \multicolumn{3}{c}{\textbf{Ionised Gas}} \\
    \hline
    H$\alpha$ luminosity & $(3.00 \pm 0.06)\times10^{40}$ erg s$^{-1}$ & from MUSE datacube \\
    \hline
    \multicolumn{3}{c}{\textbf{Dust and Star Formation}} \\
    \hline
    Dust temperature & $29.2^{+3.5}_{-2.9}$ K & $\beta = 2$ \\
    Dust mass & $1.60^{+0.39}_{-0.29}\times10^{6}$ M$_\odot$ &  \\
    Dust-to-gas ratio & 0.008 &  \\
    Total IR luminosity & $8.08^{+0.53}_{-0.46}\times10^{42}$ erg s$^{-1}$ &  \\
    Star formation rate & 0.29 M$_\odot$ yr$^{-1}$ & FUV + IR \\
    Molecular gas depletion time & $6.74\times10^{8}$ yr &  \\
    \hline
    \multicolumn{3}{c}{\textbf{AGN}} \\
    \hline
    $L_{\rm 1.28~GHz}$ & $1.3 \times 10^{38}$ erg s$^{-1}$ & from MeerKAT continuum image \\
    $L_{\rm 2-10~keV}$ & $5.9 \times 10^{39}$ erg s$^{-1}$ & from $Chandra$ data \\
    M$_{\rm BH}$ & $1.3 \times 10^9$ M$_\odot$ & from $M-\sigma$ relation (see Appendix~\ref{sec:stellar}) \\
    \hline
    \end{tabular}
    \end{table*}

\section{Discussion}

    \subsection{Origin of the cold molecular gas reservoir}
    \label{sec:h2-origin}

    NGC 4936 is a rare system which hosts a substantial reservoir of molecular gas ($M_{\rm mol} \approx 2 \times 10^8$ M$_\odot$) accompanied by extended H$\alpha$ emission and prominent dust filaments with minimum $t_{\rm cool}/t_{\rm ff} \gtrsim 25$ in a relatively shallow group potential. Therefore, the origin of this multiphase gas reservoir remains uncertain. Two mechanisms are likely to contribute: condensation from the hot intragroup medium, and external acquisition through past or ongoing interactions within the group. In the following sections, we discuss these two scenarios.

    \subsubsection{Atmospheric cooling and the precipitation threshold}
    
    The presence of a cool X-ray core with central cooling times below $\sim1$ Gyr supports atmospheric cooling as a plausible source of the molecular and ionised gas. In the standard condensation scenario in analytical studies and simulations, cold gas forms when minimum $t_{\rm cool}/t_{\rm ff} \sim 10-20$ \citep[e.g.,][]{voit17,sharma12}. \citet{pulido18} found that central galaxies containing molecular gas predominantly occupy the range $10<\min(t_{\rm cool}/t_{\rm ff})<20$. However, they also discussed the role of observational selection effects, and cautioned against interpreting this range as a universal condensation threshold. NGC~4936 lies close to or above its upper end. A relevant exception in the \citet{pulido18} sample is A1060, which contains molecular gas despite its long central cooling time and very high $t_{\rm cool}/t_{\rm ff}$. They suggested that gas may have been acquired by this galaxy through a merger or condensed at an earlier epoch when the atmosphere was denser or AGN feedback was less effective. This ratio therefore reflects the present day, azimuthally averaged thermodynamic state of the hot atmosphere and may not capture the conditions under which the cold gas originally formed \citep[e.g.,][]{prasad22}.

    The observed azimuthally averaged thermodynamic profiles of NGC~4936 likewise need not represent the local conditions under which its molecular gas may have formed. As argued in Section~\ref{sec:tcool}, the molecular disk has an estimated age of $\sim10^8$ yr, comparable to the sound-crossing time across the inner group atmosphere of $\sim(7$--$9)\times10^7$ yr. Therefore, the atmosphere could have responded to disturbances since the gas condensed, although this timescale comparison alone does not establish that its thermodynamic state has changed. The ongoing interaction provides a possible source of disturbance, but the available \emph{Chandra} observations are too shallow to determine its effect on the hot atmosphere. Thus, atmospheric cooling remains a plausible origin of the molecular gas, given the short central cooling time and the innermost $t_{\rm cool}/t_{\rm ff}\sim25$ close to the upper end of the range commonly observed in molecular gas rich central galaxies. The ongoing interaction may additionally be perturbing the surrounding hot atmosphere.

    \subsubsection{The effect of rotating atmosphere on cooling}
    If the hot group atmosphere of NGC~4936 is rotationally supported, this may promote gas condensation even at an elevated $t_{\rm cool}/t_{\rm ff}$ threshold.  Analytical studies and numerical simulations have shown that rotation suppresses buoyant mixing and modifies the standard condensation criterion, promoting gas condensation well above the $t_{\rm cool}/t_{\rm ff} \sim 10$ threshold \citep[e.g.,][]{gaspari15,voit17,sobacchi19}. In these models, condensing gas conserves angular momentum and settles into rotating cold disks rather than forming extended radial filaments. The morphology of the molecular disk observed in NGC 4936 is broadly consistent with this picture. Rotating molecular disks coexisting with hot atmospheres at elevated $t_{\rm cool}/t_{\rm ff}$ has been reported in other early-type galaxies and interpreted in this context \citep[e.g.,][]{juranova19,juravnova20}.

    MUSE H$\alpha$ velocity map in Figure~\ref{fig:alma} shows that extended filaments on the east and west side of the nucleus 4--5 kpc away are blueshifted and redshifted to velocities of $\sim 150$ km s$^{-1}$ compared to the systemic velocity of the BCG, respectively. If they cooled from the hot X-ray gas and preserved its kinematic signature, it could indicate rotation on large scales. However, the observed velocity structure may also arise from inflow, outflow, or tidal motions, and therefore does not provide definitive evidence for a rotating atmosphere.

    The current minor merger could have transferred angular momentum to the group atmosphere and perturbed the intragroup medium over a longer timescale and promoted condensation if this is not the first passage of the companion galaxy. However, the orbital history of the current minor merger is unknown. The distinction between a first infall and a decaying orbit cannot be established from the available data.

    A rotating atmosphere would thus provide an additional explanation for why substantial molecular gas can coexist with a globally elevated $t_{\rm cool}/t_{\rm ff}$, complementary to the scenarios discussed in Section~\ref{sec:h2-origin}. Deeper X-ray observations capable of resolving the ellipticity or kinematic structure of the hot gas would be required to test this scenario directly.
    
    \subsubsection{Role of interactions and external gas acquisition}
    \label{sec:merger}
    
    NGC 4936 has previously been classified as a shell galaxy undergoing a minor merger \citep{wilkinson87b,reid94}. Our continuum subtracted DECaLS $g$-band images provide
    additional evidence for a disturbed stellar morphology. In particular, a diffuse
    low surface brightness plume extends to the N--NW approximately along the projected
    axis connecting the nucleus of NGC~4936 and the nearby companion, and is recovered
    using two independent stellar-light subtraction methods (see Appendix~\ref{sec:tidal}).
    Additional broad arc-like residual structures are also visible to the SW and NE, although some of these residuals may arise from imperfect modelling of the stellar continuum.
    Nevertheless, the extended plume, together with the small projected separation
    ($\sim2.4$ kpc) and velocity offset ($\sim365$ km s$^{-1}$) of the companion,
    supports a recent or ongoing tidal interaction. However, the morphology alone
    is not sufficient to establish that all of these structures were produced by the
    present companion. Detailed modelling of the low surface brightness structure and reconstruction of the interaction history are beyond the scope of this work.
    
    The large-scale dust distribution provides further evidence that the interaction
    has affected the cold interstellar medium. Dust filaments extend well beyond the
    central region and appear to wrap around the companion before continuing outward
    (Figure~\ref{fig:hst}). In addition, asymmetries in the molecular and ionised gas
    distributions, together with the kinematically disturbed molecular filament
    identified in the residual velocity field, are consistent with dynamical perturbations
    and/or ongoing gas accretion.
    
    The broader group environment also appears dynamically active. In addition to NGC 4936, the nearby galaxy ESO 443-55, located at a projected distance of $\sim 100$ kpc, exhibits prominent shell structures in DECaLS imaging (see Figures~\ref{fig:HI_sources} and~\ref{fig:HI_kin}), suggesting a previous merger or interaction. Notably, the large reservoir (few $10^9$ M$_\odot$) of spatially extended H\,\textsc{i} within the group atmosphere suggests substantial redistribution of atomic gas within the group and ongoing tidal interactions or ram pressure stripping. Similar large-scale H\,\textsc{i} disturbances have been observed in the NGC~3100 group, where minor interactions were proposed to redistribute atomic gas and potentially contributed to the build-up of a central molecular reservoir \citep{maccagni23}. It supports the possibility that external acquisition may have contributed to the molecular gas reservoir of NGC~4936. However, there is no continuous spatial or kinematic connection between the extended H\,\textsc{i} and the central molecular disc. Therefore, the presently observed H\,\textsc{i} cannot be identified as the direct progenitor of the molecular gas. However, we cannot rule out the possibility that an earlier interaction supplied some cold gas that subsequently settled towards the centre where most of H\,\textsc{i} is converted into H$_2$ or gets ionised, thus remaining consistent with atmospheres of central galaxies where the cold gas reservoir is predominantly molecular \citep{babyk23,tamhane26}.
    
    A second candidate source for the molecular gas is direct donation from the companion galaxy itself if the two galaxies are interacting. The companion galaxy lies only $\sim2.5$ kpc away in projection and has a velocity offset of 360 km s$^{-1}$ relative to NGC 4936 (see Appendix~\ref{sec:stellar}). Stellar population modelling of the MUSE spectrum yields an $r$-band stellar mass-to-light ratio of 7.75 M$_\odot$/$L_\odot$ and a stellar mass of $1.47\times10^{10}$ M$_\odot$, although this should be considered as an upper limit because the underlying light from NGC 4936 was not removed. For comparison, NGC 4936 has an $r$-band mass-to-light ratio of 8.43 M$_\odot$/$L_\odot$ and stellar mass of $2.78\times10^{11}$ M$_\odot$, implying a stellar mass ratio of approximately 20:1, consistent with a minor interaction if the companion is physically associated with NGC 4936. Although the companion's stellar mass is far in excess of the molecular gas mass it would need to have supplied, it appears morphologically similar to an early-type galaxy, and there is no direct evidence that it contains, or recently contained, sufficient cold gas to account for the molecular gas observed in NGC 4936 through tidal stripping alone. Moreover, a scenario in which the present interaction directly transferred the molecular gas, would more naturally be expected to show stronger asymmetry, or strongly non-circular kinematics. Instead, the strong central concentration and relatively ordered rotation of the molecular gas, together with the similar central concentration of the warm ionised gas and their broad correspondence with the X-ray emitting atmosphere, favour an origin linked primarily to cooling and condensation of the hot phase. Some gas may have been acquired externally, but its contribution is likely secondary.

    \subsubsection{High $t_{\rm cool}/t_{\rm ff}$ and dynamical timescales}
    \label{sec:tcool}

    The molecular gas has formed a rotating disk with a radius of $\sim 1$ kpc inclined at $61.8^\circ$. The observed line-of-sight velocity amplitude of $\sim 200$ km s$^{-1}$ implies an intrinsic rotational velocity of $\sim 227$ km s$^{-1}$ after correcting for inclination ($v_{\rm rot} = v_{\rm los}/\sin\,61.8^\circ$), corresponding to an orbital period of approximately $2.7 \times 10^7$ yr. Rotating molecular gas disks are commonly observed in merger remnants, suggesting that interactions can efficiently redistribute cold gas into ordered rotational structures once the gas has dissipated its orbital energy \citep{ueda14}. If several orbital periods are required for initially disturbed gas to settle into rotation, the observed disk is likely $\sim10^8$ yr old.
    
    The companion galaxy is separated by only $\sim2.5$ kpc in projection and has a relative velocity of 360 km s$^{-1}$, corresponding to a characteristic crossing time of $\sim 7 \times 10^6$ yr, although the true timescale is likely longer due to projection effects. The sound-crossing time across the central $\sim 30$--40 kpc of the group is $\sim(7$--$9) \times 10^7$ yr, indicating that interaction-driven pressure disturbances can propagate through the inner atmosphere on timescales comparable to those over which the molecular disk has been assembling. Therefore, the atmosphere could plausibly have transitioned from a more precipitation-favourable state to its current elevated $t_{\rm cool}/t_{\rm ff}$ while the molecular gas was settling into a disk.
    
    Several interpretations are consistent with this picture.  The gas may have condensed during an earlier epoch when the atmosphere was more susceptible to precipitation, with subsequent heating raising the global ratio to its current value.  Alternatively, the merger may have generated localised reductions in $t_{\rm cool}/t_{\rm ff}$ through gas compression or turbulent dissipation, without producing a significant change in the azimuthally averaged X-ray profiles. Some of the molecular gas may also have been externally acquired during an earlier interaction and subsequently settled into the present disc, in which case its existence need not be directly related to the current $t_{\rm cool}/t_{\rm ff}$. A further possibility is that some fraction of the cold gas formed through AGN-driven uplift of low-entropy material, \citep[e.g.,][]{mcnamara16}. However, we did not find sufficient evidence for recent or ongoing AGN-driven feedback to assess its role in the thermodynamic history of the atmosphere. The merger-perturbation interpretations are therefore physically motivated by the morphological and kinematic evidence in the cold gas and dust. Deeper X-ray observations are required to test them directly.
    
    \subsubsection{Contribution from Stellar Mass Loss}
    The old evolved population of stars in NGC 4936 can contribute to the cold gas reservoir through continuous stellar mass loss. Our pPXF modelling of the MUSE spectrum gives a stellar age of $\sim 10^{10}$ yr in NGC 4936. For a galaxy of this stellar mass and age, the expected stellar mass-loss rate is of order 1 M$_\odot$ yr$^{-1}$ \citep{canning13,ciotti91}, comparable to or somewhat exceeding both the present-day star formation rate and the X-ray cooling rate.
    
    Over $10^8$ years, stellar mass loss could inject a significant amount of cold gas into the galaxy and may be an important contributor to the current cool and cold gas being generated in the galaxy. However, stellar ejecta in massive early-type galaxies are generally expected to thermalise and mix with the hot atmosphere rather than cool directly into molecular clouds \citep{parriott08,bregman09}. Therefore, while stellar mass loss likely contributes to replenishing the hot gas reservoir, it is unlikely to be the dominant origin of the observed $2\times10^8$ M$_\odot$ molecular disk unless the cooling efficiency is exceptionally high. Furthermore, the stellar velocity field derived from the MUSE data (see Figure~\ref{fig:stellar}) shows no evidence of ordered rotation in the old stellar population. The presence of a rotating molecular disk is therefore difficult to reconcile with stellar mass loss as the dominant gas source, which would be expected to produce a more kinematically disordered cold gas distribution.
    
    \subsection{Star formation and gas consumption}

    The star formation rate of $\sim 0.3$ M$_\odot$ yr$^{-1}$ implies a molecular gas depletion time of $\sim 7 \times 10^8$ yr, comparable to values measured in cool-core groups and clusters. The ratio of SFR to X-ray gas cooling rate  of $\sim0.63$, places NGC 4936 among low cooling rate systems with relatively efficient star formation. However, this ratio should not be interpreted as direct evidence that the hot atmosphere is cooling with unusually high efficiency. As discussed by \citet{mcdonald18}, systems with $\dot{M}_{\rm Xray} \lesssim 10$ M$_\odot$ yr$^{-1}$ show a relatively narrow range of star formation rates despite spanning a much wider range in cooling rates, suggesting that additional processes may contribute to the cold gas reservoir or regulate star formation. In the case of NGC 4936, the disturbed dust morphology, nearby companion galaxy, and asymmetric gas distribution point to a minor interaction or a merger that may have both perturbed the intragroup medium and contributed to the cold gas supply. Consequently, the observed star formation is likely sustained by a combination of cooling from the hot atmosphere and processes related to galaxy interactions rather than by steady-state cooling alone.

    \subsection{Origin of the extended H\,\textsc{i} emission}

    The extended, asymmetric, and partially optically dark H\,\textsc{i} distribution suggests that at least some of the atomic gas has been redistributed through tidal interactions and/or stripping within the group. The sources could be pre-existing gas-rich dwarf galaxies or gas removed from interacting galaxies. The compact blue counterparts, raise the possibility that stars are forming within this displaced gas. However, blue colours and UV emission alone do not distinguish such condensations from pre-existing star forming dwarfs.
    
    If the roughly east--west H\,\textsc{i} structure originated in a common interaction, its projected extent of $\sim180$ kpc provides an approximate dispersal timescale. Adopting the observed speed of $\sim200$ km s$^{-1}$ gives $\sim0.9$ Gyr, suggesting that the atomic gas could have originated from an interaction within the last Gyr timescale.
    
    No clear progenitor has been identified for H\,\textsc{i} clouds within 300 kpc of NGC 4936. The absence of H\,\textsc{i} detections in nearby catalogued galaxies and of obvious stellar tidal tails also prevents an association with a particular encounter, but does not exclude stripping or a tidal origin.

    \subsection{Implications for cooling thresholds in galaxy groups}
    \label{sec:implications}
     
    NGC 4936 is an excellent example of systems in which substantial cold gas reservoirs coexist with globally elevated $t_{\rm cool}/t_{\rm ff}$ ratios. While the minimum $t_{\rm cool}/t_{\rm ff} \lesssim 10$--$20$ condensation threshold has considerable empirical and theoretical support \citep[e.g.,][]{voit15a,voit18,sharma12,pulido18,babyk18,lakhchaura18}, NGC 4936 shows that this threshold should not be interpreted as a universal boundary. Instead, the observed molecular gas may reflect the cumulative thermodynamic history of the system, including atmospheric condensation during an earlier epoch, locally enhanced cooling driven by environmental perturbations, external acquisition during past interactions or additional effects such as angular momentum in the hot atmosphere that modify the conditions for condensation.
     
    This is particularly relevant in the group regime, where the gravitational potentials are shallow and intragroup medium temperatures are low. Therefore, group atmospheres are more susceptible to perturbations from galaxy interactions and AGN activity compared to massive clusters \citep{sun12,eckert21}. Therefore, the combination of a cool X-ray core, a dynamically active environment, and a substantial multiphase gas reservoir in NGC 4936 may represent a mode of cold gas accumulation that is more common in groups but has been under-characterised. A broader survey of group-central galaxies at intermediate $t_{\rm cool}/t_{\rm ff}$ values, coupled with morphological indicators of recent interactions, would help establish how common this pathway is and whether it leaves a distinctive imprint on the thermodynamic and kinematic state of the multiphase medium.
    
\section{Summary}
    
    We presented a multiwavelength analysis of NGC 4936, a group central galaxy in a weak X-ray environment. Its 0.2--2 keV luminosity within the radius where $t_{\rm cool} = 3$ Gyr is only about $\sim 10^{41}$ erg s$^{-1}$, an order of magnitude below NGC 5044 and lower than the brightest groups in \cite{sun09}, yet it hosts a large reservoir of molecular gas. We used ALMA, MUSE, MeerKAT, HST, SOAR and Chandra data to investigate the origin of the cold gas and its connection to the surrounding group environment and a nearby companion galaxy. Our main results are as follows:

    \begin{itemize}
        \item The available shallow X-ray data does not show any evidence of AGN activity in the form of central point source or cavities and shocks. The thermodynamic profiles indicate the presence of a weak cool-core with central $t_{\rm cool} \lesssim 1$ Gyr and minimum $t_{\rm cool}/t_{\rm ff} \gtrsim 25$.
        
        \item We detected a molecular gas reservoir of $M_{\rm H_2} = (1.93 \pm 0.05) \times 10^8$ M$_\odot$ distributed in a rotating disk of size $\sim 1$ kpc, with asymmetries and extended filamentary structure.
    
        \item The ionised gas traced by H$\alpha$ extends to $\sim 16$ kpc and shows complex kinematics broadly consistent with rotation observed in the molecular gas in the inner $\sim 1$ kpc regions but with complex filamentary emission at larger radii. Given the correlation between the ionised and molecular gas in the central region, they likely trace a common origin and reservoir of cold gas.

        \item We did not detect H\,\textsc{i} emission in the centre of NGC 4936, placing a strong $3\sigma$ upper limit of $M_{\rm HI} < 1.13 \times 10^7$ M$_\odot$, indicating that the central cold gas reservoir is strongly dominated by molecular gas. The resulting $M_{\rm H_2}/M_{\rm H I} > 17$ is the largest lower limit in X-ray cool cores. Such a high molecular fraction is consistent with efficient conversion of atomic gas into H$_2$ in the relatively dense, high pressure central environment.
        
        \item In contrast, we detect multiple morphologically complex H\,\textsc{i} structures with masses of $\sim4\times10^7$--$6\times10^8$ M$_\odot$ throughout the wider group environment. All but one lie within $\pm300$ km s$^{-1}$ of the systemic velocity of NGC~4936. Their irregular, asymmetric morphologies, together with the absence of obvious optical counterparts for several clouds, suggest that at least some of the atomic gas has been redistributed by tidal interactions and/or stripping processes.

        \item The shallow slope ($\alpha = 0.57$) of the radio to submm nuclear continuum and the detection of spatially coincident hard X-ray nuclear source with $L_{\rm 2-10 \, keV} \approx 6 \times 10^{39}$ erg s$^{-1}$ suggests the presence of a low-luminosity AGN. The AGN shows variability in both X-ray and sub-mm flux on timescales of several days.
    
        \item Deep optical imaging reveals dust filaments extending beyond the central galaxy and wrapping around a nearby companion, indicating a strong large-scale morphological connection between the two systems.

        \item NGC 4936 has a star formation rate of 0.29 M$_\odot$ yr$^{-1}$ with molecular gas depletion time of $\sim 7 \times 10^8$ yr. Comparison with the X-ray gas cooling rate gives a star formation efficiency of $\sim 63\%$.
    
        \item The companion galaxy may be undergoing a minor merger ($\sim$20:1 stellar mass ratio) with a velocity offset of $\sim$360 km s$^{-1}$ and projected separation of 2.5 kpc, but shows no clear evidence of hosting a large cold gas reservoir that could explain the molecular gas in NGC 4936 via direct stripping. Earlier interactions may nevertheless have contributed some externally acquired cold gas.
    
        \item Despite elevated $t_{\rm cool}/t_{\rm ff}$ ratio, NGC 4936 hosts a substantial molecular gas reservoir, indicating that the present thermodynamic state does not reflect the conditions under which the cold gas originally formed.
    \end{itemize}

    These results suggest that the present-day multiphase medium in NGC 4936 reflects the combined effects of atmospheric cooling and galaxy interactions operating over different evolutionary stages. NGC 4936 demonstrates that substantial cold gas reservoirs can exist in galaxy groups with relatively high $t_{\rm cool}/t_{\rm ff}$, highlighting the importance of a system's thermodynamic and dynamical history when interpreting the origin of multiphase gas.

\section*{Acknowledgement}
We thank Tom\'a\v{s} Pl\v{s}ek and Norbert Werner on discussion about the \cite{lakhchaura18} cooling time. We also thank the anonymous referee for their useful comments.
Support for this work was provided by the National Aeronautics and Space Administration through {\em Chandra} Award Number GO2-23114X and GO3-24115X
issued by the {\em Chandra} X-ray Center, which is operated by the Smithsonian Astrophysical Observatory for and on behalf of the National Aeronautics Space Administration under contract NAS8-03060. We also acknowledge support from the NSF grant 2407821.
Based on observations with the NASA/ESA \emph{Hubble Space Telescope} obtained from the Mikulski Archive for Space Telescopes (MAST). This work also makes use of observations obtained with the MUSE instrument mounted on UT4 of the ESO Very Large Telescope under programme 094.A-0859, data obtained from the \emph{Chandra} Data Archive, observations with the MeerKAT radio telescope,
and imaging from the DESI Legacy Imaging Surveys. This paper makes use of the following ALMA data: ADS/JAO.ALMA.2024.1.01206.S. ALMA is a partnership of ESO (representing its member states), NSF (USA), and NINS (Japan), together with NRC (Canada), MOST and ASIAA (Taiwan), and KASI (Republic of Korea), in cooperation with the Republic of Chile. The Joint ALMA Observatory is operated by ESO, AUI/NRAO, and NAOJ. 
The MeerKAT telescope is operated by the South African Radio Astronomy Observatory, which is a facility of the National Research Foundation, an agency of the Department of Science and Innovation. (Part of) the data published here have been reduced using the CARACal pipeline, partially supported by ERC Starting grant number 679627 “FORNAX”, MAECI Grant Number ZA18GR02, DST-NRF Grant Number 113121 as part of the ISARP Joint Research Scheme, and BMBF project 05A17PC2 for D-MeerKAT. Information about CARACal can be obtained online under the URL (https://caracal.readthedocs.io).
We made use of OpenAI's ChatGPT (GPT-5) and Anthropic's Claude (Sonnet 4.6) to assist with language editing and polishing of the manuscript. All scientific content and conclusions are the responsibility of the authors.

\section*{Data Availability}

All archival data used in this work are publicly available through their respective mission archives. Reduced data products and analysis products generated in this study will be made available upon reasonable request to the corresponding author.

\bibliographystyle{paslike_pasa}
\bibliography{ref}

@ARTICLE{cappellari23,
       author = {{Cappellari}, Michele},
        title = "{Full spectrum fitting with photometry in PPXF: stellar population versus dynamical masses, non-parametric star formation history and metallicity for 3200 LEGA-C galaxies at redshift z {\ensuremath{\approx}} 0.8}",
      journal = {\mnras},
         year = 2023,
        month = dec,
       volume = {526},
       number = {3},
        pages = {3273-3300},
          doi = {10.1093/mnras/stad2597},
archivePrefix = {arXiv},
       eprint = {2208.14974},
 primaryClass = {astro-ph.GA},
       adsurl = {https://ui.adsabs.harvard.edu/abs/2023MNRAS.526.3273C}
}

@ARTICLE{miles,
       author = {{Vazdekis}, A. and {S{\'a}nchez-Bl{\'a}zquez}, P. and {Falc{\'o}n-Barroso}, J. and {Cenarro}, A.~J. and {Beasley}, M.~A. and {Cardiel}, N. and {Gorgas}, J. and {Peletier}, R.~F.},
        title = "{Evolutionary stellar population synthesis with MILES - I. The base models and a new line index system}",
      journal = {\mnras},
         year = 2010,
        month = jun,
       volume = {404},
       number = {4},
        pages = {1639-1671},
          doi = {10.1111/j.1365-2966.2010.16407.x},
archivePrefix = {arXiv},
       eprint = {1004.4439},
 primaryClass = {astro-ph.CO},
       adsurl = {https://ui.adsabs.harvard.edu/abs/2010MNRAS.404.1639V}
}

@ARTICLE{wilkinson87,
       author = {{Wilkinson}, A. and {Browne}, I.~W.~A. and {Wolstencroft}, R.~D.},
        title = "{Shell galaxies detected with IRAS.}",
      journal = {\mnras},
         year = 1987,
        month = oct,
       volume = {228},
        pages = {933-940},
          doi = {10.1093/mnras/228.4.933},
       adsurl = {https://ui.adsabs.harvard.edu/abs/1987MNRAS.228..933W}
}

@ARTICLE{grossova22,
       author = {{Grossov{\'a}}, Romana and {Werner}, Norbert and {Massaro}, Francesco and {Lakhchaura}, Kiran and {Pl{\v{s}}ek}, Tom{\'a}{\v{s}} and {Gab{\'a}nyi}, Krisztina and {Rajpurohit}, Kamlesh and {Canning}, Rebecca E.~A. and {Nulsen}, Paul and {O'Sullivan}, Ewan and {Allen}, Steven W. and {Fabian}, Andrew},
        title = "{Very Large Array Radio Study of a Sample of Nearby X-Ray and Optically Bright Early-type Galaxies}",
      journal = {\apjs},
         year = 2022,
        month = feb,
       volume = {258},
       number = {2},
          eid = {30},
        pages = {30},
          doi = {10.3847/1538-4365/ac366c},
archivePrefix = {arXiv},
       eprint = {2111.02430},
 primaryClass = {astro-ph.HE},
       adsurl = {https://ui.adsabs.harvard.edu/abs/2022ApJS..258...30G}
}

@ARTICLE{reid94,
       author = {{Reid}, N. and {Boisson}, C. and {Sansom}, A.~E.},
        title = "{Elliptical galaxies at low surface brightness.}",
      journal = {\mnras},
         year = 1994,
        month = aug,
       volume = {269},
        pages = {713-736},
          doi = {10.1093/mnras/269.3.713},
       adsurl = {https://ui.adsabs.harvard.edu/abs/1994MNRAS.269..713R}
}

@ARTICLE{oosterloo07,
       author = {{Oosterloo}, T.~A. and {Morganti}, R. and {Sadler}, E.~M. and {van der Hulst}, T. and {Serra}, P.},
        title = "{Extended, regular HI structures around early-type galaxies}",
      journal = {\aap},
         year = 2007,
        month = apr,
       volume = {465},
       number = {3},
        pages = {787-798},
          doi = {10.1051/0004-6361:20066384},
archivePrefix = {arXiv},
       eprint = {astro-ph/0701716},
 primaryClass = {astro-ph},
       adsurl = {https://ui.adsabs.harvard.edu/abs/2007A&A...465..787O}
}

@ARTICLE{babyk18,
       author = {{Babyk}, Iu. V. and {McNamara}, B.~R. and {Nulsen}, P.~E.~J. and {Hogan}, M.~T. and {Vantyghem}, A.~N. and {Russell}, H.~R. and {Pulido}, F.~A. and {Edge}, A.~C.},
        title = "{X-Ray Scaling Relations of Early-type Galaxies}",
      journal = {\apj},
         year = 2018,
        month = apr,
       volume = {857},
       number = {1},
          eid = {32},
        pages = {32},
          doi = {10.3847/1538-4357/aab3c9},
archivePrefix = {arXiv},
       eprint = {1803.00020},
 primaryClass = {astro-ph.GA},
       adsurl = {https://ui.adsabs.harvard.edu/abs/2018ApJ...857...32B}
}

@ARTICLE{weilbacher20,
       author = {{Weilbacher}, Peter M. and {Palsa}, Ralf and {Streicher}, Ole and {Bacon}, Roland and {Urrutia}, Tanya and {Wisotzki}, Lutz and {Conseil}, Simon and {Husemann}, Bernd and {Jarno}, Aur{\'e}lien and {Kelz}, Andreas and {P{\'e}contal-Rousset}, Arlette and {Richard}, Johan and {Roth}, Martin M. and {Selman}, Fernando and {Vernet}, Jo{\"e}l},
        title = "{The data processing pipeline for the MUSE instrument}",
      journal = {\aap},
         year = 2020,
        month = sep,
       volume = {641},
          eid = {A28},
        pages = {A28},
          doi = {10.1051/0004-6361/202037855},
archivePrefix = {arXiv},
       eprint = {2006.08638},
 primaryClass = {astro-ph.IM},
       adsurl = {https://ui.adsabs.harvard.edu/abs/2020A&A...641A..28W}
}

@ARTICLE{emsellem22,
       author = {{Emsellem}, Eric and {Schinnerer}, Eva and {Santoro}, Francesco and {Belfiore}, Francesco and {Pessa}, Ismael and {McElroy}, Rebecca and {Blanc}, Guillermo A. and {Congiu}, Enrico and {Groves}, Brent and {Ho}, I.-Ting and {Kreckel}, Kathryn and {Razza}, Alessandro and {Sanchez-Blazquez}, Patricia and {Egorov}, Oleg and {Faesi}, Chris and {Klessen}, Ralf S. and {Leroy}, Adam K. and {Meidt}, Sharon and {Querejeta}, Miguel and {Rosolowsky}, Erik and {Scheuermann}, Fabian and {Anand}, Gagandeep S. and {Barnes}, Ashley T. and {Be{\v{s}}li{\'c}}, Ivana and {Bigiel}, Frank and {Boquien}, M{\'e}d{\'e}ric and {Cao}, Yixian and {Chevance}, M{\'e}lanie and {Dale}, Daniel A. and {Eibensteiner}, Cosima and {Glover}, Simon C.~O. and {Grasha}, Kathryn and {Henshaw}, Jonathan D. and {Hughes}, Annie and {Koch}, Eric W. and {Kruijssen}, J.~M. Diederik and {Lee}, Janice and {Liu}, Daizhong and {Pan}, Hsi-An and {Pety}, J{\'e}r{\^o}me and {Saito}, Toshiki and {Sandstrom}, Karin M. and {Schruba}, Andreas and {Sun}, Jiayi and {Thilker}, David A. and {Usero}, Antonio and {Watkins}, Elizabeth J. and {Williams}, Thomas G.},
        title = "{The PHANGS-MUSE survey. Probing the chemo-dynamical evolution of disc galaxies}",
      journal = {\aap},
         year = 2022,
        month = mar,
       volume = {659},
          eid = {A191},
        pages = {A191},
          doi = {10.1051/0004-6361/202141727},
archivePrefix = {arXiv},
       eprint = {2110.03708},
 primaryClass = {astro-ph.GA},
       adsurl = {https://ui.adsabs.harvard.edu/abs/2022A&A...659A.191E}
}

@ARTICLE{cappellari03,
       author = {{Cappellari}, Michele and {Copin}, Yannick},
        title = "{Adaptive spatial binning of integral-field spectroscopic data using Voronoi tessellations}",
      journal = {\mnras},
         year = 2003,
        month = jun,
       volume = {342},
       number = {2},
        pages = {345-354},
          doi = {10.1046/j.1365-8711.2003.06541.x},
archivePrefix = {arXiv},
       eprint = {astro-ph/0302262},
 primaryClass = {astro-ph},
       adsurl = {https://ui.adsabs.harvard.edu/abs/2003MNRAS.342..345C}
}

@software{photutils,
  author       = {Bradley, Larry and
                  Sipőcz, Brigitta M. and
                  Robitaille, T. P. and
                  Tollerud, E. J. and
                  Vinícius, Zé and
                  Deil, Christoph and
                  Barbary, Kyle and
                  Wilson, Tom J. and
                  Busko, Ivo and
                  Donath, Axel and
                  Günther, Hans Moritz and
                  Cara, Mihai and
                  Lim, P. L. and
                  Meßlinger, Sebastian and
                  Conseil, Simon and
                  Droettboom, Michael and
                  Bostroem, K. Azalee and
                  Bray, E. M. and
                  Bratholm, Lars Andersen and
                  Burnett, Zach and
                  Jamieson, William and
                  Ginsburg, Adam and
                  Taranu, Dan and
                  Barentsen, Geert and
                  Craig, Matthew W. and
                  Morris, Brett M. and
                  Perrin, Marshall and
                  Rathi, Shivangee},
  title        = {Photutils},
  month        = apr,
  year         = 2026,
  publisher    = {Zenodo},
  version      = {3.0.0},
  doi          = {10.5281/zenodo.19636730},
  url          = {https://doi.org/10.5281/zenodo.19636730},
  swhid        = {swh:1:dir:5ea90432cd987336a26af201c1b50343b12e2daa
                   ;origin=https://doi.org/10.5281/zenodo.596036;visi
                   t=swh:1:snp:25dd84d95353ec3813baa8785c248017da7192
                   33;anchor=swh:1:rel:253cbde27398f9aea9d7368f386661
                   8c00fe16f9;path=astropy-photutils-3322558
                  },
}

@ARTICLE{lakhchaura18,
       author = {{Lakhchaura}, K. and {Werner}, N. and {Sun}, M. and {Canning}, R.~E.~A. and {Gaspari}, M. and {Allen}, S.~W. and {Connor}, T. and {Donahue}, M. and {Sarazin}, C.},
        title = "{Thermodynamic properties, multiphase gas, and AGN feedback in a large sample of giant ellipticals}",
      journal = {\mnras},
         year = 2018,
        month = dec,
       volume = {481},
       number = {4},
        pages = {4472-4504},
          doi = {10.1093/mnras/sty2565},
archivePrefix = {arXiv},
       eprint = {1806.00455},
 primaryClass = {astro-ph.GA},
       adsurl = {https://ui.adsabs.harvard.edu/abs/2018MNRAS.481.4472L}
}

@ARTICLE{disney77,
       author = {{Disney}, M.~J. and {Wall}, J.~V.},
        title = "{A 5-GHz survey of bright Southern elliptical and S0 galaxies.}",
      journal = {\mnras},
         year = 1977,
        month = apr,
       volume = {179},
        pages = {235-254},
          doi = {10.1093/mnras/179.2.235},
       adsurl = {https://ui.adsabs.harvard.edu/abs/1977MNRAS.179..235D}
}

@ARTICLE{wilkinson87b,
       author = {{Wilkinson}, A. and {Browne}, I.~W.~A. and {Kotanyi}, C. and {Christiansen}, W.~A. and {Williams}, R. and {Sparks}, W.~B.},
        title = "{Radio emission from shell elliptical galaxies.}",
      journal = {\mnras},
         year = 1987,
        month = feb,
       volume = {224},
        pages = {895-910},
          doi = {10.1093/mnras/224.4.895},
       adsurl = {https://ui.adsabs.harvard.edu/abs/1987MNRAS.224..895W}
}

@ARTICLE{schlegel98,
       author = {{Schlegel}, David J. and {Finkbeiner}, Douglas P. and {Davis}, Marc},
        title = "{Maps of Dust Infrared Emission for Use in Estimation of Reddening and Cosmic Microwave Background Radiation Foregrounds}",
      journal = {\apj},
         year = 1998,
        month = jun,
       volume = {500},
       number = {2},
        pages = {525-553},
          doi = {10.1086/305772},
archivePrefix = {arXiv},
       eprint = {astro-ph/9710327},
 primaryClass = {astro-ph},
       adsurl = {https://ui.adsabs.harvard.edu/abs/1998ApJ...500..525S}
}

@ARTICLE{vaddi16,
       author = {{Vaddi}, Sravani and {O'Dea}, Christopher P. and {Baum}, Stefi A. and {Whitmore}, Samantha and {Ahmed}, Rabeea and {Pierce}, Katherine and {Leary}, Sara},
        title = "{Constraints on Feedback in the Local Universe: The Relation between Star Formation and AGN Activity in Early-type Galaxies}",
      journal = {\apj},
         year = 2016,
        month = feb,
       volume = {818},
       number = {2},
          eid = {182},
        pages = {182},
          doi = {10.3847/0004-637X/818/2/182},
archivePrefix = {arXiv},
       eprint = {1601.00344},
 primaryClass = {astro-ph.GA},
       adsurl = {https://ui.adsabs.harvard.edu/abs/2016ApJ...818..182V}
}

@ARTICLE{hao11,
       author = {{Hao}, Cai-Na and {Kennicutt}, Robert C. and {Johnson}, Benjamin D. and {Calzetti}, Daniela and {Dale}, Daniel A. and {Moustakas}, John},
        title = "{Dust-corrected Star Formation Rates of Galaxies. II. Combinations of Ultraviolet and Infrared Tracers}",
      journal = {\apj},
         year = 2011,
        month = nov,
       volume = {741},
       number = {2},
          eid = {124},
        pages = {124},
          doi = {10.1088/0004-637X/741/2/124},
archivePrefix = {arXiv},
       eprint = {1108.2837},
 primaryClass = {astro-ph.CO},
       adsurl = {https://ui.adsabs.harvard.edu/abs/2011ApJ...741..124H}
}

@ARTICLE{mcdonald18,
       author = {{McDonald}, M. and {Gaspari}, M. and {McNamara}, B.~R. and {Tremblay}, G.~R.},
        title = "{Revisiting the Cooling Flow Problem in Galaxies, Groups, and Clusters of Galaxies}",
      journal = {\apj},
         year = 2018,
        month = may,
       volume = {858},
       number = {1},
          eid = {45},
        pages = {45},
          doi = {10.3847/1538-4357/aabace},
archivePrefix = {arXiv},
       eprint = {1803.04972},
 primaryClass = {astro-ph.HE},
       adsurl = {https://ui.adsabs.harvard.edu/abs/2018ApJ...858...45M}
}

@ARTICLE{solomon05,
       author = {{Solomon}, P.~M. and {Vanden Bout}, P.~A.},
        title = "{Molecular Gas at High Redshift}",
      journal = {\araa},
         year = 2005,
        month = sep,
       volume = {43},
       number = {1},
        pages = {677-725},
          doi = {10.1146/annurev.astro.43.051804.102221},
archivePrefix = {arXiv},
       eprint = {astro-ph/0508481},
 primaryClass = {astro-ph},
       adsurl = {https://ui.adsabs.harvard.edu/abs/2005ARA&A..43..677S}
}

@ARTICLE{bolatto13,
       author = {{Bolatto}, Alberto D. and {Wolfire}, Mark and {Leroy}, Adam K.},
        title = "{The CO-to-H$_{2}$ Conversion Factor}",
      journal = {\araa},
         year = 2013,
        month = aug,
       volume = {51},
       number = {1},
        pages = {207-268},
          doi = {10.1146/annurev-astro-082812-140944},
archivePrefix = {arXiv},
       eprint = {1301.3498},
 primaryClass = {astro-ph.GA},
       adsurl = {https://ui.adsabs.harvard.edu/abs/2013ARA&A..51..207B}
}

@ARTICLE{edge01,
       author = {{Edge}, A.~C.},
        title = "{The detection of molecular gas in the central galaxies of cooling flow clusters}",
      journal = {\mnras},
         year = 2001,
        month = dec,
       volume = {328},
       number = {3},
        pages = {762-782},
          doi = {10.1046/j.1365-8711.2001.04802.x},
archivePrefix = {arXiv},
       eprint = {astro-ph/0106225},
 primaryClass = {astro-ph},
       adsurl = {https://ui.adsabs.harvard.edu/abs/2001MNRAS.328..762E}
}

@ARTICLE{russell19,
       author = {{Russell}, H.~R. and {McNamara}, B.~R. and {Fabian}, A.~C. and {Nulsen}, P.~E.~J. and {Combes}, F. and {Edge}, A.~C. and {Madar}, M. and {Olivares}, V. and {Salom{\'e}}, P. and {Vantyghem}, A.~N.},
        title = "{Driving massive molecular gas flows in central cluster galaxies with AGN feedback}",
      journal = {\mnras},
         year = 2019,
        month = dec,
       volume = {490},
       number = {3},
        pages = {3025-3045},
          doi = {10.1093/mnras/stz2719},
archivePrefix = {arXiv},
       eprint = {1902.09227},
 primaryClass = {astro-ph.GA},
       adsurl = {https://ui.adsabs.harvard.edu/abs/2019MNRAS.490.3025R}
}

@ARTICLE{di15,
       author = {{Di Teodoro}, E.~M. and {Fraternali}, F.},
        title = "{$^{3D}$ BAROLO: a new 3D algorithm to derive rotation curves of galaxies}",
      journal = {\mnras},
         year = 2015,
        month = aug,
       volume = {451},
       number = {3},
        pages = {3021-3033},
          doi = {10.1093/mnras/stv1213},
archivePrefix = {arXiv},
       eprint = {1505.07834},
 primaryClass = {astro-ph.GA},
       adsurl = {https://ui.adsabs.harvard.edu/abs/2015MNRAS.451.3021D}
}

@ARTICLE{mulchaey00,
       author = {{Mulchaey}, John S.},
        title = "{X-ray Properties of Groups of Galaxies}",
      journal = {\araa},
         year = 2000,
        month = jan,
       volume = {38},
        pages = {289-335},
          doi = {10.1146/annurev.astro.38.1.289},
archivePrefix = {arXiv},
       eprint = {astro-ph/0009379},
 primaryClass = {astro-ph},
       adsurl = {https://ui.adsabs.harvard.edu/abs/2000ARA&A..38..289M}
}

@ARTICLE{giodini09,
       author = {{Giodini}, S. and {Pierini}, D. and {Finoguenov}, A. and {Pratt}, G.~W. and {Boehringer}, H. and {Leauthaud}, A. and {Guzzo}, L. and {Aussel}, H. and {Bolzonella}, M. and {Capak}, P. and {Elvis}, M. and {Hasinger}, G. and {Ilbert}, O. and {Kartaltepe}, J.~S. and {Koekemoer}, A.~M. and {Lilly}, S.~J. and {Massey}, R. and {McCracken}, H.~J. and {Rhodes}, J. and {Salvato}, M. and {Sanders}, D.~B. and {Scoville}, N.~Z. and {Sasaki}, S. and {Smolcic}, V. and {Taniguchi}, Y. and {Thompson}, D. and {COSMOS Collaboration}},
        title = "{Stellar and Total Baryon Mass Fractions in Groups and Clusters Since Redshift 1}",
      journal = {\apj},
         year = 2009,
        month = sep,
       volume = {703},
       number = {1},
        pages = {982-993},
          doi = {10.1088/0004-637X/703/1/982},
archivePrefix = {arXiv},
       eprint = {0904.0448},
 primaryClass = {astro-ph.CO},
       adsurl = {https://ui.adsabs.harvard.edu/abs/2009ApJ...703..982G}
}

@ARTICLE{sun12,
       author = {{Sun}, M.},
        title = "{Hot gas in galaxy groups: recent observations}",
      journal = {New Journal of Physics},
         year = 2012,
        month = apr,
       volume = {14},
       number = {4},
          eid = {045004},
        pages = {045004},
          doi = {10.1088/1367-2630/14/4/045004},
archivePrefix = {arXiv},
       eprint = {1203.4228},
 primaryClass = {astro-ph.CO},
       adsurl = {https://ui.adsabs.harvard.edu/abs/2012NJPh...14d5004S}
}

@ARTICLE{osullivan18,
       author = {{O'Sullivan}, E. and {Combes}, F. and {Salom{\'e}}, P. and {David}, L.~P. and {Babul}, A. and {Vrtilek}, J.~M. and {Lim}, J. and {Olivares}, V. and {Raychaudhury}, S. and {Schellenberger}, G.},
        title = "{Cold gas in a complete sample of group-dominant early-type galaxies}",
      journal = {\aap},
         year = 2018,
        month = oct,
       volume = {618},
          eid = {A126},
        pages = {A126},
          doi = {10.1051/0004-6361/201833580},
archivePrefix = {arXiv},
       eprint = {1807.09110},
 primaryClass = {astro-ph.GA},
       adsurl = {https://ui.adsabs.harvard.edu/abs/2018A&A...618A.126O}
}

@ARTICLE{olivares22,
       author = {{Olivares}, V. and {Salom{\'e}}, P. and {Hamer}, S.~L. and {Combes}, F. and {Gaspari}, M. and {Kolokythas}, K. and {O'Sullivan}, E. and {Beckmann}, R.~S. and {Babul}, A. and {Polles}, F.~L. and {Lehnert}, M. and {Loubser}, S.~I. and {Donahue}, M. and {Gendron-Marsolais}, M.-L. and {Lagos}, P. and {Pineau des Forets}, G. and {Godard}, B. and {Rose}, T. and {Tremblay}, G. and {Ferland}, G. and {Guillard}, P.},
        title = "{Gas condensation in brightest group galaxies unveiled with MUSE. Morphology and kinematics of the ionized gas}",
      journal = {\aap},
         year = 2022,
        month = oct,
       volume = {666},
          eid = {A94},
        pages = {A94},
          doi = {10.1051/0004-6361/202142475},
archivePrefix = {arXiv},
       eprint = {2201.07838},
 primaryClass = {astro-ph.GA},
       adsurl = {https://ui.adsabs.harvard.edu/abs/2022A&A...666A..94O}
}

@ARTICLE{voit15a,
       author = {{Voit}, G. Mark and {Donahue}, Megan},
        title = "{Cooling Time, Freefall Time, and Precipitation in the Cores of ACCEPT Galaxy Clusters}",
      journal = {\apjl},
         year = 2015,
        month = jan,
       volume = {799},
       number = {1},
          eid = {L1},
        pages = {L1},
          doi = {10.1088/2041-8205/799/1/L1},
archivePrefix = {arXiv},
       eprint = {1409.1601},
 primaryClass = {astro-ph.GA},
       adsurl = {https://ui.adsabs.harvard.edu/abs/2015ApJ...799L...1V}
}

@ARTICLE{sharma12,
       author = {{Sharma}, Prateek and {McCourt}, Michael and {Quataert}, Eliot and {Parrish}, Ian J.},
        title = "{Thermal instability and the feedback regulation of hot haloes in clusters, groups and galaxies}",
      journal = {\mnras},
         year = 2012,
        month = mar,
       volume = {420},
       number = {4},
        pages = {3174-3194},
          doi = {10.1111/j.1365-2966.2011.20246.x},
archivePrefix = {arXiv},
       eprint = {1106.4816},
 primaryClass = {astro-ph.CO},
       adsurl = {https://ui.adsabs.harvard.edu/abs/2012MNRAS.420.3174S}
}

@ARTICLE{gaspari17,
       author = {{Gaspari}, M. and {Temi}, P. and {Brighenti}, F.},
        title = "{Raining on black holes and massive galaxies: the top-down multiphase condensation model}",
      journal = {\mnras},
         year = 2017,
        month = apr,
       volume = {466},
       number = {1},
        pages = {677-704},
          doi = {10.1093/mnras/stw3108},
archivePrefix = {arXiv},
       eprint = {1608.08216},
 primaryClass = {astro-ph.GA},
       adsurl = {https://ui.adsabs.harvard.edu/abs/2017MNRAS.466..677G}
}

@ARTICLE{gaspari12,
       author = {{Gaspari}, M. and {Ruszkowski}, M. and {Sharma}, P.},
        title = "{Cause and Effect of Feedback: Multiphase Gas in Cluster Cores Heated by AGN Jets}",
      journal = {\apj},
         year = 2012,
        month = feb,
       volume = {746},
       number = {1},
          eid = {94},
        pages = {94},
          doi = {10.1088/0004-637X/746/1/94},
archivePrefix = {arXiv},
       eprint = {1110.6063},
 primaryClass = {astro-ph.CO},
       adsurl = {https://ui.adsabs.harvard.edu/abs/2012ApJ...746...94G}
}

@ARTICLE{pulido18,
       author = {{Pulido}, F.~A. and {McNamara}, B.~R. and {Edge}, A.~C. and {Hogan}, M.~T. and {Vantyghem}, A.~N. and {Russell}, H.~R. and {Nulsen}, P.~E.~J. and {Babyk}, I. and {Salom{\'e}}, P.},
        title = "{The Origin of Molecular Clouds in Central Galaxies}",
      journal = {\apj},
         year = 2018,
        month = feb,
       volume = {853},
       number = {2},
          eid = {177},
        pages = {177},
          doi = {10.3847/1538-4357/aaa54b},
archivePrefix = {arXiv},
       eprint = {1710.04664},
 primaryClass = {astro-ph.GA},
       adsurl = {https://ui.adsabs.harvard.edu/abs/2018ApJ...853..177P}
}

@ARTICLE{olivares19,
       author = {{Olivares}, V. and {Salome}, P. and {Combes}, F. and {Hamer}, S. and {Guillard}, P. and {Lehnert}, M.~D. and {Polles}, F.~L. and {Beckmann}, R.~S. and {Dubois}, Y. and {Donahue}, M. and {Edge}, A. and {Fabian}, A.~C. and {McNamara}, B. and {Rose}, T. and {Russell}, H.~R. and {Tremblay}, G. and {Vantyghem}, A. and {Canning}, R.~E.~A. and {Ferland}, G. and {Godard}, B. and {Peirani}, S. and {Pineau des Forets}, G.},
        title = "{Ubiquitous cold and massive filaments in cool core clusters}",
      journal = {\aap},
         year = 2019,
        month = nov,
       volume = {631},
          eid = {A22},
        pages = {A22},
          doi = {10.1051/0004-6361/201935350},
archivePrefix = {arXiv},
       eprint = {1902.09164},
 primaryClass = {astro-ph.GA},
       adsurl = {https://ui.adsabs.harvard.edu/abs/2019A&A...631A..22O}
}

@ARTICLE{kewley01,
       author = {{Kewley}, L.~J. and {Dopita}, M.~A. and {Sutherland}, R.~S. and {Heisler}, C.~A. and {Trevena}, J.},
        title = "{Theoretical Modeling of Starburst Galaxies}",
      journal = {\apj},
         year = 2001,
        month = jul,
       volume = {556},
       number = {1},
        pages = {121-140},
          doi = {10.1086/321545},
archivePrefix = {arXiv},
       eprint = {astro-ph/0106324},
 primaryClass = {astro-ph},
       adsurl = {https://ui.adsabs.harvard.edu/abs/2001ApJ...556..121K}
}

@ARTICLE{kauffmann03,
       author = {{Kauffmann}, Guinevere and {Heckman}, Timothy M. and {Tremonti}, Christy and {Brinchmann}, Jarle and {Charlot}, St{\'e}phane and {White}, Simon D.~M. and {Ridgway}, Susan E. and {Brinkmann}, Jon and {Fukugita}, Masataka and {Hall}, Patrick B. and {Ivezi{\'c}}, {\v{Z}}eljko and {Richards}, Gordon T. and {Schneider}, Donald P.},
        title = "{The host galaxies of active galactic nuclei}",
      journal = {\mnras},
         year = 2003,
        month = dec,
       volume = {346},
       number = {4},
        pages = {1055-1077},
          doi = {10.1111/j.1365-2966.2003.07154.x},
archivePrefix = {arXiv},
       eprint = {astro-ph/0304239},
 primaryClass = {astro-ph},
       adsurl = {https://ui.adsabs.harvard.edu/abs/2003MNRAS.346.1055K}
}

@ARTICLE{schawinski07,
       author = {{Schawinski}, Kevin and {Thomas}, Daniel and {Sarzi}, Marc and {Maraston}, Claudia and {Kaviraj}, Sugata and {Joo}, Seok-Joo and {Yi}, Sukyoung K. and {Silk}, Joseph},
        title = "{Observational evidence for AGN feedback in early-type galaxies}",
      journal = {\mnras},
         year = 2007,
        month = dec,
       volume = {382},
       number = {4},
        pages = {1415-1431},
          doi = {10.1111/j.1365-2966.2007.12487.x},
archivePrefix = {arXiv},
       eprint = {0709.3015},
 primaryClass = {astro-ph},
       adsurl = {https://ui.adsabs.harvard.edu/abs/2007MNRAS.382.1415S}
}

@ARTICLE{meyer17,
       author = {{Meyer}, Martin and {Robotham}, Aaron and {Obreschkow}, Danail and {Westmeier}, Tobias and {Duffy}, Alan R. and {Staveley-Smith}, Lister},
        title = "{Tracing HI Beyond the Local Universe}",
      journal = {\pasa},
         year = 2017,
        month = nov,
       volume = {34},
        pages = {52},
          doi = {10.1017/pasa.2017.31},
archivePrefix = {arXiv},
       eprint = {1705.04210},
 primaryClass = {astro-ph.CO},
       adsurl = {https://ui.adsabs.harvard.edu/abs/2017PASA...34...52M}
}

@ARTICLE{cardelli89,
       author = {{Cardelli}, Jason A. and {Clayton}, Geoffrey C. and {Mathis}, John S.},
        title = "{The Relationship between Infrared, Optical, and Ultraviolet Extinction}",
      journal = {\apj},
         year = 1989,
        month = oct,
       volume = {345},
        pages = {245},
          doi = {10.1086/167900},
       adsurl = {https://ui.adsabs.harvard.edu/abs/1989ApJ...345..245C}
}

@ARTICLE{donahue26,
       author = {{Donahue}, Megan and {Koeppe}, Dana and {Frisbie}, Rachel and {Baldi}, Allesandro and {Voit}, G. Mark},
        title = "{ACCEPT 2.0: A Compilation of X-Ray Gas Properties for Clusters of Galaxies Observed by the Chandra X-Ray Observatory}",
      journal = {\apjs},
         year = 2026,
        month = feb,
       volume = {282},
       number = {2},
          eid = {61},
        pages = {61},
          doi = {10.3847/1538-4365/ae2c5f},
       adsurl = {https://ui.adsabs.harvard.edu/abs/2026ApJS..282...61D}
}

@ARTICLE{ueda14,
       author = {{Ueda}, Junko and {Iono}, Daisuke and {Yun}, Min S. and {Crocker}, Alison F. and {Narayanan}, Desika and {Komugi}, Shinya and {Espada}, Daniel and {Hatsukade}, Bunyo and {Kaneko}, Hiroyuki and {Matsuda}, Yuichi and {Tamura}, Yoichi and {Wilner}, David J. and {Kawabe}, Ryohei and {Pan}, Hsi-An},
        title = "{Cold Molecular Gas in Merger Remnants. I. Formation of Molecular Gas Disks}",
      journal = {\apjs},
         year = 2014,
        month = sep,
       volume = {214},
       number = {1},
          eid = {1},
        pages = {1},
          doi = {10.1088/0067-0049/214/1/1},
archivePrefix = {arXiv},
       eprint = {1407.6873},
 primaryClass = {astro-ph.GA},
       adsurl = {https://ui.adsabs.harvard.edu/abs/2014ApJS..214....1U}
}

@ARTICLE{voit18,
       author = {{Voit}, G. Mark},
        title = "{A Role for Turbulence in Circumgalactic Precipitation}",
      journal = {\apj},
         year = 2018,
        month = dec,
       volume = {868},
       number = {2},
          eid = {102},
        pages = {102},
          doi = {10.3847/1538-4357/aae8e2},
archivePrefix = {arXiv},
       eprint = {1803.06036},
 primaryClass = {astro-ph.GA},
       adsurl = {https://ui.adsabs.harvard.edu/abs/2018ApJ...868..102V}
}

@ARTICLE{eckert21,
       author = {{Eckert}, Dominique and {Gaspari}, Massimo and {Gastaldello}, Fabio and {Le Brun}, Amandine M.~C. and {O'Sullivan}, Ewan},
        title = "{Feedback from Active Galactic Nuclei in Galaxy Groups}",
      journal = {Universe},
         year = 2021,
        month = may,
       volume = {7},
       number = {5},
          eid = {142},
        pages = {142},
          doi = {10.3390/universe7050142},
archivePrefix = {arXiv},
       eprint = {2106.13259},
 primaryClass = {astro-ph.GA},
       adsurl = {https://ui.adsabs.harvard.edu/abs/2021Univ....7..142E}
}

@ARTICLE{mcnamara16,
       author = {{McNamara}, B.~R. and {Russell}, H.~R. and {Nulsen}, P.~E.~J. and {Hogan}, M.~T. and {Fabian}, A.~C. and {Pulido}, F. and {Edge}, A.~C.},
        title = "{A Mechanism for Stimulating AGN Feedback by Lifting Gas in Massive Galaxies}",
      journal = {\apj},
         year = 2016,
        month = oct,
       volume = {830},
       number = {2},
          eid = {79},
        pages = {79},
          doi = {10.3847/0004-637X/830/2/79},
archivePrefix = {arXiv},
       eprint = {1604.04629},
 primaryClass = {astro-ph.GA},
       adsurl = {https://ui.adsabs.harvard.edu/abs/2016ApJ...830...79M}
}

@ARTICLE{sobacchi19,
       author = {{Sobacchi}, Emanuele and {Sormani}, Mattia C.},
        title = "{The effect of rotation on the thermal instability of stratified galactic atmospheres - I. Local analysis}",
      journal = {\mnras},
         year = 2019,
        month = jun,
       volume = {486},
       number = {1},
        pages = {205-214},
          doi = {10.1093/mnras/stz792},
archivePrefix = {arXiv},
       eprint = {1903.06172},
 primaryClass = {astro-ph.GA},
       adsurl = {https://ui.adsabs.harvard.edu/abs/2019MNRAS.486..205S}
}

@ARTICLE{juranova19,
       author = {{Jur{\'a}{\v{n}}ov{\'a}}, A. and {Werner}, N. and {Gaspari}, M. and {Lakhchaura}, K. and {Nulsen}, P.~E.~J. and {Sun}, M. and {Canning}, R.~E.~A. and {Allen}, S.~W. and {Simionescu}, A. and {Oonk}, J.~B.~R. and {Connor}, T. and {Donahue}, M.},
        title = "{Cooling in the X-ray halo of the rotating, massive early-type galaxy NGC 7049}",
      journal = {\mnras},
         year = 2019,
        month = apr,
       volume = {484},
       number = {2},
        pages = {2886-2895},
          doi = {10.1093/mnras/stz185},
archivePrefix = {arXiv},
       eprint = {1808.05761},
 primaryClass = {astro-ph.GA},
       adsurl = {https://ui.adsabs.harvard.edu/abs/2019MNRAS.484.2886J}
}

@ARTICLE{gaspari15,
       author = {{Gaspari}, M. and {Brighenti}, F. and {Temi}, P.},
        title = "{Chaotic cold accretion on to black holes in rotating atmospheres}",
      journal = {\aap},
         year = 2015,
        month = jul,
       volume = {579},
          eid = {A62},
        pages = {A62},
          doi = {10.1051/0004-6361/201526151},
archivePrefix = {arXiv},
       eprint = {1407.7531},
 primaryClass = {astro-ph.GA},
       adsurl = {https://ui.adsabs.harvard.edu/abs/2015A&A...579A..62G}
}

@ARTICLE{canning13,
       author = {{Canning}, R.~E.~A. and {Sun}, M. and {Sanders}, J.~S. and {Clarke}, T.~E. and {Fabian}, A.~C. and {Giacintucci}, S. and {Lal}, D.~V. and {Werner}, N. and {Allen}, S.~W. and {Donahue}, M. and {Edge}, A.~C. and {Johnstone}, R.~M. and {Nulsen}, P.~E.~J. and {Salom{\'e}}, P. and {Sarazin}, C.~L.},
        title = "{A multiwavelength view of cooling versus AGN heating in the X-ray luminous cool-core of Abell 3581}",
      journal = {\mnras},
         year = 2013,
        month = oct,
       volume = {435},
       number = {2},
        pages = {1108-1125},
          doi = {10.1093/mnras/stt1345},
archivePrefix = {arXiv},
       eprint = {1305.0050},
 primaryClass = {astro-ph.CO},
       adsurl = {https://ui.adsabs.harvard.edu/abs/2013MNRAS.435.1108C}
}

@ARTICLE{parriott08,
       author = {{Parriott}, Joel R. and {Bregman}, Joel N.},
        title = "{Mass Loss from Evolved Stars in Elliptical Galaxies}",
      journal = {\apj},
         year = 2008,
        month = jul,
       volume = {681},
       number = {2},
        pages = {1215-1232},
          doi = {10.1086/588033},
archivePrefix = {arXiv},
       eprint = {0805.4331},
 primaryClass = {astro-ph},
       adsurl = {https://ui.adsabs.harvard.edu/abs/2008ApJ...681.1215P}
}

@ARTICLE{bregman09,
       author = {{Bregman}, Joel N. and {Parriott}, Joel R.},
        title = "{Mass Loss from Planetary Nebulae in Elliptical Galaxies}",
      journal = {\apj},
         year = 2009,
        month = jul,
       volume = {699},
       number = {2},
        pages = {923-932},
          doi = {10.1088/0004-637X/699/2/923},
       adsurl = {https://ui.adsabs.harvard.edu/abs/2009ApJ...699..923B}
}

@ARTICLE{ciotti91,
       author = {{Ciotti}, Luca and {D'Ercole}, Annibale and {Pellegrini}, Silvia and {Renzini}, Alvio},
        title = "{Winds, Outflows, and Inflows in X-Ray Elliptical Galaxies. I.}",
      journal = {\apj},
         year = 1991,
        month = aug,
       volume = {376},
        pages = {380},
          doi = {10.1086/170289},
       adsurl = {https://ui.adsabs.harvard.edu/abs/1991ApJ...376..380C}
}

@ARTICLE{juravnova20,
       author = {{Jur{\'a}{\v{n}}ov{\'a}}, A. and {Werner}, N. and {Nulsen}, P.~E.~J. and {Gaspari}, M. and {Lakhchaura}, K. and {Canning}, R.~E.~A. and {Donahue}, M. and {Hroch}, F. and {Voit}, G.~M.},
        title = "{Hot gaseous atmospheres of rotating galaxies observed with XMM-Newton}",
      journal = {\mnras},
         year = 2020,
        month = dec,
       volume = {499},
       number = {4},
        pages = {5163-5174},
          doi = {10.1093/mnras/staa3182},
archivePrefix = {arXiv},
       eprint = {2008.01161},
 primaryClass = {astro-ph.GA},
       adsurl = {https://ui.adsabs.harvard.edu/abs/2020MNRAS.499.5163J}
}

@INPROCEEDINGS{jozsa20a,
       author = {{J{\'o}zsa}, G.~I.~G. and {White}, S.~V. and {Thorat}, K. and {Smirnov}, O.~M. and {Serra}, P. and {Ramatsoku}, M. and {Ramaila}, A.~J.~T. and {Perkins}, S.~J. and {Maccagni}, F.~M. and {Makhathini}, S. and {Moln{\'a}r}, D.~C. and {Kamphuis}, P. and {Kleiner}, D. and {Hugo}, B.~V. and {de Blok}, W.~J.~G. and {Andati}, L.~A.~L.},
        title = "{MeerKATHI - an End-to-End Data Reduction Pipeline for MeerKAT and Other Radio Telescopes}",
    booktitle = {Astronomical Data Analysis Software and Systems XXIX},
         year = 2020,
       editor = {{Pizzo}, R. and {Deul}, E.~R. and {Mol}, J.~D. and {de Plaa}, J. and {Verkouter}, H.},
       series = {Astronomical Society of the Pacific Conference Series},
       volume = {527},
        month = jan,
        pages = {635},
          doi = {10.48550/arXiv.2006.02955},
archivePrefix = {arXiv},
       eprint = {2006.02955},
 primaryClass = {astro-ph.IM},
       adsurl = {https://ui.adsabs.harvard.edu/abs/2020ASPC..527..635J}
}

@software{jozsa20b,
       author = {{J{\'o}zsa}, Gyula I.~G. and {White}, Sarah V. and {Thorat}, Kshitij and {Smirnov}, Oleg M. and {Serra}, Paolo and {Ramatsoku}, Mpati and {Ramaila}, Athanaseus J.~T. and {Perkins}, Simon J. and {Moln{\'a}r}, D{\'a}niel Cs. and {Makhathini}, Sphesihle and {Maccagni}, Filippo M. and {Kleiner}, Dane and {Kamphuis}, Peter and {Hugo}, Benjamin V. and {de Blok}, W.~J.~G. and {Andati}, Lexy A.~L.},
        title = "{CARACal: Containerized Automated Radio Astronomy Calibration pipeline}",
 howpublished = {Astrophysics Source Code Library, record ascl:2006.014},
         year = 2020,
        month = jun,
          eid = {ascl:2006.014},
archivePrefix = {ascl},
       eprint = {2006.014},
       adsurl = {https://ui.adsabs.harvard.edu/abs/2020ascl.soft06014J}
}

@INPROCEEDINGS{fruscione06,
       author = {{Fruscione}, Antonella and {McDowell}, Jonathan C. and {Allen}, Glenn E. and {Brickhouse}, Nancy S. and {Burke}, Douglas J. and {Davis}, John E. and {Durham}, Nick and {Elvis}, Martin and {Galle}, Elizabeth C. and {Harris}, Daniel E. and {Huenemoerder}, David P. and {Houck}, John C. and {Ishibashi}, Bish and {Karovska}, Margarita and {Nicastro}, Fabrizio and {Noble}, Michael S. and {Nowak}, Michael A. and {Primini}, Frank A. and {Siemiginowska}, Aneta and {Smith}, Randall K. and {Wise}, Michael},
        title = "{CIAO: Chandra's data analysis system}",
    booktitle = {Observatory Operations: Strategies, Processes, and Systems},
         year = 2006,
       editor = {{Silva}, David R. and {Doxsey}, Rodger E.},
       series = {Society of Photo-Optical Instrumentation Engineers (SPIE) Conference Series},
       volume = {6270},
        month = jun,
          eid = {62701V},
        pages = {62701V},
          doi = {10.1117/12.671760},
       adsurl = {https://ui.adsabs.harvard.edu/abs/2006SPIE.6270E..1VF}
}

@ARTICLE{casa22,
       author = {{CASA Team} and {Bean}, Ben and {Bhatnagar}, Sanjay and {Castro}, Sandra and {Donovan Meyer}, Jennifer and {Emonts}, Bjorn and {Garcia}, Enrique and {Garwood}, Robert and {Golap}, Kumar and {Gonzalez Villalba}, Justo and {Harris}, Pamela and {Hayashi}, Yohei and {Hoskins}, Josh and {Hsieh}, Mingyu and {Jagannathan}, Preshanth and {Kawasaki}, Wataru and {Keimpema}, Aard and {Kettenis}, Mark and {Lopez}, Jorge and {Marvil}, Joshua and {Masters}, Joseph and {McNichols}, Andrew and {Mehringer}, David and {Miel}, Renaud and {Moellenbrock}, George and {Montesino}, Federico and {Nakazato}, Takeshi and {Ott}, Juergen and {Petry}, Dirk and {Pokorny}, Martin and {Raba}, Ryan and {Rau}, Urvashi and {Schiebel}, Darrell and {Schweighart}, Neal and {Sekhar}, Srikrishna and {Shimada}, Kazuhiko and {Small}, Des and {Steeb}, Jan-Willem and {Sugimoto}, Kanako and {Suoranta}, Ville and {Tsutsumi}, Takahiro and {van Bemmel}, Ilse M. and {Verkouter}, Marjolein and {Wells}, Akeem and {Xiong}, Wei and {Szomoru}, Arpad and {Griffith}, Morgan and {Glendenning}, Brian and {Kern}, Jeff},
        title = "{CASA, the Common Astronomy Software Applications for Radio Astronomy}",
      journal = {\pasp},
         year = 2022,
        month = nov,
       volume = {134},
       number = {1041},
          eid = {114501},
        pages = {114501},
          doi = {10.1088/1538-3873/ac9642},
archivePrefix = {arXiv},
       eprint = {2210.02276},
 primaryClass = {astro-ph.IM},
       adsurl = {https://ui.adsabs.harvard.edu/abs/2022PASP..134k4501C}
}

@ARTICLE{prasad22,
       author = {{Prasad}, Deovrat and {Voit}, G. Mark and {O'Shea}, Brian W.},
        title = "{Atmospheric Circulation in Simulations of the AGN-CGM Connection at Halo Masses 10$^{13.5}$ M $_{☉}$}",
      journal = {\apj},
         year = 2022,
        month = jun,
       volume = {932},
       number = {1},
          eid = {18},
        pages = {18},
          doi = {10.3847/1538-4357/ac69ee},
archivePrefix = {arXiv},
       eprint = {2110.15382},
 primaryClass = {astro-ph.GA},
       adsurl = {https://ui.adsabs.harvard.edu/abs/2022ApJ...932...18P}
}

@ARTICLE{kauffman08,
       author = {{Kauffmann}, J. and {Bertoldi}, F. and {Bourke}, T.~L. and {Evans}, II, N.~J. and {Lee}, C.~W.},
        title = "{MAMBO mapping of Spitzer c2d small clouds and cores}",
      journal = {\aap},
         year = 2008,
        month = sep,
       volume = {487},
       number = {3},
        pages = {993-1017},
          doi = {10.1051/0004-6361:200809481},
archivePrefix = {arXiv},
       eprint = {0805.4205},
 primaryClass = {astro-ph},
       adsurl = {https://ui.adsabs.harvard.edu/abs/2008A&A...487..993K}
}

@ARTICLE{lovisari15,
       author = {{Lovisari}, L. and {Reiprich}, T.~H. and {Schellenberger}, G.},
        title = "{Scaling properties of a complete X-ray selected galaxy group sample}",
      journal = {\aap},
         year = 2015,
        month = jan,
       volume = {573},
          eid = {A118},
        pages = {A118},
          doi = {10.1051/0004-6361/201423954},
archivePrefix = {arXiv},
       eprint = {1409.3845},
 primaryClass = {astro-ph.CO},
       adsurl = {https://ui.adsabs.harvard.edu/abs/2015A&A...573A.118L}
}

@ARTICLE{temi09,
       author = {{Temi}, Pasquale and {Brighenti}, Fabrizio and {Mathews}, William G.},
        title = "{Spitzer Observations of Passive and Star-Forming Early-Type Galaxies: An Infrared Color-Color Sequence}",
      journal = {\apj},
         year = 2009,
        month = dec,
       volume = {707},
       number = {2},
        pages = {890-902},
          doi = {10.1088/0004-637X/707/2/890},
archivePrefix = {arXiv},
       eprint = {0911.0720},
 primaryClass = {astro-ph.CO},
       adsurl = {https://ui.adsabs.harvard.edu/abs/2009ApJ...707..890T}
}

@ARTICLE{malin83,
       author = {{Malin}, D.~F. and {Carter}, D.},
        title = "{A catalog of elliptical galaxies with shells.}",
      journal = {\apj},
         year = 1983,
        month = nov,
       volume = {274},
        pages = {534-540},
          doi = {10.1086/161467},
       adsurl = {https://ui.adsabs.harvard.edu/abs/1983ApJ...274..534M}
}

@ARTICLE{babyk23,
       author = {{Babyk}, Iu. V. and {McNamara}, B.~R. and {Nulsen}, P.~E.~J. and {Russell}, H.~R. and {Edge}, A.~C. and {Blitz}, Leo},
        title = "{Atmospheric Pressure and Molecular Cloud Formation in Early-type Galaxies}",
      journal = {\apj},
         year = 2023,
        month = feb,
       volume = {944},
       number = {1},
          eid = {69},
        pages = {69},
          doi = {10.3847/1538-4357/acb53f},
archivePrefix = {arXiv},
       eprint = {2110.15410},
 primaryClass = {astro-ph.GA},
       adsurl = {https://ui.adsabs.harvard.edu/abs/2023ApJ...944...69B}
}

@ARTICLE{mcconnell13,
       author = {{McConnell}, Nicholas J. and {Ma}, Chung-Pei},
        title = "{Revisiting the Scaling Relations of Black Hole Masses and Host Galaxy Properties}",
      journal = {\apj},
         year = 2013,
        month = feb,
       volume = {764},
       number = {2},
          eid = {184},
        pages = {184},
          doi = {10.1088/0004-637X/764/2/184},
archivePrefix = {arXiv},
       eprint = {1211.2816},
 primaryClass = {astro-ph.CO},
       adsurl = {https://ui.adsabs.harvard.edu/abs/2013ApJ...764..184M}
}

@ARTICLE{voit17,
       author = {{Voit}, G. Mark and {Meece}, Greg and {Li}, Yuan and {O'Shea}, Brian W. and {Bryan}, Greg L. and {Donahue}, Megan},
        title = "{A Global Model for Circumgalactic and Cluster-core Precipitation}",
      journal = {\apj},
         year = 2017,
        month = aug,
       volume = {845},
       number = {1},
          eid = {80},
        pages = {80},
          doi = {10.3847/1538-4357/aa7d04},
archivePrefix = {arXiv},
       eprint = {1607.02212},
 primaryClass = {astro-ph.GA},
       adsurl = {https://ui.adsabs.harvard.edu/abs/2017ApJ...845...80V}
}

@ARTICLE{serra15,
       author = {{Serra}, Paolo and {Westmeier}, Tobias and {Giese}, Nadine and {Jurek}, Russell and {Fl{\"o}er}, Lars and {Popping}, Attila and {Winkel}, Benjamin and {van der Hulst}, Thijs and {Meyer}, Martin and {Koribalski}, B{\"a}rbel S. and {Staveley-Smith}, Lister and {Courtois}, H{\'e}l{\`e}ne},
        title = "{SOFIA: a flexible source finder for 3D spectral line data}",
      journal = {\mnras},
         year = 2015,
        month = apr,
       volume = {448},
       number = {2},
        pages = {1922-1929},
          doi = {10.1093/mnras/stv079},
archivePrefix = {arXiv},
       eprint = {1501.03906},
 primaryClass = {astro-ph.IM},
       adsurl = {https://ui.adsabs.harvard.edu/abs/2015MNRAS.448.1922S}
}

@ARTICLE{westmeier21,
       author = {{Westmeier}, T. and {Kitaeff}, S. and {Pallot}, D. and {Serra}, P. and {van der Hulst}, J.~M. and {Jurek}, R.~J. and {Elagali}, A. and {For}, B.-Q. and {Kleiner}, D. and {Koribalski}, B.~S. and {Lee-Waddell}, K. and {Mould}, J.~R. and {Reynolds}, T.~N. and {Rhee}, J. and {Staveley-Smith}, L.},
        title = "{SOFIA 2 - an automated, parallel H I source finding pipeline for the WALLABY survey}",
      journal = {\mnras},
         year = 2021,
        month = sep,
       volume = {506},
       number = {3},
        pages = {3962-3976},
          doi = {10.1093/mnras/stab1881},
archivePrefix = {arXiv},
       eprint = {2106.15789},
 primaryClass = {astro-ph.IM},
       adsurl = {https://ui.adsabs.harvard.edu/abs/2021MNRAS.506.3962W}
}

@ARTICLE{maccagni23,
       author = {{Maccagni}, F.~M. and {Ruffa}, I. and {Loni}, A. and {Prandoni}, I. and {Ragusa}, R. and {Kleiner}, D. and {Serra}, P. and {Iodice}, E. and {Spavone}, M.},
        title = "{The AGN fuelling/feedback cycle in nearby radio galaxies. V. The cold atomic gas of NGC 3100 and its group}",
      journal = {\aap},
         year = 2023,
        month = jul,
       volume = {675},
          eid = {A59},
        pages = {A59},
          doi = {10.1051/0004-6361/202346521},
archivePrefix = {arXiv},
       eprint = {2305.01075},
 primaryClass = {astro-ph.GA},
       adsurl = {https://ui.adsabs.harvard.edu/abs/2023A&A...675A..59M}
}

@software{conley16,
       author = {{Conley}, Alexander},
        title = "{mbb\_emcee: Modified Blackbody MCMC}",
 howpublished = {Astrophysics Source Code Library, record ascl:1602.020},
         year = 2016,
        month = feb,
          eid = {ascl:1602.020},
archivePrefix = {ascl},
       eprint = {1602.020},
       adsurl = {https://ui.adsabs.harvard.edu/abs/2016ascl.soft02020C}
}

@ARTICLE{tamhane26,
       author = {{Tamhane}, Prathamesh and {Sun}, Ming and {Waldron}, William and {Hosogi}, Kokoro and {da Silva}, Patricia and {Le}, Huan and {Gaspari}, Massimo and {Combes}, Francoise and {Werner}, Norbert and {Schellenberger}, Gerrit and {Fabian}, Andrew and {Canning}, Rebecca and {David}, Laurence and {Donahue}, Megan and {Voit}, Mark},
        title = "{HST view of NGC 5044: Constraints on filament widths, magnetic support, multiphase structure, and comparison with cluster environments}",
      journal = {\pasa},
         year = 2026,
        month = mar,
       volume = {43},
          eid = {e036},
        pages = {e036},
          doi = {10.1017/pasa.2026.10172},
archivePrefix = {arXiv},
       eprint = {2603.00249},
 primaryClass = {astro-ph.GA},
       adsurl = {https://ui.adsabs.harvard.edu/abs/2026PASA...43...36T}
}

@ARTICLE{asplund09,
       author = {{Asplund}, Martin and {Grevesse}, Nicolas and {Sauval}, A. Jacques and {Scott}, Pat},
        title = "{The Chemical Composition of the Sun}",
      journal = {\araa},
         year = 2009,
        month = sep,
       volume = {47},
       number = {1},
        pages = {481-522},
          doi = {10.1146/annurev.astro.46.060407.145222},
archivePrefix = {arXiv},
       eprint = {0909.0948},
 primaryClass = {astro-ph.SR},
       adsurl = {https://ui.adsabs.harvard.edu/abs/2009ARA&A..47..481A}
}

@ARTICLE{anders89,
       author = {{Anders}, E. and {Grevesse}, N.},
        title = "{Abundances of the elements: Meteoritic and solar}",
      journal = {\gca},
         year = 1989,
        month = jan,
       volume = {53},
       number = {1},
        pages = {197-214},
          doi = {10.1016/0016-7037(89)90286-X},
       adsurl = {https://ui.adsabs.harvard.edu/abs/1989GeCoA..53..197A}
}

@ARTICLE{kumari19,
       author = {{Kumari}, Nimisha and {Maiolino}, Roberto and {Belfiore}, Francesco and {Curti}, Mirko},
        title = "{Metallicity calibrations for diffuse ionized gas and low-ionization emission regions}",
      journal = {\mnras},
         year = 2019,
        month = may,
       volume = {485},
       number = {1},
        pages = {367-381},
          doi = {10.1093/mnras/stz366},
archivePrefix = {arXiv},
       eprint = {1902.01408},
 primaryClass = {astro-ph.GA},
       adsurl = {https://ui.adsabs.harvard.edu/abs/2019MNRAS.485..367K}
}

@ARTICLE{blitz06,
       author = {{Blitz}, Leo and {Rosolowsky}, Erik},
        title = "{The Role of Pressure in GMC Formation II: The H$_{2}$-Pressure Relation}",
      journal = {\apj},
         year = 2006,
        month = oct,
       volume = {650},
       number = {2},
        pages = {933-944},
          doi = {10.1086/505417},
archivePrefix = {arXiv},
       eprint = {astro-ph/0605035},
 primaryClass = {astro-ph},
       adsurl = {https://ui.adsabs.harvard.edu/abs/2006ApJ...650..933B}
}

@ARTICLE{leroy08,
       author = {{Leroy}, Adam K. and {Walter}, Fabian and {Brinks}, Elias and {Bigiel}, Frank and {de Blok}, W.~J.~G. and {Madore}, Barry and {Thornley}, M.~D.},
        title = "{The Star Formation Efficiency in Nearby Galaxies: Measuring Where Gas Forms Stars Effectively}",
      journal = {\aj},
         year = 2008,
        month = dec,
       volume = {136},
       number = {6},
        pages = {2782-2845},
          doi = {10.1088/0004-6256/136/6/2782},
archivePrefix = {arXiv},
       eprint = {0810.2556},
 primaryClass = {astro-ph},
       adsurl = {https://ui.adsabs.harvard.edu/abs/2008AJ....136.2782L}
}

@ARTICLE{schellenberger24,
       author = {{Schellenberger}, Gerrit and {O'Sullivan}, Ewan and {David}, Laurence P. and {Vrtilek}, Jan and {Romero}, Charles and {Petitpas}, Glen and {Forman}, William and {Giacintucci}, Simona and {Gurwell}, Mark and {Jones}, Christine and {Rajpurohit}, Kamlesh and {Ubertosi}, Francesco and {Venturi}, Tiziana},
        title = "{Probing the High-frequency Variability of NGC 5044: The Key to Active Galactic Nucleus Feedback}",
      journal = {\apj},
         year = 2024,
        month = dec,
       volume = {976},
       number = {2},
          eid = {246},
        pages = {246},
          doi = {10.3847/1538-4357/ad89bc},
archivePrefix = {arXiv},
       eprint = {2409.06039},
 primaryClass = {astro-ph.GA},
       adsurl = {https://ui.adsabs.harvard.edu/abs/2024ApJ...976..246S}
}

@ARTICLE{grevesse98,
       author = {{Grevesse}, N. and {Sauval}, A.~J.},
        title = "{Standard Solar Composition}",
      journal = {\ssr},
         year = 1998,
        month = may,
       volume = {85},
        pages = {161-174},
          doi = {10.1023/A:1005161325181},
       adsurl = {https://ui.adsabs.harvard.edu/abs/1998SSRv...85..161G}
}

@ARTICLE{schure09,
       author = {{Schure}, K.~M. and {Kosenko}, D. and {Kaastra}, J.~S. and {Keppens}, R. and {Vink}, J.},
        title = "{A new radiative cooling curve based on an up-to-date plasma emission code}",
      journal = {\aap},
         year = 2009,
        month = dec,
       volume = {508},
       number = {2},
        pages = {751-757},
          doi = {10.1051/0004-6361/200912495},
archivePrefix = {arXiv},
       eprint = {0909.5204},
 primaryClass = {astro-ph.GA},
       adsurl = {https://ui.adsabs.harvard.edu/abs/2009A&A...508..751S}
}

@ARTICLE{stofanova21,
       author = {{{\v{S}}tofanov{\'a}}, L. and {Kaastra}, J. and {Mehdipour}, M. and {de Plaa}, J.},
        title = "{New radiative loss curve from updates to collisional excitation in the low-density, optically thin plasmas in SPEX}",
      journal = {\aap},
         year = 2021,
        month = nov,
       volume = {655},
          eid = {A2},
        pages = {A2},
          doi = {10.1051/0004-6361/202037768},
archivePrefix = {arXiv},
       eprint = {2109.04184},
 primaryClass = {astro-ph.HE},
       adsurl = {https://ui.adsabs.harvard.edu/abs/2021A&A...655A...2S}
}

@ARTICLE{sun09,
       author = {{Sun}, M.},
        title = "{Every BCG with a Strong Radio Agn has an X-Ray Cool Core: Is the Cool Core-Noncool Core Dichotomy Too Simple?}",
      journal = {\apj},
         year = 2009,
        month = oct,
       volume = {704},
       number = {2},
        pages = {1586-1604},
          doi = {10.1088/0004-637X/704/2/1586},
archivePrefix = {arXiv},
       eprint = {0904.2006},
 primaryClass = {astro-ph.CO},
       adsurl = {https://ui.adsabs.harvard.edu/abs/2009ApJ...704.1586S}
}

@ARTICLE{scott22,
       author = {{Scott}, T.~C. and {Cortese}, L. and {Lagos}, P. and {Brinks}, E. and {Finoguenov}, A. and {Coccato}, L.},
        title = "{FGC 1287 and its enigmatic 250 kpc long HI tail in the outskirts of Abell 1367}",
      journal = {\mnras},
         year = 2022,
        month = mar,
       volume = {511},
       number = {1},
        pages = {980-993},
          doi = {10.1093/mnras/stac118},
archivePrefix = {arXiv},
       eprint = {2201.04763},
 primaryClass = {astro-ph.GA},
       adsurl = {https://ui.adsabs.harvard.edu/abs/2022MNRAS.511..980S}
}

@ARTICLE{pak24,
       author = {{Pak}, Mina and {Baek}, Junhyun and {Lee}, Joon Hyeop and {Chung}, Aeree and {Owers}, Matt and {Jeong}, Hyunjin and {Sung}, Eon-Chang and {Sheen}, Yun-Kyeong},
        title = "{Discovery of a Large-scale H I Plume in the NGC 7194 Group}",
      journal = {\apjl},
         year = 2024,
        month = jan,
       volume = {961},
       number = {1},
          eid = {L11},
        pages = {L11},
          doi = {10.3847/2041-8213/ad0ed2},
archivePrefix = {arXiv},
       eprint = {2312.09567},
 primaryClass = {astro-ph.GA},
       adsurl = {https://ui.adsabs.harvard.edu/abs/2024ApJ...961L..11P}
}

@ARTICLE{serra13,
       author = {{Serra}, Paolo and {Koribalski}, B{\"a}rbel and {Duc}, Pierre-Alain and {Oosterloo}, Tom and {McDermid}, Richard M. and {Michel-Dansac}, Leo and {Emsellem}, Eric and {Cuillandre}, Jean-Charles and {Alatalo}, Katherine and {Blitz}, Leo and {Bois}, Maxime and {Bournaud}, Fr{\'e}d{\'e}ric and {Bureau}, Martin and {Cappellari}, Michele and {Crocker}, Alison F. and {Davies}, Roger L. and {Davis}, Timothy A. and {de Zeeuw}, P.~T. and {Khochfar}, Sadegh and {Krajnovi{\'c}}, Davor and {Kuntschner}, Harald and {Lablanche}, Pierre-Yves and {Morganti}, Raffaella and {Naab}, Thorsten and {Sarzi}, Marc and {Scott}, Nicholas and {Weijmans}, Anne-Marie and {Young}, Lisa M.},
        title = "{Discovery of a giant HI tail in the galaxy group HCG 44}",
      journal = {\mnras},
         year = 2013,
        month = jan,
       volume = {428},
       number = {1},
        pages = {370-380},
          doi = {10.1093/mnras/sts033},
archivePrefix = {arXiv},
       eprint = {1209.4107},
 primaryClass = {astro-ph.CO},
       adsurl = {https://ui.adsabs.harvard.edu/abs/2013MNRAS.428..370S}
}

\appendix
\counterwithout{figure}{section}     
\setcounter{figure}{0}               
\renewcommand{\thefigure}{A\arabic{figure}}

\counterwithout{table}{section}      
\setcounter{table}{0}                
\renewcommand{\thetable}{A\arabic{table}}

    \section{Discrepancy in cooling time estimates}
    \label{sec:discrepancy}

    The thermodynamic properties of the hot atmosphere of NGC~4936 have previously been presented by \citetalias{babyk18}, \citetalias{lakhchaura18}, and \citetalias{donahue26}. The deprojected temperature profiles are broadly consistent across the three studies, while the density, pressure, and entropy profiles show a modest offset in \citetalias{lakhchaura18} relative to \citetalias{babyk18} and \citetalias{donahue26}. This offset appears to arise from a difference in the density convention. \citetalias{lakhchaura18} label their profiles as total particle density, $n$, but the reported values agree with the electron densities, $n_e$, of \citetalias{babyk18} and \citetalias{donahue26} without conversion. Interpreting these values as total densities and converting them to electron densities instead introduces a systematic offset. We therefore tentatively interpret the \citetalias{lakhchaura18} values as electron densities and under this assumption, the density, pressure, and entropy profiles agree with those of \citetalias{babyk18} and \citetalias{donahue26}.

    The published cooling time profiles from these works nevertheless differ appreciably. An additional source of uncertainty is the adopted metallicity and solar abundance table, as discussed in Section~\ref{sec:xray}. To compare the derived thermodynamic properties on a common basis, we recalculated the pressure, entropy, cooling time, and $t_{\rm cool}/t_{\rm ff}$ profiles from the published deprojected temperature and electron density measurements, adopting the density interpretation described above and a common set of assumptions. This comparison isolates differences introduced when deriving these quantities from the input profiles, although the published densities retain their dependence on the assumptions used in the original spectral fits.
    
    We then calculated the thermal pressure and entropy using the standard relations
    \begin{equation}
        P = n kT \simeq 1.92\,n_{\rm e}kT,
        \label{eq:pressure_recalc}
    \end{equation}
    and
    \begin{equation}
        K = kT n_{\rm e}^{-2/3},
        \label{eq:entropy_recalc}
    \end{equation}
    respectively, where $n$ is the total particle number density. For a fully ionised plasma with approximately cosmic H/He composition, we adopt $n=1.92n_{\rm e}$ and $n_{\rm H}=0.83n_{\rm e}$, where $n_{\rm H}$ is the number density of hydrogen nuclei. The resulting pressure and entropy profiles are consistent among the three studies when calculated directly from their published $kT$ and $n_{\rm e}$ measurements.
    
    The radiative cooling time was calculated as
    \begin{equation}
        t_{\rm cool} =
        \frac{3}{2}
        \frac{n kT}
        {n_{\rm e}n_{\rm H}\Lambda(T,Z)},
        \label{eq:tcool_recalc}
    \end{equation}
    where $\Lambda(T,Z)$ is the plasma cooling function. We calculated $\Lambda(T,Z)$ directly with \textsc{xspec} using the \textsc{apec} plasma emission model with AtomDB v3.1.3 and the \citet{anders89} solar abundance table. For consistency between the three profiles, we adopted a fixed metallicity of $Z=0.2Z_\odot$.
    This is close to the metallicity of 0.22Z$_\odot$ assumed by \citetalias{babyk18}. \citetalias{lakhchaura18} state that they measured a metallicity within the central 10 kpc and applied this value to all radial bins, but do not report the adopted value. We therefore use $0.2Z_\odot$ for their profile as well. The same metallicity is adopted for the \citetalias{donahue26} profile so that differences between the recalculated cooling times are driven by the published temperature and density profiles rather than by different abundance assumptions.
    
    To evaluate $\Lambda(T,Z)$, we generated a dummy response in \textsc{xspec} v12.15.1 from 0.01 to 100 keV with 10,000 linearly spaced energy bins using \texttt{dummyrsp}. At the temperature of each radial bin, we evaluated an \textsc{apec} model with the abundance fixed at $0.2Z_\odot$, $z = 0.010742$, and the model normalisation fixed to 1. We then integrated the model flux over 0.01--100 keV. The \textsc{apec} normalisation is defined as
    \begin{equation}
        {\rm norm} =
        \frac{10^{-14}}
        {4\pi [D_{\rm A}(1+z)]^2}
        \int n_{\rm e}n_{\rm H}\,dV ,
        \label{eq:apec_norm}
    \end{equation}
    where $D_{\rm A}$ is the angular-diameter distance. For a unit normalisation, the bolometric flux $F_{\rm bol}$ therefore gives the cooling function through
    \begin{equation}
        \Lambda(T,Z)
        =
        10^{-14}(1+z)^2 F_{\rm bol},
        \label{eq:lambda_apec}
    \end{equation}
    in units of erg cm$^{3}$ s$^{-1}$. This procedure directly provides the bolometric emissivity per $n_{\rm e}n_{\rm H}$ for the adopted temperature and
    metallicity.

    This resulted in approximately 40\% lower $t_{\rm cool}$ compared to those estimated by \citetalias{babyk18} and \citetalias{lakhchaura18}, and $\sim67$\% lower values compared to \citetalias{donahue26}. One possible explanation for the remaining discrepancy is the treatment of the bolometric cooling luminosity. If the earlier analyses evaluated the thermal model only over the energy range covered by the instrumental response (without running ``\texttt{dummyrsp 0.01 100}'' or ``\texttt{energies 0.01 100}'' before calculating the bolometric flux in \textsc{xspec}) while integrating the nominal bolometric flux over a much broader band, the bolometric luminosity would be underestimated and the corresponding cooling time overestimated. This effect would be most significant for lower temperature systems, where a larger fraction of the thermal emission lies below the \textit{Chandra} response range. This was indeed confirmed for the \citetalias{donahue26} results.
    If the previous studies for other sources were affected in this way, their reported cooling times may require systematic corrections.

    The adopted cooling function in different works thus introduces an additional systematic uncertainty. At the sub-keV temperatures relevant to NGC~4936, metal-line emission contributes substantially to radiative cooling, making $\Lambda(T,Z)$ sensitive to the elemental abundances and the underlying atomic data. Comparisons of cooling calculations illustrate the dependence on the plasma code and its implementation \citep{schure09,stofanova21}. 
    At $kT=0.5-1$ keV, the difference can be up to 30\%.
    Since $t_{\rm cool}\propto\Lambda^{-1}$ at fixed temperature and density, these differences propagate directly into the inferred cooling times and $t_{\rm cool}/t_{\rm ff}$ ratios. Our use of a common cooling function ensures consistency between the recalculated profiles, but does not remove this model dependence. Comparisons with simulations should therefore account for differences in the cooling prescriptions and abundance assumptions.

    \section{Stellar velocity and velocity dispersion maps}
    \label{sec:stellar}
\begin{figure*}
        \centering
        \includegraphics[width=0.9\textwidth]{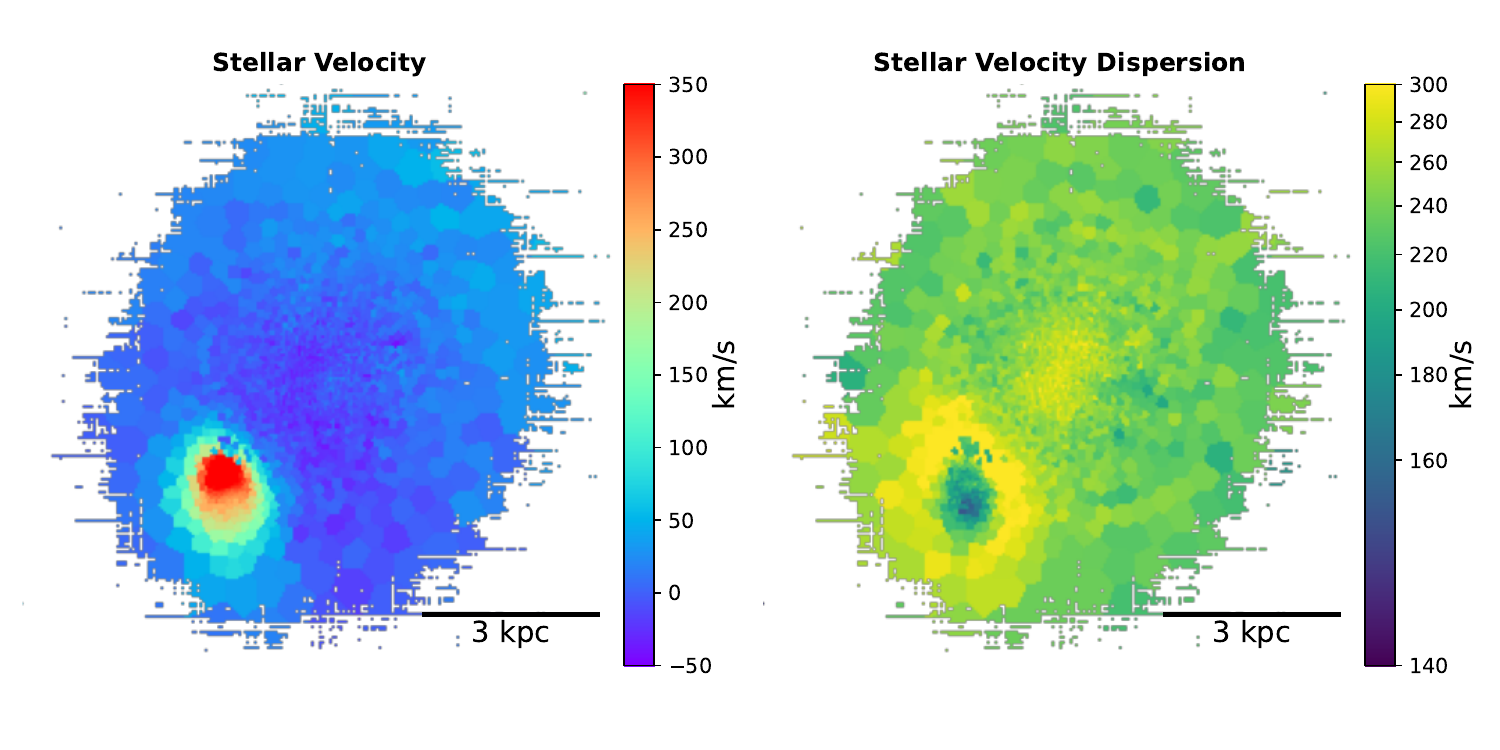}
        \caption{Maps of stellar velocity and velocity dispersion in the voronoi binned datacube. This data shows the $\sim 360$ km s$^{-1}$ velocity difference between NGC 4936 and the neighbouring galaxy.}
        \label{fig:stellar}
    \end{figure*}

    In the Figure~\ref{fig:stellar}, we show stellar velocity and velocity dispersion maps obtained from the MUSE datacube. The maps clearly show the $\sim 360$ km s$^{-1}$ separation in velocity space between NGC 4936 and the companion galaxy. The companion galaxy has a much smaller velocity dispersion ($180 \pm 16$ km s$^{-1}$) compared to NGC 4936 ($241 \pm 17$ km s$^{-1}$). Within a 1$''$ aperture centred on the nucleus of NGC 4936, we measure a stellar velocity dispersion of 275$\pm$11 km s$^{-1}$. Adopting the $M_{\rm BH}-\sigma$ relation of \citet{mcconnell13}, this corresponds to a central SMBH mass of $\sim 1.3 \times 10^9$ M$_\odot$.

\begin{figure*}
        \centering
        \includegraphics[width=\textwidth]{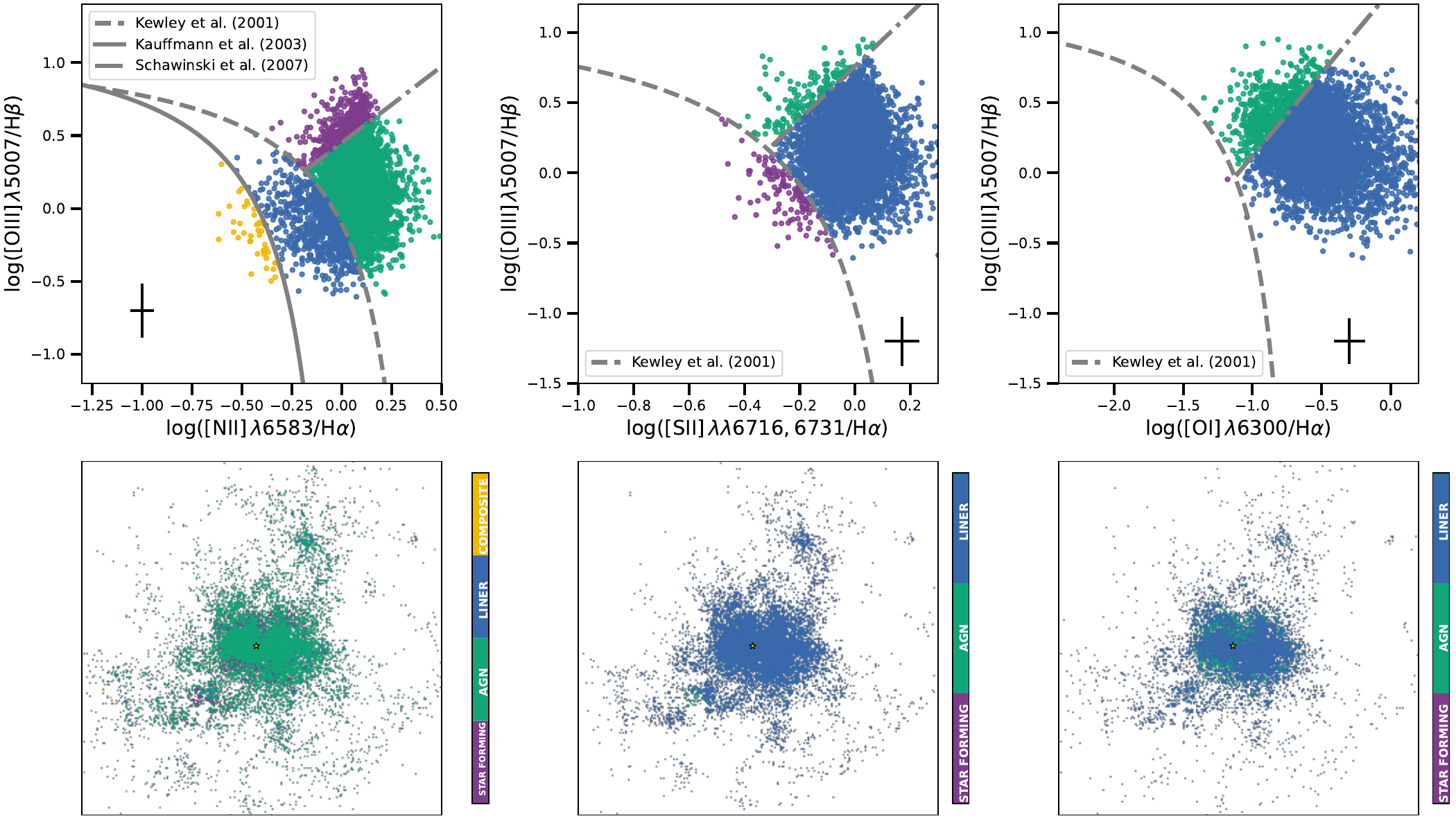}
        \caption{Standard BPT diagnostic diagrams based on the [N\,II]/H$\alpha$ (left), [S\,II]/H$\alpha$ (middle), and [O\,I]/H$\alpha$ (right) line ratios, respectively, together with the [O\,III]$\lambda5007$/H$\beta$ ratio. The top panels display the diagnostic diagrams, while the bottom panels show the spatial distribution of the corresponding spaxels within the galaxy. The spaxels are colour coded according to the classification boundaries defined by \citet{kewley01}, \citet{kauffmann03}, and \citet{schawinski07}. The right panels have a field of view of $10.7 \times 10.8$ kpc centred on the nucleus, which is marked by a yellow star. North is up and east is to the left. The error bars shown in the left panels represent the average uncertainties of all data points.}
        \label{fig:bpt}
    \end{figure*}

    Additionally, we also show BPT diagram derived from MUSE maps in Figure~\ref{fig:bpt}.

    \section{Extended H\,\textsc{i} emission and Blue Sources}
    \label{sec:blue_sources}

\begin{figure*}
        \centering
        \includegraphics[width=\textwidth]{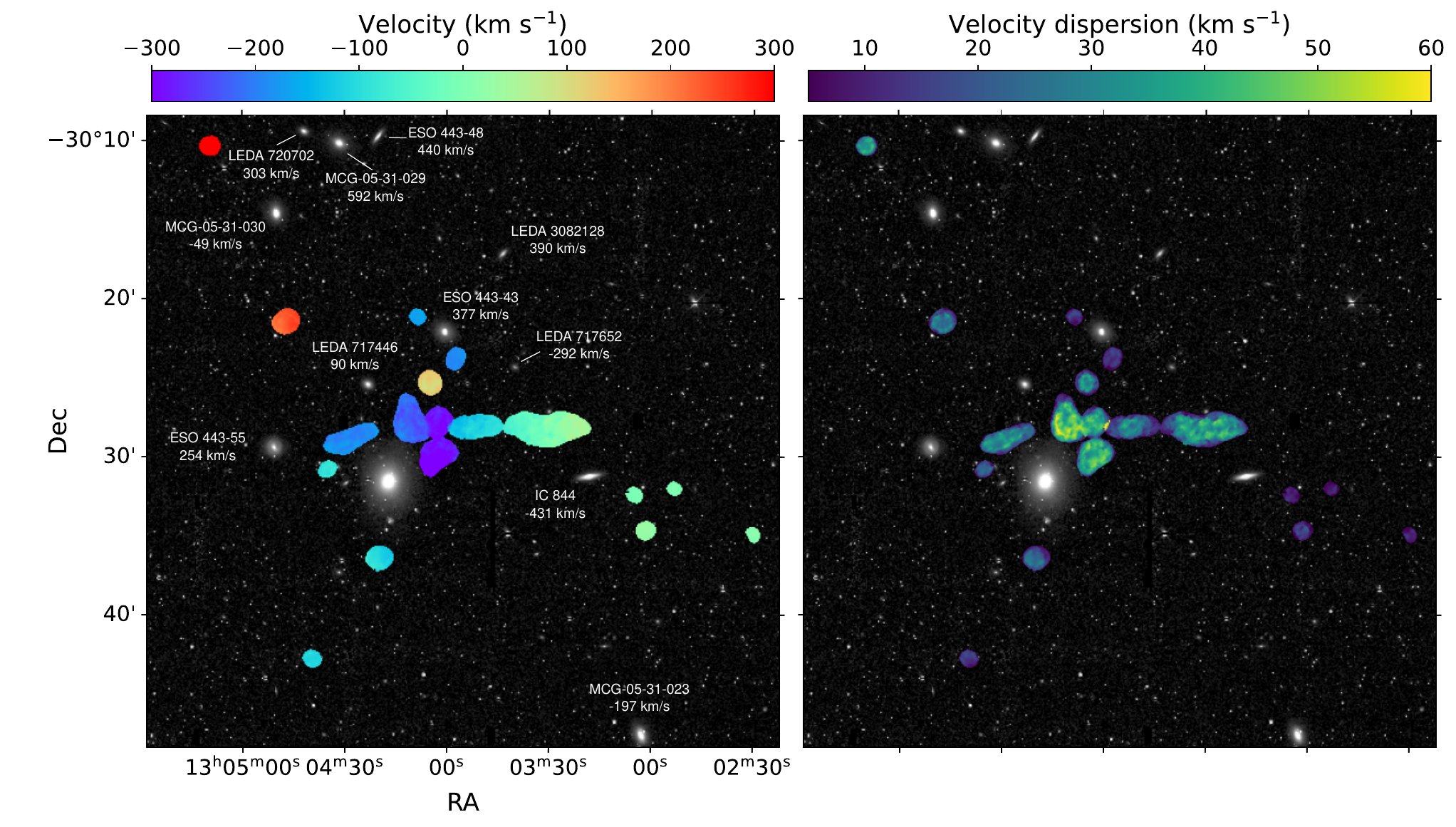}
        \caption{Velocity and velocity dispersion of the H\,\textsc{i} gas are shown in the left and right panels, respectively, in the rest frame of NGC 4936. The velocities of galaxies identified in the field of view in the rest frame of NGC 4936 are also shown in the left panel for comparison. The background in both panels is the DECaLS $g$-band image.
        }
        \label{fig:HI_kin}
    \end{figure*}

\begin{figure*}
    \centering
        \includegraphics[width=\textwidth]{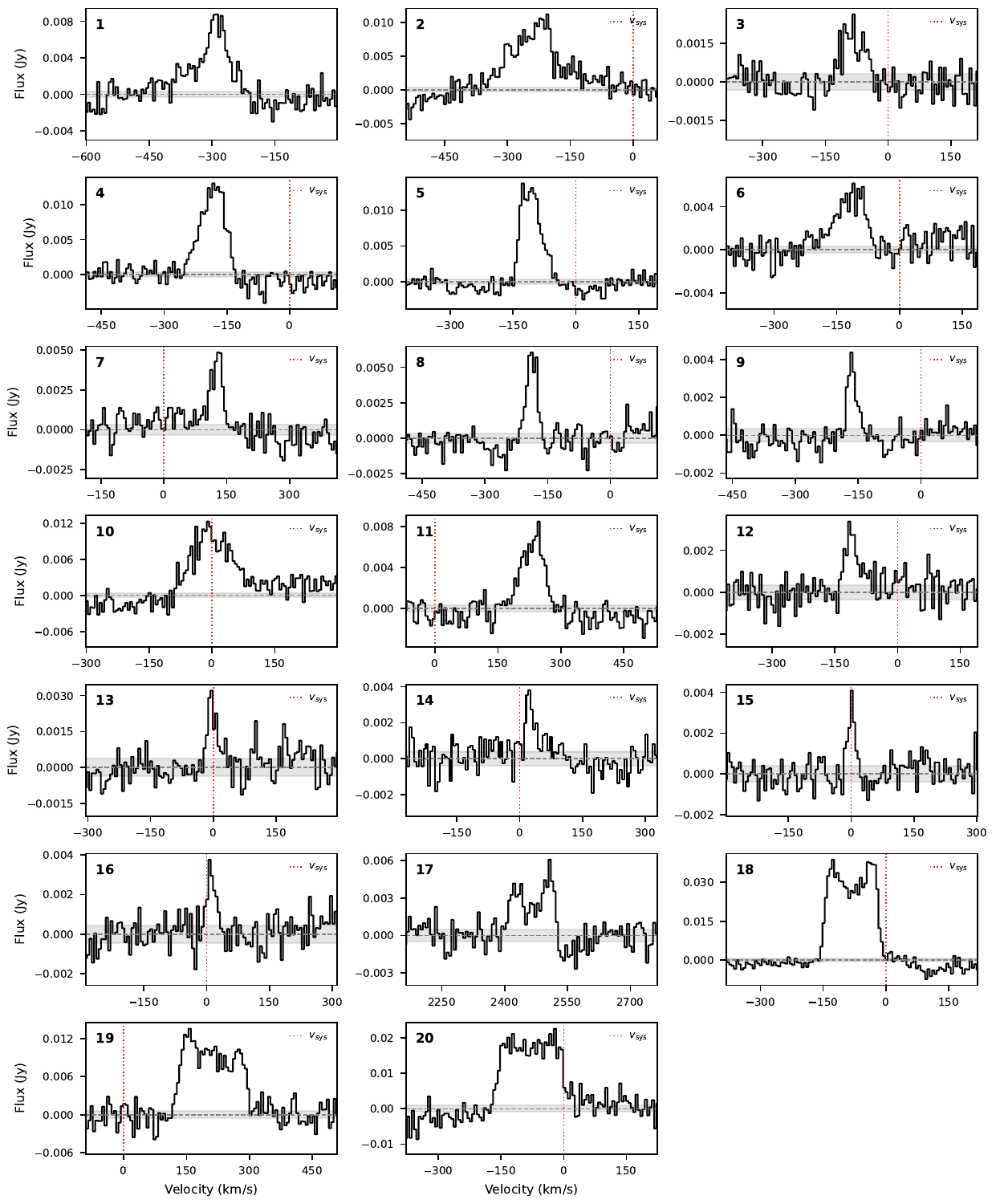}
        \label{fig:HI_spectra}
        \caption{H\,\textsc{i} spectra of the 20 sources identified with \textsc{SoFiA}, extracted from the main H\,\textsc{i} cube using each source's spatial mask and summed over the spatial axes. Source numbers correspond to those in Table~\ref{tab:hi_sources} and are ordered by projected distance from NGC~4936 (nearest to farthest). Velocity is shown relative to the systemic velocity of NGC~4936 ($v_{\rm sys} = 3220$~km\,s$^{-1}$), with each panel spanning $\pm 300$~km\,s$^{-1}$ around the source's own line centre. The red dotted line marks $v_{\rm sys}$ where it falls within the plotted range. The shaded gray region shows the 1$\sigma$ rms noise. Velocities are in the optical convention.
        }
    \end{figure*}

    The blind \textsc{SoFiA} search identified 35 candidate detections, which were visually inspected to remove spurious sources associated with noise and imaging artefacts. We retain 20 H\,\textsc{i} detections. The spatial distribution and optical counterparts of the H\,\textsc{i} detections are shown in Figure~\ref{fig:HI_sources}, while their velocity and velocity-dispersion structure are presented in Figure~\ref{fig:HI_kin}. The individual H\,\textsc{i} spectra are shown in Figure~\ref{fig:HI_spectra}, and the measured properties of all retained detections are listed in Table~\ref{tab:hi_sources}. With the exception of Source~17, their line-of-sight velocities lie between approximately $-300$ and $+230$ km s$^{-1}$ relative to NGC~4936, consistent with possible association with the wider group environment.

    \begin{figure*}
        \centering
        \includegraphics[width=0.8\textwidth]{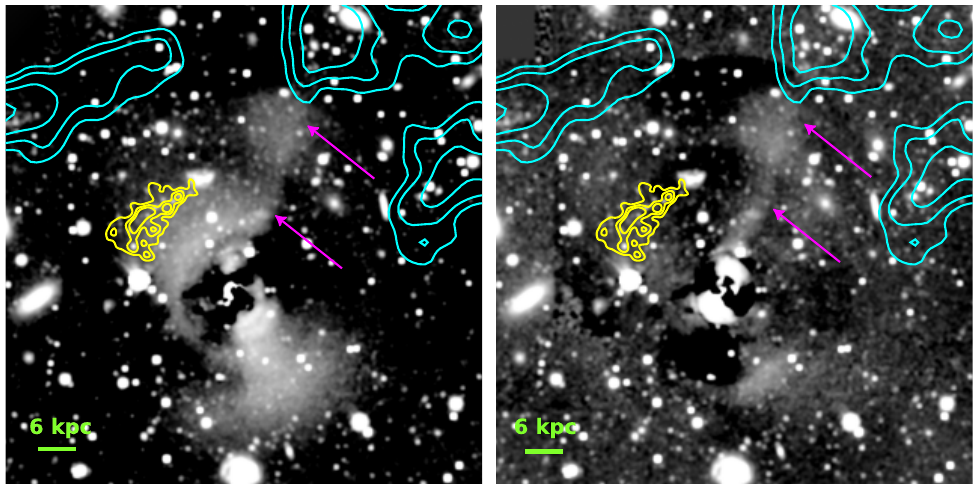}
        \caption{Left: DECaLS $g$-band continuum-subtracted image of NGC~4936 produced with \textsc{GALFIT}. Right: corresponding continuum-subtracted image obtained using \textsc{photutils}. Both images are smoothed by a Gaussian kernel with $\sigma = 5$ pixels. A diffuse low surface brightness plume extending to the N--NW is visible in both residual images and is highlighted by the magenta arrows. Cyan contours show the H\,\textsc{i} emission and yellow contours trace the radio halo. The persistence of the plume under two independent stellar light subtraction methods suggests that it may represent a genuine tidal feature rather than a subtraction artifact. North is up and east is left.
        }
        \label{fig:tidal}
    \end{figure*}

    Several H\,\textsc{i} structures have plausible optical counterparts. Sources~2, 11, 18, 19, and 20 coincide with catalogued galaxies, while Sources 4, 5, 7 and 12 are associated with blue optical sources identified in DECaLS imaging. The nature of these blue sources is uncertain. However, Sources 4, 5 and 7 are detected in GALEX ultraviolet imaging, suggesting that they host recent or ongoing star formation. They could represent low-mass or dwarf galaxies, compact star-forming condensations formed in tidally stripped gas, or stellar systems produced or displaced during interactions. In Source 4, the H\,\textsc{i} extends asymmetrically away from the optical counterpart in a tail-like morphology, suggestive of gas stripping or tidal disturbance. Some other extended H\,\textsc{i} clouds, including Sources~1, 3, 6, 8--10 and 13--16, have no optical counterpart, indicating that a substantial fraction of the atomic gas may reside in stripped or displaced intragroup structures rather than in regular galactic disks. Sources~18, 19, and 20 lie beyond $R_{500}$, although their velocities remain close to that of NGC~4936 and they may trace the wider group environment. Source~17 is offset by $\sim2444$ km s$^{-1}$ and is therefore unlikely to be associated with the NGC~4936 group. Overall, the spatially extended, asymmetric, and partially optically dark H\,\textsc{i} distribution suggests substantial redistribution of atomic gas through tidal interactions and/or stripping within the group. Similar structures have also been found in other group and cluster environments \citep[e.g.,][]{serra13,scott22,pak24}.

    \begin{sidewaystable*}
    \centering
    \begin{tabular}{c c c c c c c c c c}
    \hline
    ID & RA$_{\rm peak}$ & Dec$_{\rm peak}$ & $v$ & $\sigma_v$ & $M_{\rm HI}$ & Area & $d_{\rm proj}$ & SNR & Optical counterpart \\
     & & & (km s$^{-1})$ & (km s$^{-1})$ & ($M_\odot$) & & (kpc) & & \\
    \hline
    1 & 13:04:04.1 & -30:30:09.5 & -301.4 & 23.4 & $(3.31 \pm 0.17)\times10^{8}$ & 66.9$''\times$44.0$''$ & 42.9 & 13.5 & No \\ 
    2 & 13:04:11.1 & -30:28:04.5 & -242.5 & 47.2 & $(5.87 \pm 0.30)\times10^{8}$ & 104.5$''$$\times$66.2$''$ & 51.0 & 17.5 & WISEA J130409.29-302752.5 (spiral, no $z$) $^{\rm \S}$ \\ 
    3 & 13:04:35.1 & -30:30:49.4 & -87.0 & 21.0 & $(5.76 \pm 0.34)\times10^{7}$ & 33.6$''$$\times$24.7$''$ & 53.8 & 6.4 & No \\ 
    4 & 13:04:32.4 & -30:29:09.4 & -187.1 & 25.9 & $(4.81 \pm 0.25)\times10^{8}$ & 103.4$''$$\times$29.6$''$ & 55.7 & 21.4 & Irregular knotty blue source (no $z$) \\ 
    5 & 13:04:20.4 & -30:36:29.5 & -105.6 & 23.9 & $(4.35 \pm 0.22)\times10^{8}$ & 36.2$''$$\times$29.8$''$ & 67.8 & 28.6 & Blue source, no $z$ \\ 
    6 & 13:03:52.5 & -30:28:24.4 & -114.4 & 34.2 & $(2.04 \pm 0.11)\times10^{8}$ & 98.3$''$$\times$36.0$''$ & 84.4 & 10.3 & No \\ 
    7 & 13:04:04.9 & -30:25:19.5 & 113.7 & 13.8 & $(1.04 \pm 0.06)\times10^{8}$ & 36.8$''$$\times$33.3$''$ & 92.7 & 7.2 & Faint  blue source, no $z$\\ 
    8 & 13:03:57.2 & -30:23:39.4 & -188.3 & 12.2 & $(9.78 \pm 0.53)\times10^{7}$ & 34.9$''$$\times$24.7$''$ & 123.1 & 10.0 & No \\ 
    9 & 13:04:08.0 & -30:21:14.5 & -165.1 & 9.2 & $(5.81 \pm 0.33)\times10^{7}$ & 25.5$''$$\times$22.7$''$ & 143.8 & 7.6 & No \\ 
    10 & 13:03:27.0 & -30:28:03.9 & -1.2 & 45.7 & $(6.11 \pm 0.31)\times10^{8}$ & 158.8$''$$\times$52.3$''$ & 155.1 & 17.1 & No \\ 
    11 & 13:04:47.0 & -30:21:34.3 & 230.0 & 24.5 & $(2.38 \pm 0.12)\times10^{8}$ & 36.5$''$$\times$30.0$''$ & 162.6 & 13.8 & FLASH J130447.24-302136.8 ($v_{\rm rel} = 246$ km/s) \\ 
    12 & 13:04:39.8 & -30:42:44.4 & -109.6 & 12.7 & $(5.41 \pm 0.33)\times10^{7}$ & 31.0$''$$\times$24.5$''$ & 166.4 & 6.3 & Potential very faint blue source, no $z$ \\ 
    13 & 13:03:04.9 & -30:32:23.3 & -4.4 & 10.0 & $(3.89 \pm 0.25)\times10^{7}$ & 27.9$''$$\times$25.0$''$ & 212.8 & 5.1 & No\\ 
    14 & 13:03:01.0 & -30:34:38.1 & 28.5 & 8.8 & $(5.45 \pm 0.35)\times10^{7}$ & 34.5$''$$\times$29.8$''$ & 227.8 & 5.2 & No\\ 
    15 & 13:02:53.3 & -30:31:57.8 & 1.3 & 6.3 & $(3.89 \pm 0.24)\times10^{7}$ & 23.6$''$$\times$19.3$''$ & 246.7 & 5.7 & No\\ 
    16 & 13:02:30.0 & -30:34:51.8 & 11.8 & 8.5 & $(4.53 \pm 0.28)\times10^{7}$ & 27.4$''$$\times$22.2$''$ & 318.2 & 5.5 & No\\ 
    17 & 13:05:09.7 & -30:10:23.8 & 2444.3 $^\dagger$ & 43.8 & $(1.97 \pm 0.11)\times10^{8}$ & 29.6$''$$\times$25.2$''$ & 328.3 & 10.1 & Faint blue source, no $z$ \\ 
    18 & 13:02:10.8 & -30:26:25.7 & -81.8 & 54.6 & $(2.42 \pm 0.12)\times10^{9}$ & 62.6$''$$\times$54.9$''$ & 378.3 $^\star$ & 60.3 & ESO 443-36 (spiral, $v_{\rm rel} = -96$ km/s)\\ 
    19 & 13:05:23.5 & -30:06:08.4 & 209.4 & 64.4 & $(8.82 \pm 0.45)\times10^{8}$ & 42.0$''$$\times$35.4$''$ & 398.9 $^\star$ & 24.5 & MCG-05-31-031 (spiral, $v_{\rm rel}=263$ km/s) \\ 
    20 & 13:06:09.3 & -30:01:56.4 & -76.6 & 64.6 & $(1.58 \pm 0.08)\times10^{9}$ & 51.9$''$$\times$36.1$''$ & 523.0 $^\star$ & 24.8 & ESO 443-64 (spiral, $v_{\rm rel}=-52$ km/s)\\
    \hline
    \end{tabular}
    \vspace{0.5mm}
    \caption{Properties of the valid H\,\textsc{i} detections identified with \textsc{SoFiA} and retained after visual inspection. Columns give the source ID; coordinates of the peak flux pixel; line-of-sight velocity $v$ relative to the systemic velocity of NGC~4936; Gaussian-equivalent velocity dispersion $\sigma_v=W_{50}/2.355$, where $W_{50}$ is the \textsc{SoFiA} measured line width; H\,\textsc{i} mass $M_{\rm HI}$ estimated using Equation~\ref{eqn:hi_mass}; projected H\,\textsc{i} size from the \textsc{SoFiA} $3\sigma$ major and minor axes; projected distance ($d_{\rm proj}$) from NGC~4936; signal-to-noise ratio (SNR) derived by \textsc{SoFiA}; and the presence/absence of a spatially coincident optical counterpart. Optical counterparts were identified by visually comparing the H\,\textsc{i} source positions with DECaLS DR10 imaging, with galaxy identifications obtained from NED and SIMBAD. Where available, galaxy redshifts were converted to velocities in the rest frame of NGC 4936 ($v_{\rm rel}$).\protect\footnotemark\\
    $^{\rm \S}$: This galaxy lacks a spectroscopic redshift and lies close to the projected edge of the H\,\textsc{i} cloud. Its morphology and apparent  size resemble those of another spiral galaxy FLASH J130416.57-303402.9 ($z=0.06480$) to the south of NGC 4936. Therefore, despite its projected coincidence with the H\,\textsc{i} emission, it may instead be an unrelated background galaxy.\\
    $^\dagger$: Likely not part of the group environment.\\
    $^\star$: Outside $R_{500}$ radius.\\
    }
    \label{tab:hi_sources}
    \end{sidewaystable*}
    \footnotetext{\url{https://ned.ipac.caltech.edu/}}

    \section{Tidal features in NGC 4936?}
    \label{sec:tidal}
    To investigate whether the nearby companion is physically interacting with NGC~4936, we examined continuum subtracted DECaLS $g$-band images using two independent stellar continuum models constructed with \textsc{GALFIT} and \textsc{photutils}, shown in Figure~\ref{fig:tidal}. Both residual images reveal a diffuse low surface brightness plume extending $\sim$35 kpc to the N--NW of NGC~4936. Some shell like residuals closer to the galaxy are also visible, however, they are not consistent in the two images and may arise from imperfect subtraction. However, extended plume is recovered with both methods and is therefore more likely to represent genuine stellar substructure. The plume provides evidence that NGC~4936 has experienced a recent or ongoing tidal interaction. However, morphology alone does not uniquely establish that the plume was produced by the present companion, and hence does not by itself confirm an ongoing merger. Interestingly, the plume extends toward H\,\textsc{i} structures with comparatively enhanced velocity dispersion, which may provide additional evidence for disturbance in the surrounding group environment.

\end{document}